%% file: 0_main.tex
\documentclass[sigconf]{acmart}
\AtBeginDocument{%
  \providecommand\BibTeX{{%
    \normalfont B\kern-0.5em{\scshape i\kern-0.25em b}\kern-0.8em\TeX}}}

\acmConference[ArXiv]{Conference}{preprint}{2024}
\setcopyright{none}
\acmDOI{}
\acmPrice{}
\acmISBN{}

\usepackage[normalem]{ulem}

\usepackage{colortbl}

\begin{document}

\title{Designing LLM Chains by Adapting Techniques from Crowdsourcing Workflows}

\author{Madeleine Grunde-McLaughlin$^{\dagger}$, Michelle S. Lam$^{\S}$, Ranjay Krishna$^{\dagger\ddagger}$, Daniel S. Weld$^{\dagger\ddagger}$, Jeffrey Heer$^{\dagger}$}
\affiliation{
    \institution{$^{\dagger}$University of Washington \quad $^{\S}$Stanford University \quad $^{\ddagger}$Allen Institute for AI}
    \city{}
    \state{}
    \country{}
}
\email{{mgrunde,ranjay,jheer}@cs.washington.edu, mlam4@cs.stanford.edu, danw@allenai.org}

\renewcommand{\shortauthors}{Grunde-McLaughlin, et al.}

\begin{abstract}
\input{sections/00_abstract}
\end{abstract}

\begin{teaserfigure}
  \includegraphics[width=\textwidth]{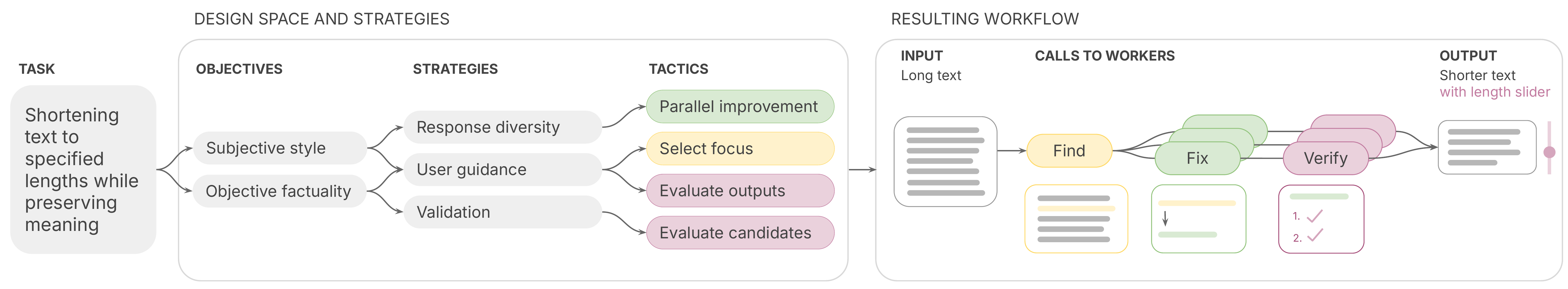}
  \caption{We contribute (1) a design space, (2) case studies, and (3) a discussion of techniques for LLM chains informed by crowdsourcing workflows. This scaffolding can help designers navigate the large possible space of LLM chains. For example, (Left) given a task of shortening text, as in Soylent~\cite{bernstein2010soylent}, our design space aids an LLM chain designer in identifying relevant high-level \emph{objectives}. These objectives incorporate elements of \emph{subjective} and \emph{objective} quality i.e., creatively shortening input text while verifying its faithfulness to the original. To support these \emph{objectives}, the designer can lean upon concrete \emph{strategies}, such as \emph{validation} and \emph{user guidance}. Strategies are high level design patterns encapsulating how prior work has manipulated workflow design to achieve \emph{objectives}.  These strategies in turn point to lower-level \emph{tactics} that define the chain’s construction, such as a \emph{subtask} to \emph{evaluate} text edit candidates. (Right) The designer can produce LLM chains to support the \emph{objectives} by implementing \emph{strategies} using \emph{tactics}, as outlined in our design space.  For example, promoting a high quality \emph{subjective} style can be supported by the \emph{user guidance} strategy, implemented by giving users control to \emph{evaluate} and select the content of the shortened text via a directly manipulable slider.
  }
  \Description{Three components: input task, design space, and resulting workflow. Input task is shortening text to specified lengths while preserving meaning. Design space flows through 3 main steps 1) Objectives which are to have a subjective style while maintaining objective factuality, 2) strategies, which are for response diversity, user guidance, and validation, and 3) tactics of: parallel generation, step to select focus, evaluate outputs, and evaluate candidates. After design space comes the resulting workflow. This workflow inputs a long text, finds areas to edit, fixes those areas, verifies those fixes, and outputs a shorter text with a length slider. The fix and verify steps have parallel generation.}
  \label{fig:pull}
\end{teaserfigure}

\maketitle

\input{sections/01_intro}
\input{sections/02_related}

\input{sections/03_methods}

\input{sections/04_design}

\input{sections/04.4_vignettes}

\input{sections/05_case}

\input{sections/06_discussion}

\input{sections/07_conclusion}

\input{sections/08_acks}

\bibliographystyle{ACM-Reference-Format}
\bibliography{0_main}

\newpage
\input{sections/09_appendix}

\end{document}

%% file: sections/00_abstract.tex
LLM chains enable complex tasks by decomposing work into a sequence of subtasks.
Similarly, the more established techniques of crowdsourcing workflows decompose complex tasks into smaller tasks for human crowdworkers. 
Chains address LLM errors analogously to the way crowdsourcing workflows address human error.
To characterize opportunities for LLM chaining, we survey 107 papers across the crowdsourcing and chaining literature to construct a design space for chain development. 
The design space covers a designer's \textit{objectives} and the \textit{tactics} used to build workflows. We then surface \textit{strategies} that mediate how workflows use tactics to achieve objectives.
To explore how techniques from crowdsourcing may apply to chaining, we adapt crowdsourcing workflows to implement LLM chains across three case studies: creating a taxonomy, shortening text, and writing a short story. From the design space and our case studies, we identify takeaways for effective chain design and raise implications for future research and development.

%% file: sections/01_intro.tex
\section{Introduction}
\label{sec:intro}

People use Large Language Models (LLMs) for assistance with both single-step tasks, like suggesting synonyms, and more complex multi-step tasks, like writing stories, editing text, and organizing ideas. 
Despite their widespread adoption, LLMs suffer from quality deficits like hallucinations~\cite{Dziri2022OnTO,nair2023dera,maynez2020faithfulness}, brittleness to prompt changes~\cite{zamfirescu2023johnny,nie2022improving,holtzman2021surface}, and user interventions limited to prompting~\cite{wu2023llms,parameswaran2023revisiting}. 
These deficits are exacerbated for complex tasks, which require multiple steps in which errors can arise and propagate~\cite{dziri2023faith}. 
Unfortunately, many common use cases require multiple reasoning steps. For example, the use case of shortening text requires first deciding which portions of the text to edit and then determining how to revise each one~\cite{bernstein2010soylent}. 

To tackle complex tasks, recent research has turned to LLM chaining techniques. Chaining decomposes a task into multiple calls to an LLM, in which the output of one call affects the input to the next call~\cite{Wu2021AICT}.
For example, when shortening text, an LLM chain could identify verbose sentences, edit each one, and propose outputs of variable lengths by composing multiple edits.
Chaining has many benefits: this strategy can enable new abilities like long-form story writing~\cite{yang2022re3,mirowski2023co}, increase accuracy on logical reasoning questions~\cite{xie2023decomposition,zhang2022automatic,creswell2022faithful}, and enable greater transparency and debuggability~\cite{huang2023pcr,wu2022promptchainer,reppert2023iterated}. 
LLM chaining is a nascent subfield, with seminal papers emerging in only the past few years~\cite{Wu2021AICT, josifoski2023flows, wu2023autogen,li2024camel}. 
Initial evidence suggests that chaining can mitigate quality deficits, but these solutions continue to face problems such as hallucination~\cite{schick2022peer}, bias~\cite{mirowski2023co}, incorrect responses~\cite{nair2023generating}, and expensiveness~\cite{zharikova2023deeppavlov}. 
There is a seemingly intractable design space of possible chaining combinations from the suggested operators of this early work, so realizing the potential of chaining is challenging due to the design space's size and complexity.

The challenge of decomposing complex work, however, is not new. In the more established field of crowdsourcing, researchers have faced a similar problem setup. Groups of people (\emph{requesters}) outsource tasks to other people (\emph{crowdworkers}) on online platforms.
Crowdworkers have their own limitations: crowdworkers may output low quality responses~\cite{bernstein2010soylent,alshaibani2020privacy,butler2017more,lasecki2013chorus}, small changes in instructions affect crowdworker responses~\cite{kittur2012crowdweaver,mcdonnell2016relevant}, and requesters have limited modes of interaction with crowdworkers~\cite{salehi2017communicating}.
To compensate for these limitations, crowdsourcing researchers have developed a variety of structured workflows. 
Crowdsourcing workflows decompose a task into a series of smaller microtasks for crowdworkers to complete. These workflows account for crowdworker limitations by reducing task complexity and leveraging redundancy, aggregation, and explicit validation~\cite{bernstein2010soylent,cheng2015flock,ambati2012collaborative,zaidan2011crowdsourcing}.

Given the parallels between these fields, \textbf{we investigate how crowdsourcing techniques may be adapted to design LLM chains} (Figure~\ref{fig:pull}).
Selecting relevant strategies from over a decade of crowdsourcing research is challenging, especially as LLMs and crowdworkers have salient differences~\cite{wu2023llms}. 
For example, crowdworkers may be slow to generate large text amounts, whereas LLMs may lack common sense and factual grounding, making them more sensitive to instructions and likely to ``hallucinate'' information.

To guide LLM chain development, \textbf{we first construct a design space based on a systematic review of the crowdsourcing workflow and LLM chaining literatures}. We analyze 107 papers and perform open coding to identify core design space dimensions. 
Our design space is organized into the \textbf{objectives} for which a designer can optimize a workflow (Figure~\ref{fig:ds-obj}) and the \textbf{tactics} used to build the workflow (Figure~\ref{fig:ds-tact}).

\textbf{From this review, we also surface the strategies used to marshal and combine tactics to achieve objectives}. For example, the objective to encourage a subjective output quality can be achieved by strategies that produce \emph{diverse responses}, allow \emph{user guidance}, and utilize \emph{adaptable architectures}. We gather from the literature how tactics have been used in prior work to support strategies. For example, workflows and chains encourage \emph{diverse responses} with \emph{redundant} or \emph{communicative} tactical architectures.
\textbf{Strategies} constrain the space and evaluation of tactic choices for both crowds and LLMs (Figure~\ref{fig:ds-actions}).

To better assess the impact of similarities and differences between crowdworkers and LLMs on the implementation of strategies,
\textbf{we implement three case studies adapting crowdsourcing workflows to use LLMs}. 
We chose workflows that enable direct manipulation of the output, an instance of the \emph{user guidance} strategy for quality control that has been explored in crowdsourcing, but is underexplored in chaining.
The case studies implement tasks that vary in their degrees of subjectivity, adapting Cascade~\cite{chilton2013cascade} to control the precision of a generated taxonomy, Soylent~\cite{bernstein2010soylent} to control the length of shortened text, and Mechanical Novel~\cite{kim2017mechanical} to control the narrative elements of a fictional story. 

We evaluate these chains in terms of the quality and controllability of their outputs. For example, compared to zero-shot prompting, chain-generated stories are preferred by crowdsourced raters and the story elements can be more precisely controlled. 
These case study implementations provide insights into how differences between LLMs and crowdworkers impact chain design to achieve the same objectives. In Mechanical Novel, for instance, we find that the LLM chain requires more explicit support than the crowdsourcing workflow to create creative and varied outputs.

Informed by both our design space and the three case studies, \textbf{we discuss how differences between LLMs and crowdworkers impact effective tactic designs for LLM chains}. For example, because LLM stochasticity differs from human diversity, tactic choices for LLMs must induce \emph{response diversity} more explicitly and account for limitations in LLM \emph{self-validation}. We find that our strategies help focus workflow adaptation and suggest areas of promising future work. Such areas include how the speed and reduced cost of LLMs relative to crowdworkers presents an opportunity to create higher quality chains, and we recommend improved tools and methods to augment this design process. 

%% file: sections/02_related.tex
\section{Our approach within the Crowdsourcing-LLM intersection}

In this paper, we integrate a broad survey of literature on LLM chaining and crowdsourcing workflows into a design space. We then use the case studies to explore open questions in this design space.
Our broad investigation and resulting design space inform a comprehensive set of recommendations for navigating design decisions and orienting future work.
These recommendations span from higher-level guidance, such as adhering to the tradeoff between outcome quality and resource constraints, to lower-level suggestions, such as pursuing architecture-based methods for quality validation.
These recommendations expand upon related work by several other groups investigating the intersection of crowdsourcing and LLMs.
One such paper explores the impact on requester and crowdworker stakeholders of incorporating LLMs into crowdsourcing workflows~\cite{allen2023power}. A second considers hybrid Crowd-LLM data annotation~\cite{DBLP:journals/corr/abs-2401-09760}. Another implements three strategies from crowdsourcing using LLMs for data processing tasks~\cite{parameswaran2023revisiting}. Closest to our work, the final paper has students implement six crowdsourcing workflows. The paper discusses the effectiveness of these workflows and the impact of the difference between LLM and crowdworker abilities~\cite{wu2023llms}. 
We present a wider scope of recommendations informed by the design space and our case studies.

%% file: sections/03_methods.tex
\begin{figure*}[t]
 \centering 
 \includegraphics[width=0.8\linewidth]{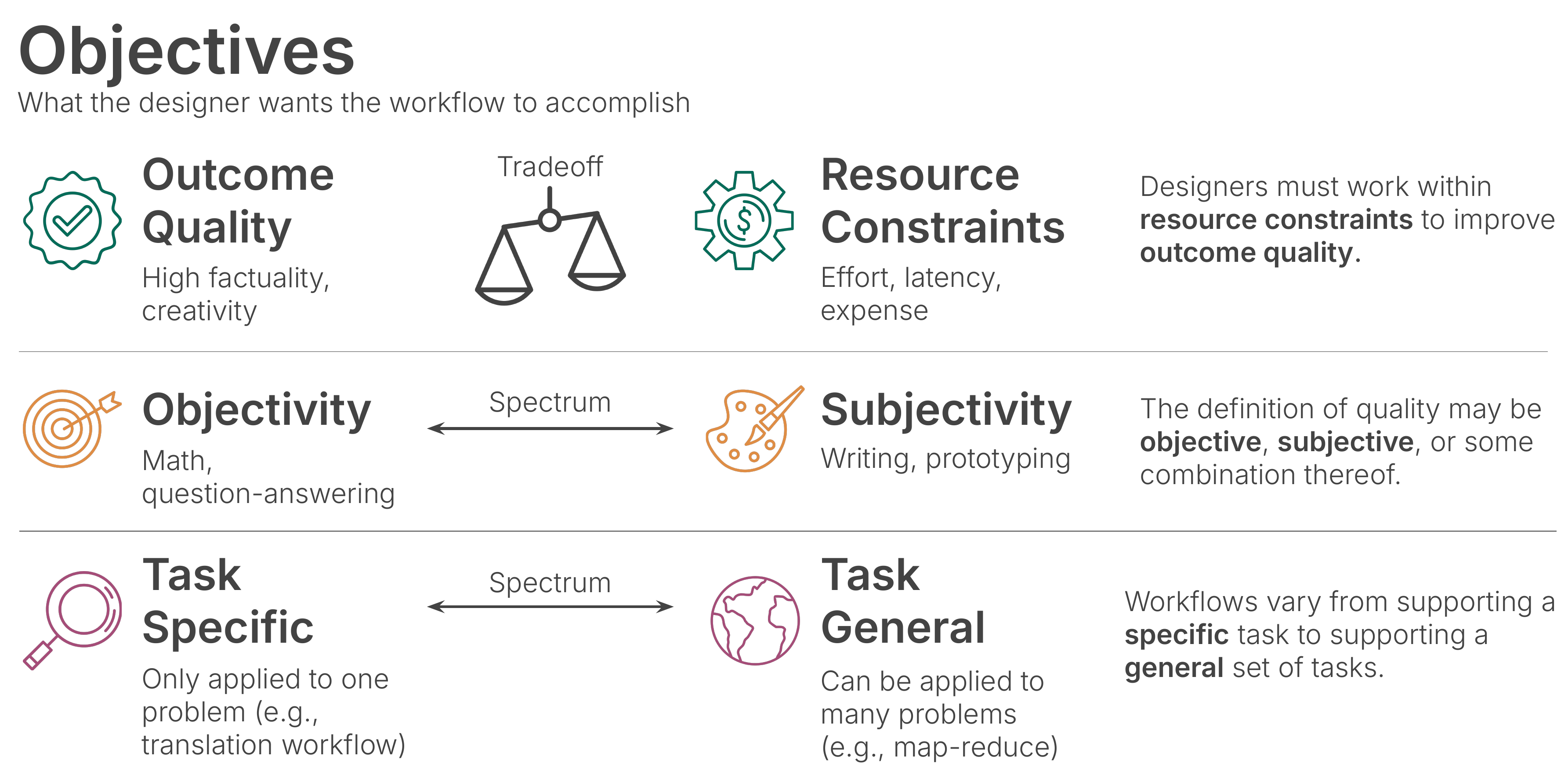}
 \caption{
 Objectives. Our design space contains three axes of workflow objectives, shown as rows above. The first regards the impact of \emph{resource constraints} on the ability to improve \emph{outcome quality}. The second axis distinguishes different notions of `quality:' whether the task objective requires \emph{objectivity}, subjectivity, or elements of both. The final axis concerns generality: if the workflow is aimed to solve a single, \emph{specific} task or to \emph{generalize} to many tasks. 
 }
 \Description{Title of "Objectives, what the designer wants the workflow to accomplish". Three rows of text each with two objectives, examples, and a description. Row 1. A tradeoff between outcome quality (ex. high factuality, creativity) and resource constraints (ex. effort, latency, expense). Description: "Designers must work within resource constraints to improve outcome quality". Row 2. A spectrum between objectivity (ex. math, question-answering) and subjectivity (ex. writing, prototyping). Description: Different tasks may require objectivity, subjectivity, or some combination thereof. Row 3. A spectrum between task specific (ex. only applied to one problem (e.g., translation workflow)) and task general (ex. can be applied to many problems(e.g., map-reduce)). Description: Workflows vary from supporting a specific task to supporting a general set of tasks. }
\label{fig:ds-obj} 
\end{figure*}

\section{Design space methodology}
  
We survey papers that involve the design and implementation of crowdsourcing workflows or LLM chains. 
Although crowdsourcing workflows and LLM chains have many overlapping concepts, the terms used to denote these concepts differ. 
In this paper, we refer to both LLMs and crowdworkers as ``workers.'' 
We use the term ``workflow'' to refer to both crowdsourcing workflows and LLM chains. 
We refer to intermediate steps of workflows as ``subtasks.'' We use the term ``user'' to describe the person receiving workflow outputs (e.g.,~the ``requester'' in crowdsourcing terms). 
The term ``designer'' in this paper refers to a person who defines a workflow to later be used by a ``user.''
Throughout the discussion of the design space and strategies, we indicate which references are from the crowdsourcing workflow literature [C] and the LLM chaining literature [L].

\subsection{Paper collection}
We run two paper collection processes, one for the crowdsourcing literature and another for the LLM chaining literature. We use snowball sampling from the relevant citations in and of these papers, including every citation of the initial AI Chains paper~\cite{wu2022ai}. We consider papers available as of August 2023.

\vspace{0.5em}
\noindent\textbf{Workflow literature.}
We collect papers through a search of the keywords: \textit{crowd(sourcing) workflow}, \textit{crowd(sourcing) pipeline}, \textit{crowdsourcing complex work} and \textit{microtask}.  We search the ACM CHI Conference on Human Factors in Computing Systems,
ACM Conference on Computer-supported Cooperative Work and
Social Computing, the ACM Symposium on User Interface Software and Technology, and the AAAI Conference on Human Computation and Crowdsourcing, as well as Semantic Scholar~\cite{fricke2018semantic} and Google Scholar. 

\vspace{0.5em}
\noindent\textbf{Chaining literature.}
We collect papers through a search of the keywords: \textit{LLM chain(ing)}, \textit{prompt chain(ing)}, \textit{LLM pipeline}, and \textit{AI chain(s/ing)}. We search the International Conference on Learning Representations, the Conference on Neural Information Processing Systems, the International Conference on Machine Learning, the Association for Computational Linguistics proceedings, the Association for Computing Machinery proceedings, as well as Semantic Scholar~\cite{fricke2018semantic} and Google Scholar. We include preprints in our literature search.

\vspace{0.5em}
\noindent\textbf{Paper inclusion criteria.}
We include papers that implement, or support the implementation of, a workflow with at least two steps; i.e.,~workflows in which the outputs from one worker inform the input to another worker. 
We include papers that focus on building an interface or domain specific language for other designers to build workflows. 
We restrict our search to academic papers, rather than software like LangChain~\cite{langchain}. We exclude works that call multiple workers in parallel but never in a chain of steps (e.g.,~majority voting of labels). This rule limits the crowdsourcing techniques considered to those that reflect the serial and dependent structure of an LLM chain. 
We also exclude techniques such as Chain of Thought~\cite{wei2022chain} which issue multiple tasks to a worker in one step instead of incorporating multiple worker calls~\cite{cheng2015break}. 
We include instances in which workers synchronously work on the same interface, viewing and editing each other's outputs.
Finally, the task at hand must have a user working towards some end goal, so we do not consider open-ended collaborative projects like Wikipedia.
We make one exception for a paper that explicitly draws from crowdsourcing strategies to complete data aggregation tasks with LLMs~\cite{parameswaran2023revisiting}. Although they do not implement chains, the connection to crowdsourcing resonates with our design space goals.

\vspace{0.5em}
\noindent\textbf{Results}.
Our initial keyword search, conducted in August 2023, found 313 papers for workflows and 180 papers for chaining. One author determined whether papers were out of scope given our collection criteria, and any ambiguous cases were discussed among multiple authors. Our final set of papers resulted in 107 papers: 68 for workflows, and 39 for chains.

\subsection{Coding strategy}
We conduct a thematic analysis~\cite{braun2006using}, following the methodologies of similar design spaces in HCI~\cite{lai2023towards,bae2022making,shi2021communicating}. 
First, one author extracted 1) the outcomes (\textbf{objectives}) and 2) the elements (\textbf{tactics}) of the workflows from the paper set. 
We then employed iterative open coding on the extracted data, in which two authors separately coded each data extraction, sorted resulting codes into sub-categories, and then discussed these categorizations. In three further rounds of iteration, the same authors refined these categorizations. 
All authors reviewed the resulting design space.

%% file: sections/04_design.tex
\section{Design Space}
\label{sec:ds}

Our design space has two parts: the \textbf{objectives} span the space of a designer's goals, and the \textbf{tactics} span the space of workflow implementations. 
These \textbf{objectives} are 1) promoting \emph{outcome qualities} while respecting \emph{resource constraints}, 2) specifying an \emph{objective} or \emph{subjective} definition of quality, and 3) defining the \emph{generality} or \emph{specificity} of the set of tasks that the designer wants the workflow to support.
For the workflow \textbf{tactics}, \emph{actors} complete various \emph{subtasks} that are linked together into the workflow's \emph{architecture}.

Although we have no guarantee that the design space is complete, it is based on an exhaustive survey of existing work. Defining this space enables understanding, classification, and comparison of existing workflows (Figure~\ref{fig:ds-hist} and Section~\ref{sec:ds-compare}), generation of new workflows (Section~\ref{sec:ds-vig}), and the discovery of design patterns (Section~\ref{sec:ds-strat}) that inform chaining practices and guide directions for future work (Section~\ref{sec:disc}).

\input{sections/04.1_objectives}

\input{sections/04.2_tactics}

\input{sections/04.25_comparison}

\input{sections/04.3_strategies}

%% file: sections/04.1_objectives.tex
\begin{figure*}[t]
 \centering 
 \includegraphics[width=0.8\linewidth]{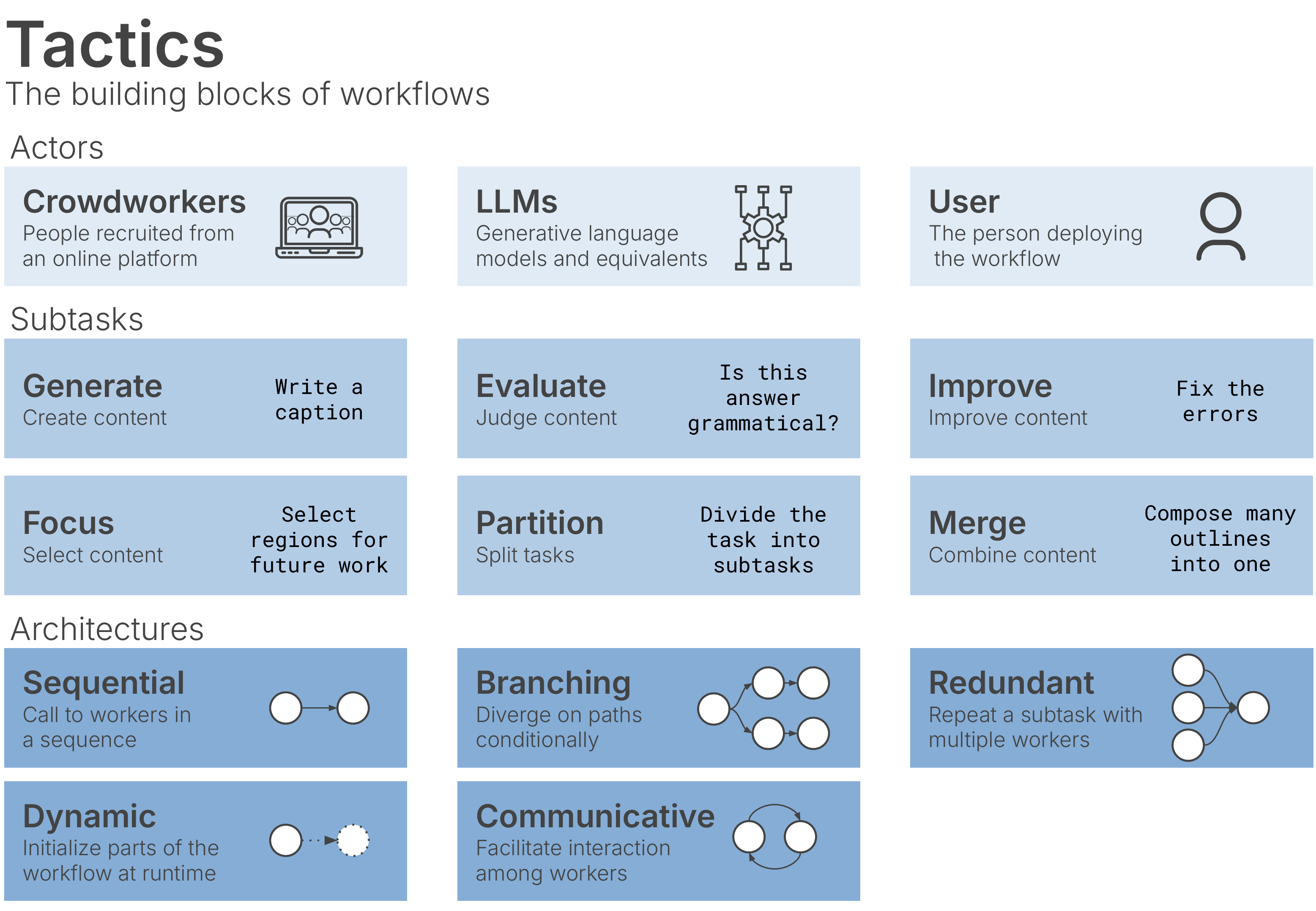}
 \caption{
 Tactics. Tactics are the building blocks of workflows. \emph{Actors} complete \emph{subtasks}, which are connected together in \emph{architectures}. There are multiple categories of actors (crowdworkers, LLMs, and the user), subtasks (generate, evaluate, improve, focus, partition and merge), and architectures (sequential, branching, redundant, dynamic, and communicative). Workflows may use multiple tactics in any of these categories.
 }
 \Description{Title: Tactics, the building blocks of workflows. One row each for actors, subtasks, and architectures, with the category and (a short description). The actor's row contains: Crowdworkers (people on an online platform), LLMs (generative large language models and equivalents), user (the person deploying the workflow). The subtasks row contains: generate (create content), evaluate (judge content), improve (improve content), focus (select content), partition (split tasks), and merge (combine content). The architecture row contains: sequential (call LLMs in sequence), branching (diverge on paths), redundant (multiple actors repeat subtask), dynamic (Initialize parts of the workflow at runtime), and communicative (facilitate interaction among actors) Each subtask category is connected with a generic example instruction. Generate: "Write a caption". Evaluate: "Is this answer grammatical?". Improve "Fix the errors" Focus: "Select regions for future work: Partition: "Divide the task into subtasks". Merge: "Compose many outlines into one".}
\label{fig:ds-tact} 
\end{figure*}

\subsection{Objectives in the design space}

The \textbf{objectives} of a workflow design define the designer's goals within several dimensions. Designers will design for some point along a \emph{outcome quality} vs \emph{resource constraints} tradeoff, in a range from \emph{objectivity} to \emph{subjectivity}, and in a range spanning between \emph{task specificity} and \emph{generality}.
(Figure~\ref{fig:ds-obj}).

\begin{sloppypar}

\subsubsection{Outcome quality vs resource constraints}
When working towards \emph{outcome qualities}, designers must be mindful of \emph{resource constraints}. 
Workflows can support quality outputs such as by improving accuracy and/or spurring creativity [C\cite{lasecki2013chorus} L\cite{kim2023metaphorian, creswell2022faithful, reppert2023iterated, Wu2021AICT}], transparency [L\cite{creswell2022selection, creswell2022faithful, wu2023llms, reppert2023iterated, wu2023autogen, kim2023metaphorian}], debuggability [L\cite{wu2022promptchainer, Wu2021AICT}], and modularity [C\cite{latoza2014microtask, lee2017sketchexpress, retelny2014expert} L\cite{huang2023pcr, zharikova2023deeppavlov, yao2023tree, wu2023autogen}]. 
\end{sloppypar}

\begin{sloppypar}
 
Workflows can also address, although imperfectly, quality concerns. Some quality concerns are induced by the workers, such as hallucinations [L\cite{nair2023generating, mirowski2023co, creswell2022faithful, nair2023dera, wu2023autogen, parameswaran2023revisiting}] and generative loops [L\cite{mirowski2023co}] in the case of LLMs, and fatigue [C\cite{dai2013pomdp, alshaibani2020privacy}] in the case of crowdworkers.
Others are induced by the workflow itself, such as cascading errors [C\cite{kittur2011crowdforge,goto2016understanding} L\cite{wu2022promptchainer, Wu2021AICT, xie2023decomposition}] and a lack of coherency [C\cite{bernstein2010soylent, kim2017mechanical, hahn2016knowledge} L\cite{yang2022re3, mirowski2023co, wu2022promptchainer}].
More concerns are induced by the broader context, such as the need to scale [C\cite{kulkarni2014wish, kulkarni2012collaboratively} L\cite{arora2022ask}], the workflow's placement within a larger system [C\cite{pavel2014video, cheng2015flock, christoforaki2014step}], unethical use of the workflow [L\cite{park2022social}], and the impact on human labor [L\cite{wu2023autogen, mirowski2023co}], sense of ownership [L\cite{mirowski2023co}] and sense of agency [L\cite{mirowski2023co, kim2023metaphorian}]. 
However, mechanisms for improving output quality can be more expensive [C\cite{huffaker2020crowdsourced} L\cite{yao2023tree}], take longer [C\cite{chang2017revolt, gouravajhala2018eureca}], and require more effort to create [C\cite{kittur2011crowdforge, latoza2014microtask}].
\end{sloppypar}

\begin{sloppypar}
 
\vspace{0.5em}
\noindent\textbf{Expense}.
The expense of a workflow refers to paying crowdworkers or paying for an API call to an LLM.
Workflows are often more expensive than simpler approaches (e.g.,~Chain of Thought prompting) [C\cite{marcus2011human, marcus2011demonstration, hahn2016knowledge, lin2014crowdsourced, dai2010decision, huffaker2020crowdsourced} L\cite{yao2023tree, wang2022iteratively, nair2023generating, du2023improving, gero2023self, zharikova2023deeppavlov}]. 
However, the cost is often less than hiring professionals [C\cite{liem2011iterative, christoforaki2014step, zaidan2011crowdsourcing, ambati2012collaborative} L\cite{fleming2023assessing}], pretraining models [L\cite{huang2023pcr, levine2022standing, kim2023metaphorian, cobbe2021training}], or completing the task yourself [L\cite{zhang2022automatic, wang2023popblends}]. 

\end{sloppypar}

\vspace{0.5em}
\noindent\textbf{Latency}.
Another constraint is the time for the workflow to complete [C\cite{kulkarni2014wish, lasecki2015apparition, marcus2011demonstration, marcus2011human}]. Many workflows have lower latency than a single person or a baseline performing the same task, as work can be parallelized across multiple workers [C\cite{pietrowicz2013crowdband, kulkarni2014wish, lasecki2015apparition, chilton2014frenzy, andre2013community, teevan2016supporting, lin2014crowdsourced, gouravajhala2018eureca, lee2017sketchexpress, guo2016vizlens} L\cite{kim2023metaphorian}].

\vspace{0.5em}
\noindent\textbf{Effort}.
Workflows can help users complete tasks with less effort [C\cite{kulkarni2014wish} L\cite{wu2023autogen, mirowski2023co, long2023tweetorial}]. However, workflows can require high effort to create [C\cite{hahn2016knowledge, kulkarni2012collaboratively, kittur2012crowdweaver}] and oversee [C\cite{latoza2014microtask} L\cite{wu2022promptchainer}].

\subsubsection{Objectivity vs subjectivity}

One element of output quality of particular importance is if the workflow should support \emph{objective} or \emph{subjective} outputs. Many tasks' definitions of output quality have elements of both objectivity and subjectivity.

\begin{sloppypar}
 
\vspace{0.5em}
\noindent\textbf{Objectivity.}
Objectivity-focused tasks have one correct answer. 
Some crowdsourcing papers in this space include TurKit [C\cite{little2010turkit}] and Second Opinion [C\cite{mohanty2019second}].
Examples of these tasks include math problems [L\cite{zhou2022least, du2023improving, wu2023empirical, liang2023encouraging, madaan2023self}], blurry text recognition  [C\cite{little2010turkit, little2010exploring}], and matching images with the same content [C\cite{kobayashi2018empirical, mohanty2019second}].
For these tasks, workflows can improve performance on benchmarks [C\cite{guo2019statelens, mcdonnell2016relevant, kobayashi2018empirical} L\cite{li2023teach, eisenstein2022honest, wu2023empirical, zhang2022automatic, schick2022peer, cobbe2021training}], enable new tasks [C\cite{little2010turkit, gouravajhala2018eureca, huffaker2020crowdsourced, guo2017facade} L\cite{huang2023ai}], classify examples for which machine learning approaches struggle [C\cite{bernstein2010soylent, alshaibani2020privacy, cheng2015flock, mohanty2019second}],
and reduce hallucinations [L\cite{wu2023autogen, creswell2022selection, nair2023generating, creswell2022faithful, parameswaran2023revisiting, nair2023dera, mirowski2023co}]. 
\end{sloppypar}

\vspace{0.5em}
\begin{sloppypar}
\noindent\textbf{Subjectivity.}
Subjective tasks have the widest degree of output freedom.
Representative crowdsourcing papers with subjective outputs include Flash Teams [C\cite{retelny2014expert}], VisiBlends [C\cite{chilton2019visiblends}], and Mechanical Novel [C\cite{kim2017mechanical}]. These papers tackle creative writing [C\cite{kim2017mechanical, andre2014effects, bernstein2010soylent} L\cite{yao2023tree, yang2022re3, mirowski2023co, zhang2023visar}], idea generation [C\cite{little2010exploring} L\cite{wu2023llms, kim2023metaphorian}], and prototyping [C\cite{kulkarni2014wish, valentine2017flash, lee2017sketchexpress, retelny2014expert} L\cite{park2022social}]. 
Prior work finds that on subjective tasks, workflows can help achieve higher quality [C\cite{kim2017mechanical, andre2014effects, lee2017sketchexpress} L\cite{yao2023tree, yang2022re3, kim2023metaphorian, mirowski2023co, park2022social}] and less constrained outputs [C\cite{liu2018conceptscape, hahn2016knowledge, drapeau2016microtalk, lee2017sketchexpress} L\cite{park2022social, wang2023popblends}]. Workflows may enable novice crowdworkers [C\cite{pietrowicz2013crowdband}] or users [C\cite{chilton2019visiblends, valentine2017flash, andre2013community, kim2016storia, lin2014crowdsourced} L\cite{wu2023llms, wang2023popblends}] to perform a subjective task with an otherwise high barrier of entry. Important qualities for subjective tasks include coherence [C\cite{kim2017mechanical} L\cite{yang2022re3, yao2023tree}] and aligning to the user's creative vision [C\cite{salehi2017communicating}]. 
On subjective tasks, a workflow is often better suited for creating and lengthening the output, rather than proofreading [C\cite{huang2017supporting, kim2017mechanical, little2010turkit} L\cite{mirowski2023co}], so these outputs are sometimes seen as a starting point for editing, rather than a finished product [C\cite{nebeling2016wearwrite, kim2016storia} L\cite{mirowski2023co, long2023tweetorial, park2022social}]. 
\end{sloppypar}

\vspace{0.5em}
\noindent\textbf{Both objectivity and subjectivity.}
Many tasks require some degree of objectivity and some degree of subjectivity. 
In some tasks, there is not one correct answer, but there are restrictions on what outputs are correct. 
Crowdsourcing papers for these tasks include Cascade [C\cite{chilton2013cascade}], Flock [C\cite{cheng2015flock}], Revolt [C\cite{chang2017revolt}], and Storia [C\cite{kim2016storia}].
Example tasks include categorizing data (categories must be accurate, but there are multiple options and unclear edge cases) [C\cite{chilton2014frenzy, andre2013community, bragg2013crowdsourcing, chilton2013cascade, cheng2015flock, chang2017revolt}], summarizing text (summaries must stay faithful, but have degrees of freedom in how to write the summary) [C\cite{kim2016storia, pavel2014video, bernstein2010soylent} L\cite{nair2023generating, nair2023dera}], and coding (programs must work, but there are multiple approaches to problem solving) [C\cite{kulkarni2014wish, latoza2014microtask} L\cite{wu2023autogen, huang2023pcr, huang2023adaptive}]. 
The goals of workflows for these tasks are to align with expert outputs [C\cite{liu2018conceptscape, kim2014crowdsourcing, chilton2013cascade, salehi2017communicating} L\cite{kim2023metaphorian, park2022social, fleming2023assessing}] or to enable users to choose among multiple correct possible output options.
Using workflows for these tasks can result in an output that is human-understandable [C\cite{cheng2015flock}], that better represents what would be important for people to know [C\cite{salisbury2017toward}], and that surfaces the uncertainty among the different possible correct options [C\cite{chang2017revolt}].

\subsubsection{Task specificity vs generality}
Some workflows are highly \emph{specialized} to the specifications of the task (e.g.,~visual blends [C\cite{chilton2019visiblends} L\cite{wang2023popblends}], masking faces in images [C\cite{alshaibani2020privacy, kaur2017crowdmask}], and programming [C\cite{latoza2014microtask} L\cite{huang2023pcr}]).
Others \emph{generalize} to a variety of tasks [C\cite{kulkarni2012collaboratively} L\cite{wu2023autogen, creswell2022selection, creswell2022faithful, nair2023dera}]. For example, AutoGen [L\cite{wu2023autogen}] uses the same conversation-based workflow for coding as well as chess games, math problems, and retrieval-based question answering.
Task-specific workflows support fewer outcome types but can incorporate more tailored support than workflows that generalize to many tasks.

%% file: sections/04.2_tactics.tex
\subsection{Tactics in the design space}

Designers use workflow \textbf{tactics} to implement workflows. Tactics need to consider the \emph{actors}, the \emph{subtasks} the actors complete, and the connections of subtasks into workflow \emph{architectures} (Figure~\ref{fig:ds-tact}).

\subsubsection{Actors}
\label{sec:ds-tools-workers}

Actors may be \emph{crowdworkers}, \emph{LLMs}, or the \emph{user}.

\vspace{0.5em}
\noindent\textbf{Crowdworkers.} These actors are usually recruited from crowdworking platforms like Amazon Mechanical Turk or UpWork (formerly known as oDesk). Platforms allow for requests of crowdworker qualifications, although these credentials are sometimes hard to validate [C\cite{kulkarni2014wish, zaidan2011crowdsourcing, valentine2017flash, ahmad2011jabberwocky}]. Methods like retainer models can reduce the latency of accessing crowdworkers on these platforms [C\cite{bernstein2011crowds}]. 
Workers can also be recruited from the public, including community members [C\cite{mahyar2018communitycrit}], conference committees [C\cite{chilton2014frenzy, andre2013community}], and online deaf communities [C\cite{bragg2021asl}].

\vspace{0.5em}
\noindent\textbf{LLMs}.
There are LLMs of many sizes and expenses [L\cite{bursztyn2022learning}]. Models have varying strengths and respond differently to prompts [L\cite{parameswaran2023revisiting, wu2023llms, mirowski2023co, liang2023encouraging}]. LLM outputs can be adjusted through parameter changes (e.g.,~temperature) [L\cite{parameswaran2023revisiting, park2022social}], fine-tuning [L\cite{li2023teach, creswell2022selection, bursztyn2022learning, creswell2022faithful, cobbe2021training}], and expressing roles through prompts [L\cite{du2023improving, park2022social}]. Recently, designers often substitute specialized or distilled models, such as embeddings, for LLM functionality at reduced cost [L\cite{mirowski2023co,li2023teach}].

\vspace{0.5em}
\noindent\textbf{Comparing crowdworkers and LLMs}. 
Some limitations of crowdworkers and LLMs overlap. First, both crowdworkers and LLMs can output low quality responses. Prior work finds that outcomes from crowdsourcing platforms include low-quality, spam, and off-topic responses [C\cite{lasecki2013chorus, alshaibani2020privacy, huang2018evorus, butler2017more, bernstein2010soylent}]. LLMs can provide erroneous outputs: they are known to confidently output incorrect answers [L\cite{parameswaran2023revisiting, du2023improving, wang2023popblends}] and to struggle with self-consistency and logical reasoning [L\cite{creswell2022selection, yao2023tree, creswell2022faithful, parameswaran2023revisiting, liang2023encouraging, xie2023decomposition}]. Second, long and complex instructions can lower performance for both crowdworkers and LLMs. Crowdworkers struggle with following long and complex instructions, as people can overly focus on the beginning or end of a given text [C\cite{hahn2016knowledge, chilton2013cascade, chen2019cicero, lasecki2015apparition}]. LLMs can ignore large fractions of long prompts or focus mostly on recent parts of the given context [L\cite{parameswaran2023revisiting, du2023improving, wu2023llms, madaan2023self}]. Third, both crowdworkers and LLMs can give biased outputs. Crowdworkers can be biased; for example, in one study, crowdworkers performed significantly better in identifying photos of their own race compared to other races [C\cite{mohanty2019second}]. LLMs also output biased results, a widespread concern [L\cite{wu2023autogen, huang2023adaptive, long2023tweetorial, park2022social}]. For example, work on screenplay generation found stereotypical outputs, with multiple instances of gender biases [L\cite{mirowski2023co}]. 

Crowdworkers have some unique limitations. Since crowdworkers are paid when they complete a task, there is an incentive to work quickly even if that means satisficing [C\cite{bernstein2010soylent} L\cite{parameswaran2023revisiting}]. Additionally, since rejected work can hurt access to future jobs on a platform [C\cite{kim2017mechanical}], crowdworkers may overcompensate to a detrimental level in an attempt to make their efforts clear [C\cite{bernstein2010soylent, kim2014crowdsourcing}]. Crowdworkers may fatigue [C\cite{alshaibani2020privacy, dai2013pomdp}], and they may lack task-relevant knowledge and delete jargon [C\cite{nebeling2016wearwrite, bernstein2010soylent}]. On synchronous tasks, other limitations become prevalent. Social blocking can make crowdworkers so cautious to change others' work that not enough work gets done [C\cite{hahn2016knowledge, lasecki2015apparition}], and territoriality can motivate edit wars in which crowdworkers' attempts to establish their own work over others degrade the final product quality [C\cite{andre2014effects, valentine2017flash}]. Crowdworkers also may be affected by diffusion of responsibility and social loafing in which there is incentive to wait for others to do tasks [C\cite{andre2014effects, lasecki2015apparition}]. 

LLMs also have unique limitations. On creative tasks, LLMs sometimes produce outputs that are perceived as robotic [L\cite{long2023tweetorial}] and that can lack the understanding, nuance, and subtext that people can provide [L\cite{mirowski2023co}]. LLM training on human data increases the risk of accidental plagiarism [L\cite{mirowski2023co}]. Furthermore, models can get stuck in generative loops, degenerating text quality [L\cite{parameswaran2023revisiting}]. LLMs also hallucinate more than crowds [L\cite{wu2023autogen, creswell2022selection, nair2023generating, creswell2022faithful, parameswaran2023revisiting, nair2023dera, mirowski2023co}] and may struggle to change outputs through self-reflection [L\cite{liang2023encouraging}].

Crowdworkers and LLMs have their comparative strengths as well. Crowdworkers excel at performing high-level structural evaluation and quickly identifying outliers [C\cite{luther2015crowdlines, huang2015guardian}]. People are also currently better at selectively choosing among subtasks than LLMs [L\cite{wu2023llms}]. LLMs may be better at divergent thinking [L\cite{wu2023autogen, wu2023llms, liang2023encouraging}]
 and comparison-based instructions [L\cite{wu2023autogen, wu2023llms}]
 than crowdworkers, who are affected more by anchoring bias. LLMs may also be better than crowdworkers at tasks requiring recollection of large amounts of data [L\cite{parameswaran2023revisiting}].

\vspace{0.5em}
\noindent\textbf{User}.
Finally, users of the workflow can also be involved in revising or guiding the process. Users can be novices using the workflow to adapt to a steep learning curve [C\cite{kulkarni2014wish, chilton2019visiblends, pietrowicz2013crowdband, lin2014crowdsourced} L\cite{wang2023popblends}], or experts using the workflow as a support tool [C\cite{nebeling2016wearwrite, mohanty2019second} L\cite{mirowski2023co}]. However, even with user involvement, prior work finds mixed results on how well tools retain the user's sense of agency and ownership [L\cite{mirowski2023co, kim2023metaphorian}].

\subsubsection{Subtasks}

Subtasks are singular tasks in a workflow. Although highly varied, they follow several thematic functions: to \emph{generate}, \emph{evaluate}, \emph{improve}, \emph{focus}, \emph{partition}, and \emph{merge}.

\begin{sloppypar}
 
\vspace{0.5em}
\noindent\textbf{Generate.}
Most workflows involve at least one step in which actors generate content. This generation may be writing words or code [C\cite{weir2015learnersourcing, pavel2014video} L\cite{li2023teach, zhang2022automatic, zhang2023visar}] brainstorming and ideating [C\cite{chilton2019visiblends, lasecki2013chorus, huang2018evorus, mahyar2018communitycrit} L\cite{yao2023tree, long2023tweetorial}], or answering a question or solving a problem [C\cite{kobayashi2018empirical} L\cite{xie2023decomposition, eisenstein2022honest, wu2023empirical}]. 
Generation tasks may also ask for labels or annotations on given content [C\cite{chilton2014frenzy, guo2019statelens, bragg2013crowdsourcing, chang2017revolt, guo2017facade}], to translate between languages [C\cite{ambati2012collaborative, zaidan2011crowdsourcing}], to contribute to a conversation with other actors [C\cite{mahyar2018communitycrit} L\cite{liang2023encouraging, du2023improving, wang2022iteratively}], to demonstrate a behavior [C\cite{bragg2021asl, lee2017sketchexpress}], or to source content from external sources [C\cite{luther2015crowdlines, pietrowicz2013crowdband, chilton2019visiblends, kim2014crowdsourcing, christoforaki2014step, kim2016storia, hahn2016knowledge, huang2015guardian}]. 
Some subtasks also require an explanation along with the generated content [C\cite{kim2017mechanical, cheng2015flock, drapeau2016microtalk, chang2017revolt} L\cite{gero2023self, schick2022peer, fleming2023assessing}]. 
\end{sloppypar}

\begin{sloppypar}
 
\vspace{0.5em}
\noindent\textbf{Evaluate.}
One of the most common evaluation tasks is choosing the best among multiple possible options [C\cite{liu2018conceptscape, weir2015learnersourcing, guo2019statelens, kittur2012crowdweaver, cheng2015flock, guo2017facade} L\cite{mirowski2023co, yang2022re3, yao2023tree}]. Other subtasks may include rating and ranking among multiple options [C\cite{marcus2011human, pavel2014video, willett2012strategies, salehi2017communicating, mahyar2018communitycrit, huffaker2020crowdsourced} L\cite{yao2023tree,li2023teach,parameswaran2023revisiting}] or a pairwise comparison [C\cite{little2010exploring, marcus2011human, little2010turkit, marcus2011demonstration, mohanty2019second} L\cite{bursztyn2022learning, parameswaran2023revisiting}]. Evaluation subtasks may also determine if another round of generation needs to occur [C\cite{latoza2014microtask} L\cite{liang2023encouraging, wang2022iteratively, Wu2021AICT}].
\end{sloppypar}

\begin{sloppypar}
Workflows can decide what moves forward through majority vote [C\cite{liu2018conceptscape, lasecki2015apparition, kim2014crowdsourcing, guo2019statelens, huang2017supporting, dai2010decision}], other aggregations of voting tasks (e.g.,~expectation maximization [C\cite{drapeau2016microtalk} L\cite{parameswaran2023revisiting}] or weighted mean [C\cite{mohanty2019second}]), returning the highest ranked or rated items [C\cite{pavel2014video, huffaker2020crowdsourced}], or automated methods. These automated methods include measuring agreement from another step [C\cite{liu2018conceptscape, hahn2016knowledge, chang2017revolt, huang2015guardian, bernstein2011crowds, huffaker2020crowdsourced} L\cite{nair2023dera}], weak supervision[L\cite{arora2022ask}], and using non-AI checks like semantic embeddings and rule-based heuristics [C\cite{ambati2012collaborative} L\cite{parameswaran2023revisiting, wang2023popblends}]. 
There are also subtasks that seek critiques and reviews from the user [C\cite{nebeling2016wearwrite} L\cite{schick2022peer}] or a worker [C\cite{kulkarni2014wish, valentine2017flash, kim2017mechanical, christoforaki2014step, mahyar2018communitycrit} L\cite{parameswaran2023revisiting, madaan2023self, liang2023encouraging, schick2022peer}]. These subtasks may ask for feedback or verify if errors or other qualities exist [C\cite{chilton2019visiblends, weir2015learnersourcing, kim2014crowdsourcing, christoforaki2014step, andre2013community, bragg2013crowdsourcing, huang2017supporting, bernstein2010soylent, huang2015guardian, kulkarni2012collaboratively, butler2017more, kittur2011crowdforge} L\cite{huang2023pcr, yang2022re3, gero2023self, liang2023encouraging, wu2023llms}].
\end{sloppypar}

\vspace{0.5em}
\noindent\textbf{Improve.}
Other subtasks improve content generated by previous steps, sometimes by incorporating feedback from evaluation steps. Improving on previously generated content can come in many forms: updating the format of the content [L\cite{huang2023pcr, wang2022iteratively, Wu2021AICT, eisenstein2022honest, arora2022ask}], pruning bad results [C\cite{liu2018conceptscape} L\cite{gero2023self}], adding missing results [C\cite{liu2018conceptscape, weir2015learnersourcing} L\cite{huang2023pcr}], correcting errors [C\cite{drapeau2016microtalk, gouravajhala2018eureca, kobayashi2018empirical} L\cite{huang2023pcr, yang2022re3}], and performing direct improvements [C\cite{liem2011iterative, kim2017mechanical, little2010turkit, nebeling2016wearwrite, guo2019statelens, mcdonnell2016relevant, christoforaki2014step, hahn2016knowledge, bernstein2010soylent, mahyar2018communitycrit, lee2017sketchexpress, dai2010decision, huffaker2020crowdsourced} L\cite{madaan2023self, schick2022peer}].

\vspace{0.5em}
\noindent\textbf{Focus.}
One set of subtasks looks to select the attention of the upcoming tasks within a broader context. For example, this step may find patches of the input that need editing, extract information from context, or focus on common keywords [C\cite{kim2017mechanical, bernstein2011crowds, huffaker2020crowdsourced} L\cite{creswell2022selection, creswell2022faithful, nair2023dera, gero2023self}].

\vspace{0.5em}
\noindent\textbf{Partition}.
Some subtasks transform a task by breaking it up into smaller pieces [C\cite{valentine2017flash, latoza2014microtask, kulkarni2012collaboratively, kittur2011crowdforge} L\cite{zhou2022least, Wu2021AICT, wu2023llms, huang2023ai}]. For example, the Divide step in the Turkomatic workflow asks crowdworkers to divide a given subtask into simpler subtasks. The workflow calls this step recursively until the subtask complexity is valued (by a separate crowdworker) at a specified cost [C\cite{kulkarni2012collaboratively}].

\vspace{0.5em}
\noindent\textbf{Merge.}
Subtasks can merge multiple streams of incoming content. This merge could be algorithmic [C\cite{chilton2019visiblends, alshaibani2020privacy, teevan2016supporting, alshaibani2021pterodactyl, kittur2011crowdforge, ahmad2011jabberwocky}] or accomplished by an actor [C\cite{luther2015crowdlines, pietrowicz2013crowdband, ambati2012collaborative, latoza2014microtask, teevan2016supporting, kittur2012crowdweaver, kulkarni2012collaboratively, lee2017sketchexpress, kittur2011crowdforge} L\cite{Wu2021AICT, wu2023llms}].

\begin{figure*}[t]
  \centering 
  \includegraphics[width=0.85\linewidth]{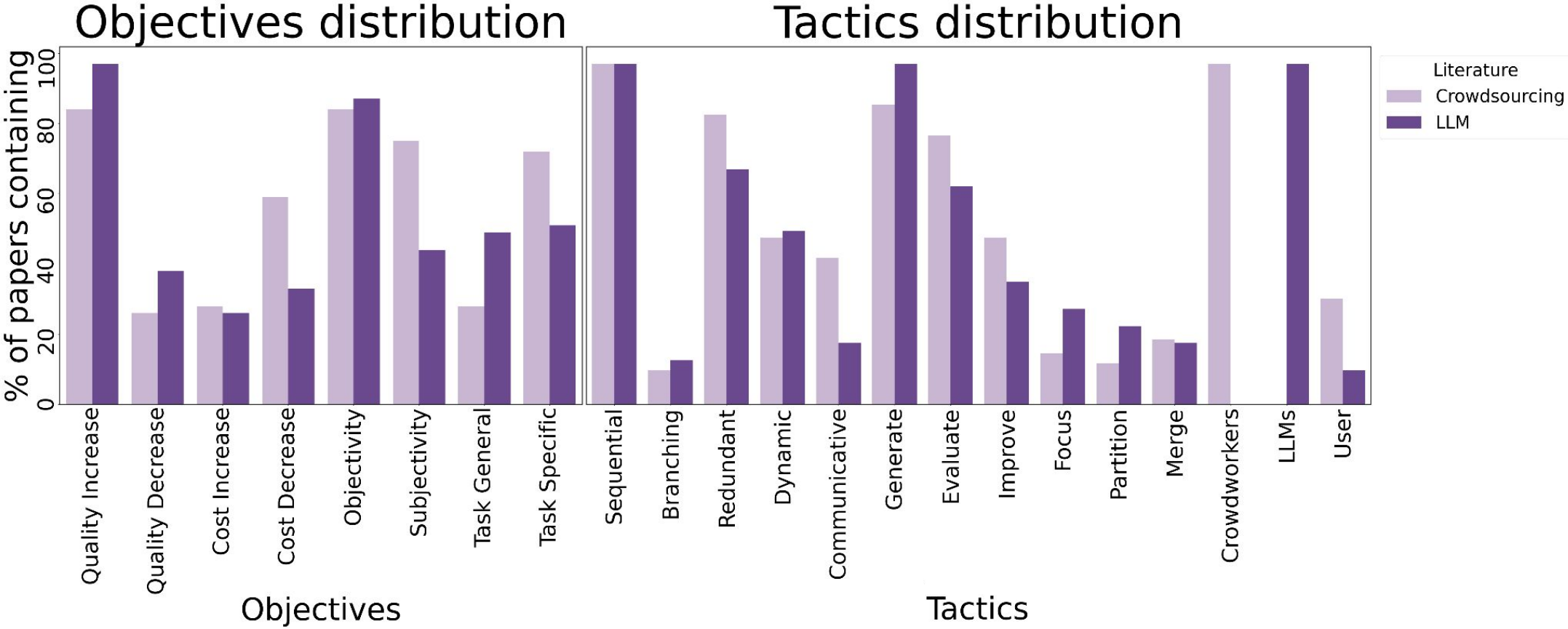}
  \caption{The distribution of the corpus in the design space. We categorize the papers in our corpus across different categories of the design space. Each set of bars is the categorization, split by the crowdsourcing and LLM literatures. There are fewer LLM papers in our corpus (39) than crowdsourcing papers (68). The length of the bar is the proportion of all papers in each literature that incorporate the design space element. We provide the categorization of each paper, as well as the criteria for our categorizations in the supplementary materials. Through this comparison, we see how the literatures similarly span the entirety of the design space, with the exception of using LLMs or crowdworkers. We surface some differences as well, such as how crowdsourcing papers are more often \emph{task-specific}, incorporate \emph{subjectivity}, and use \emph{communicative} architectures. }
  \Description{Two sets of bar graphs. The left is titled "Objectives Distribution" and contains data on the objectives. The right is titled "Tactics Distribution" and contains data on the tactics. The x-axis of each contains the design space categories, and the y-axis ranges from 0 percent to 100 percent. Each design space category on the x axis has a bar color-coded to represent the crowdsourcing or LLM literature. All categories, besides the ``LLM'' and ``Crowdworker'' categories, exist in both literatures. The bar values summarize the categorizations found in the supplementary.}
\label{fig:ds-hist} 
\end{figure*}

\subsubsection{Workflow architecture}
\label{sec:ds-tools-architecture}

Workflows combine subtasks into larger architectures that may be \emph{sequential}, \emph{branching}, \emph{redundant}, \emph{dynamic}, or \emph{communicative}. One workflow may employ multiple architectural patterns. 

\vspace{0.5em}
\noindent\textbf{Sequential} architectures order subtasks in a chain in which the output of one actor passes forward to another actor [C\cite{christoforaki2014step,mcdonnell2016relevant,huang2017supporting,huffaker2020crowdsourced} L\cite{zhou2022least, huang2023adaptive, huang2023pcr, nair2023generating, Wu2021AICT, eisenstein2022honest, gero2023self, long2023tweetorial, reppert2023iterated, wu2023llms, wang2023popblends}].

\vspace{0.5em}
\noindent\textbf{Branching} workflows apply conditionals to choose different execution paths based on the outputs of prior steps [C\cite{huang2018evorus,kittur2012crowdweaver} L\cite{huang2023pcr,wu2022promptchainer,zharikova2023deeppavlov}].

\vspace{0.5em}
\noindent\textbf{Redundant} architectures require different actors to complete the same task in parallel or iteratively. In parallel generation, multiple actors perform the same subtask separately [C\cite{weir2015learnersourcing, guo2019statelens, huang2017supporting, butler2017more, alshaibani2021pterodactyl, lin2014crowdsourced} L\cite{cobbe2021training}]. This subtask is often, but not always, followed by an aggregation or evaluation subtask to merge or select among the different parallel generations. Unlike parallelization that runs a similar subtask on different items at the same time (e.g., five crowdworkers label one image each for five labeled images total), parallel redundant architectures run the same subtask on the same item (e.g., five crowdworkers independently label the same image for one image total).
Iterative architectures run iteratively over a changing input. For example, iterative improvement takes the output of a subtask and iteratively gives it to a series of actors to improve, sometimes with interspersed decision tasks choosing between the current and prior iterations [C\cite{luther2015crowdlines, lasecki2015apparition, andre2014effects, latoza2014microtask, goto2016understanding, willett2012strategies, dai2013pomdp}]. The number of iterations can be determined to be a fixed number of steps [C\cite{kim2017mechanical, little2010turkit, zaidan2011crowdsourcing, huffaker2020crowdsourced} L\cite{wu2023llms}], by including a subtask to check for sufficiency [C\cite{kulkarni2014wish, salehi2017communicating, guo2016vizlens} L\cite{creswell2022selection, wang2022iteratively, creswell2022faithful, wu2023empirical, nair2023dera, liang2023encouraging, madaan2023self}], until the user stops calling iterations [C\cite{kulkarni2014wish, chilton2019visiblends, cheng2015flock, willett2012strategies} L\cite{reppert2023iterated}], until the outputs converge [C\cite{liem2011iterative} L\cite{schick2022peer, gero2023self, du2023improving}], until all quality checks are passed [C\cite{kim2017mechanical, marcus2011human, marcus2011demonstration, bragg2013crowdsourcing, chilton2013cascade, teevan2016supporting} L\cite{yang2022re3}], through increasing complexity levels [C\cite{alshaibani2020privacy, alshaibani2021pterodactyl}], and through decision-theoretic approaches [C\cite{dai2010decision} L\cite{park2022social}].

\begin{sloppypar}
 
\vspace{0.5em}
\noindent\textbf{Dynamic} architectures support changing the workflow architecture as it is executed. Map-reduce is a dynamic architecture in which actors break down a task recursively until the subtasks are an appropriate size [C\cite{liem2011iterative, valentine2017flash, kaur2017crowdmask, kim2016storia, kulkarni2012collaboratively, kittur2011crowdforge} L\cite{wu2023llms, huang2023ai}]. The literature defines such workflows as a hierarchy [C\cite{kittur2011crowdforge}]. Other dynamic workflow architectures include doing beam search across multiple options [L\cite{yao2023tree, creswell2022faithful, xie2023decomposition}] and allowing for actors to flexibly create tasks on a shared to-do list before choosing which of those to complete [C\cite{chilton2019visiblends, nebeling2016wearwrite, latoza2014microtask, mahyar2018communitycrit, kittur2011crowdforge, zhang2012human}]. 
Some other architecture methods may have dynamic elements, such as parallelizing an input into chunks and running the workflow for each chunk [C\cite{kaur2017crowdmask, kim2014crowdsourcing, andre2013community, pavel2014video, zaidan2011crowdsourcing, gouravajhala2018eureca}] or dynamically determining the number of iterative steps [C\cite{chilton2014frenzy, chen2019cicero} L\cite{zhou2022least, nair2023dera, madaan2023self}].
\end{sloppypar}

\begin{sloppypar}
 
\vspace{0.5em}
\noindent\textbf{Communicative} architectures can facilitate debate and conversation among actors [C\cite{teevan2016supporting, drapeau2016microtalk, chang2017revolt, kobayashi2018empirical, salisbury2017toward, chen2019cicero} L\cite{wu2023autogen, du2023improving, liang2023encouraging}] and allow workers to return feedback to the user [C\cite{nebeling2016wearwrite, kulkarni2012collaboratively, lasecki2015apparition}]. Some workflows create roles for different actors, often based on roles held in real-life organizations (e.g., journalism teams, software development teams, etc.) [C\cite{agapie2015crowdsourcing, valentine2017flash, retelny2014expert} L\cite{park2022social}]. Some papers integrate these roles into a hierarchy of communication, with managers, midlevel managers, and different forms of communication between different levels of the hierarchy [C\cite{agapie2015crowdsourcing, retelny2014expert}]. Communication may be restrained through subtasks [C\cite{chang2017revolt, chen2019cicero}], free-form though email [C\cite{agapie2015crowdsourcing, valentine2017flash}], or enabled through a shared interface [C\cite{kulkarni2014wish, lasecki2015apparition, lasecki2013chorus, nebeling2016wearwrite, mahyar2018communitycrit, lee2017sketchexpress}].
Communication among actors may be synchronous or asynchronous. Synchronous workflows help enable collaboration [C\cite{andre2014effects, chilton2019visiblends, salisbury2017toward, chilton2014frenzy}] and can reduce latency [C\cite{bernstein2011crowds}]. In asynchronous workflows or workflows with limited communication architectures, a working memory can help actors access progress reports, what other actors think are good next steps, and the overarching goals [C\cite{valentine2017flash, lasecki2013chorus, latoza2014microtask, hahn2016knowledge, zhang2012human} L\cite{wu2023autogen, yang2022re3, nair2023dera}].
\end{sloppypar}

%% file: sections/04.25_comparison.tex
\input{tables/crowd_coding}
\input{tables/llm_coding}

\subsection{Comparing the utilization of techniques}
\label{sec:ds-compare}

This section discusses the distribution of the crowdsourcing workflow and LLM chaining literature across the design space. We fit each paper in the corpus into its design space categories. The labeling procedure and outcomes are detailed in Supplementary materials and in Tables~\ref{tab:corpus-crowd} and~\ref{tab:corpus-llm}. Here we provide a snapshot of the LLM chaining sub-field as it expands, but these distributions will change over time with more published papers. 

Comparing these distributions first shows how both literatures cover all design space categories, with the natural exception of the \emph{LLM} and \emph{crowdworker} categories. The methods for creating the design space did not enforce overlap of the literatures, so the high degree of overlap between them provides a sanity check that both fields can share a design space. 

However, their distribution differs across some categories. For instance, a greater proportion of crowdsourcing papers promote the decrease of \emph{cost} compared to experts, while LLM chains' costs are usually compared to a baseline of one call to the LLM, inherently increasing the \emph{costs}. Due to the high range of tasks and sparsity of reporting, we cannot compare the scale of the relative costs of workflows and chains. Furthermore, differences in execution details hinder comparison across the literatures even for similar tasks. 

Across the objectivity-subjectivity spectrum, we find that crowdsourcing workflows more often implement a task with \emph{subjectivity}, while LLM chains more often have purely \emph{objective} tasks such as math. Finally, more chains in the LLM literature are designed to be \emph{generalizable} to many tasks, while a greater proportion of workflows in the crowdsourcing literature are designed for a \emph{specific} task function. 

The crowdsourcing workflow literature uses \emph{redundant} and especially \emph{communicative} architectures more frequently. More chaining papers use only a \emph{sequential} architecture than crowdsourcing papers (9 papers vs. 1 paper). There are slight differences in subtask use, in which the crowdsourcing workflow literature uses \emph{evaluate} and \emph{improve} subtasks more often, and the LLM chaining literature uses \emph{generate}, \emph{focus} and \emph{partition} subtasks more often. 
Finally, more papers in the crowdsourcing workflow literature explicitly incorporate the \emph{user} than those in the LLM chaining literature.

\begin{figure*}[t]
 \centering 
 \includegraphics[width=\linewidth]{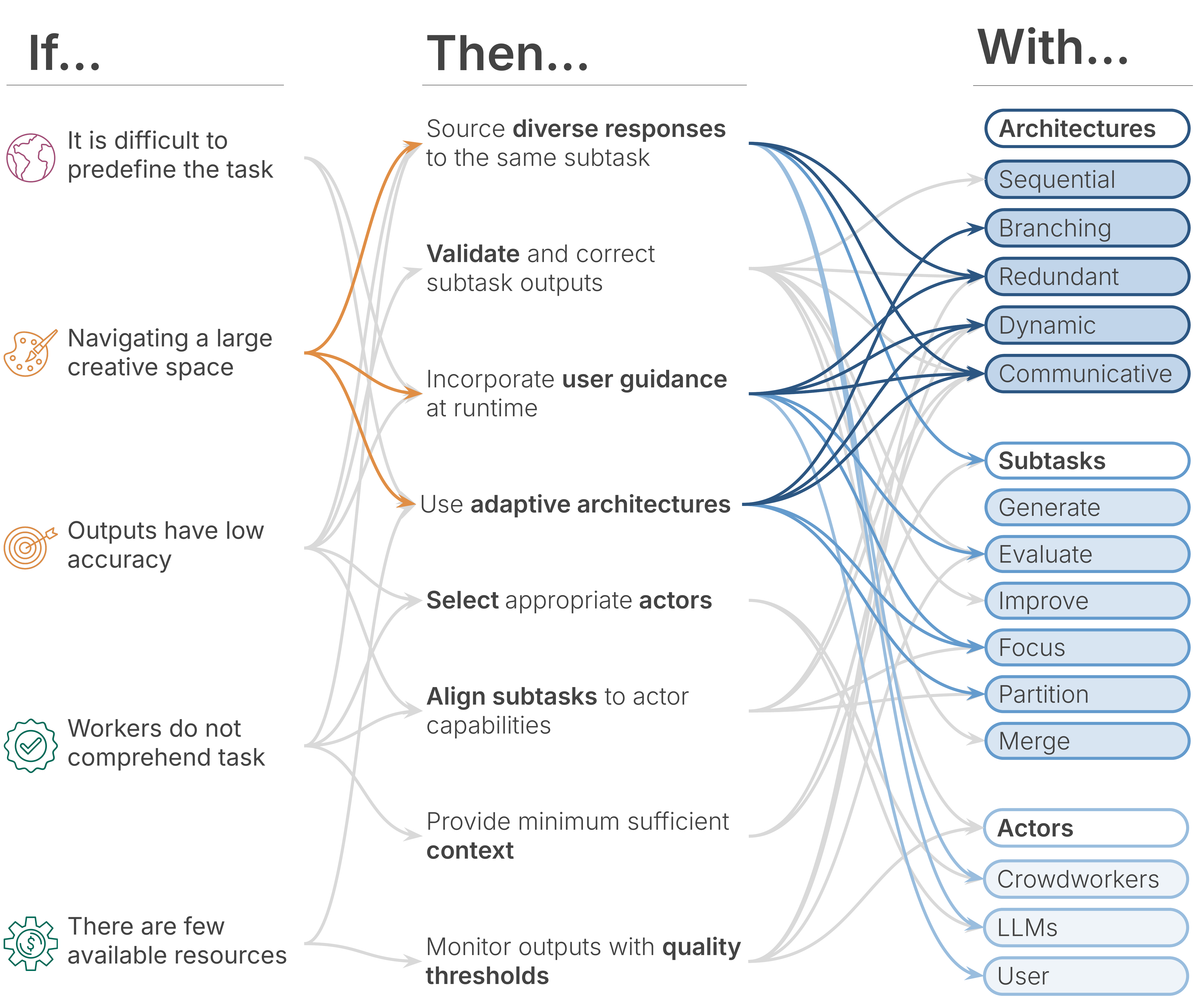}
 \caption{Design space connections. In working towards objectives, challenges can arise. This figure captures a subset of possible issues in achieving objectives. Based on our design space, these issues may be addressed by utilizing different strategies. In turn, these strategies are supported by various tactics. For example, if the objective requires navigating a large creative space of \emph{subjective} outcomes, a challenging proposition, then designers can incorporate the \emph{adaptive architectures}, \emph{diverse responses}, and \emph{user guidance} strategies. Each of these strategies has been implemented in prior work by a variety of architectures and other tactics. 
 }
 \Description{Three columns headed by "If...", "Then..." and "With...". The "If" column contains different possible challenges relating to achieving objectives. The "Then" column lists all of the strategies. There are arrows connecting the "If" column to the "Then" column, with arrows from "navigating a large creative space" objectives connect to the adaptive architectures, diverse responses, and user guidance strategies. The "With" column lists all of the tactics. There are arrows connecting the "Then" column to the "With" column. There are many arrows, and they correspond to the tactics used for strategies as defined in the text and other figures.}
\label{fig:ds-actions} 
\end{figure*}

Surfacing the differences between these distributions focuses attention on potentially informative sections of the design space.  These differences could result for a multitude of reasons. We speculate that the lower rate of ~\emph{subjective} tasks in chaining reflects a relative preference for benchmark evaluations in the LLM chaining community. The incorporation of the \emph{user} and the focus on \emph{task specificity} could reflect that crowdsourcing work was primarily housed in the Human-Computer Interaction field. The variations in these distributions could also reflect inherent differences between LLMs and crowdworkers or indicate areas of potential future work. 
 
Comparing the two distributions confirms the many similarities between crowdsourcing and chaining methods while focusing attention on the explanatory and actionable potential of the differences. We explore these differences in further detail in the case studies (Section~\ref{sec:case}) and the discussion (Section~\ref{sec:disc}).

%% file: tables/crowd_coding.tex
\begin{table*}[t]
\caption{Crowdsourcing workflow corpus categorization. Each row references a crowdsourcing workflow paper from our corpus, and each column represents a category of the design space. Colored cells indicate that the paper's workflow contains the category element. Details on the categorization process can be found in Supplementary materials. 
}
\label{tab:corpus-crowd}
\centering
\resizebox{0.89\linewidth}{!}{

\begin{tabular}{llll|llll|ll|ll|lllll|llllll|lll|}

\rotatebox{90}{Reference}        & \rotatebox{90}{Title}               & \rotatebox{90}{Venue} & \rotatebox{90}{Year} & \rotatebox{90}{Quality Increase} & \rotatebox{90}{Quality Decrease} & \rotatebox{90}{Cost Increase} & \rotatebox{90}{Cost Decrease} & \rotatebox{90}{Objectivity} & \rotatebox{90}{Subjectivity} & \rotatebox{90}{Task General} & \rotatebox{90}{Task Specific} & \rotatebox{90}{Sequential} & \rotatebox{90}{Branching}  & \rotatebox{90}{Redundant}  & \rotatebox{90}{Dynamic}    & \rotatebox{90}{Communicative} & \rotatebox{90}{Generate}   & \rotatebox{90}{Evaluate}   & \rotatebox{90}{Improve}    & \rotatebox{90}{Focus}      & \rotatebox{90}{Partition}  & \rotatebox{90}{Merge}      & \rotatebox{90}{Crowdworkers} & \rotatebox{90}{LLMs} & \rotatebox{90}{User}       \\ \hline
\cite{bernstein2010soylent}        & Soylent: a word processor with...      & UIST                     & 2010                    & \cellcolor[HTML]{006d5a}       & \cellcolor[HTML]{006d5a}       &                                  &                                  & \cellcolor[HTML]{e08e45}  & \cellcolor[HTML]{e08e45}   & \cellcolor[HTML]{a64d79}   &                                  & \cellcolor[HTML]{6b9ccd} &                               & \cellcolor[HTML]{6b9ccd} &  &                                  &                               & \cellcolor[HTML]{94b7dc} & \cellcolor[HTML]{94b7dc} & \cellcolor[HTML]{94b7dc} &                               &                               & \cellcolor[HTML]{bdd4ea}   &                         & \cellcolor[HTML]{bdd4ea} \\
\end{tabular}
}
\end{table*}

%% file: tables/llm_coding.tex
\begin{table*}[t]
\caption{LLM chain corpus categorization. The color cells indicate when a paper from the LLM chaining corpus (rows) includes an element of the design space (columns). }
\label{tab:corpus-llm}
\centering
\resizebox{0.89\linewidth}{!}{

\begin{tabular}{llll|llll|ll|ll|lllll|llllll|lll|}

\rotatebox{90}{Reference}       & \rotatebox{90}{Title}                  & \rotatebox{90}{Venue} & \rotatebox{90}{Year} & \rotatebox{90}{Quality Increase} & \rotatebox{90}{Quality Decrease} & \rotatebox{90}{Cost Increase} & \rotatebox{90}{Cost Decrease} & \rotatebox{90}{Objectivity} & \rotatebox{90}{Subjectivity} & \rotatebox{90}{Task General} & \rotatebox{90}{Task Specific} & \rotatebox{90}{Sequential} & \rotatebox{90}{Branching}  & \rotatebox{90}{Redundant}  & \rotatebox{90}{Dynamic}    & \rotatebox{90}{Communicative} & \rotatebox{90}{Generate}   & \rotatebox{90}{Evaluate}   & \rotatebox{90}{Improve}    & \rotatebox{90}{Focus}      & \rotatebox{90}{Partition}  & \rotatebox{90}{Merge}      & \rotatebox{90}{Crowdworkers} & \rotatebox{90}{LLMs}       & \rotatebox{90}{User}       \\ \hline
\cite{cobbe2021training}          & Training Verifiers to Solve Math...       & arxiv                    & 2021                    & \cellcolor[HTML]{006d5a}       &                                     &                                  & \cellcolor[HTML]{006d5a}    & \cellcolor[HTML]{e08e45}  &                                 & \cellcolor[HTML]{a64d79}   &                                  & \cellcolor[HTML]{6b9ccd} &                               & \cellcolor[HTML]{6b9ccd} &                               &                                  & \cellcolor[HTML]{94b7dc} & \cellcolor[HTML]{94b7dc} &                               &                               &                               &                               &                                 & \cellcolor[HTML]{bdd4ea} &                               \\
\end{tabular}
}
\end{table*}

%% file: sections/04.3_strategies.tex
\section{Strategies integrating the design space}
\label{sec:ds-strat}

Choosing \textbf{tactic} designs to support \textbf{objectives} is a non-trivial task, as combining different tactic types and implementations supports a huge range of possible workflows. However, through the thematic coding process we surfaced strategies by which tactics are used to support objectives (Figure~\ref{fig:ds-actions}). Specifically, these strategies (\ref{strategies:diverse}) encourage diverse responses, (\ref{strategies:validation}) validate outputs from subtasks, (\ref{strategies:guidance}) enable user guidance, (\ref{strategies:adaptable}) utilize adaptable architectures, (\ref{strategies:recruitment}) improve actor selection, (\ref{strategies:alignment}) align a subtask's required skills with an actor's capabilities, (\ref{strategies:context}) promote global context, and (\ref{strategies:threshold}) ensure outputs meet quality thresholds.

We outline the strategies that surfaced from the literature in a manner adapted from \emph{design patterns}~\cite{Alexander_1977,Gamma_1994} used in architecture and software engineering. For each, we define the strategy, explain the effects on objectives that motivate using the strategy, and how to implement the strategy with tactics. We then discuss secondary benefits, tradeoffs, examples from the literature, and connections to other strategies. 
While we note some differences between the crowdsourcing and chaining literatures throughout this section, we leave the bulk of such discussion to Section~\ref{sec:disc}.

\begin{figure}[t]
 \centering 
 \includegraphics[width=\linewidth]{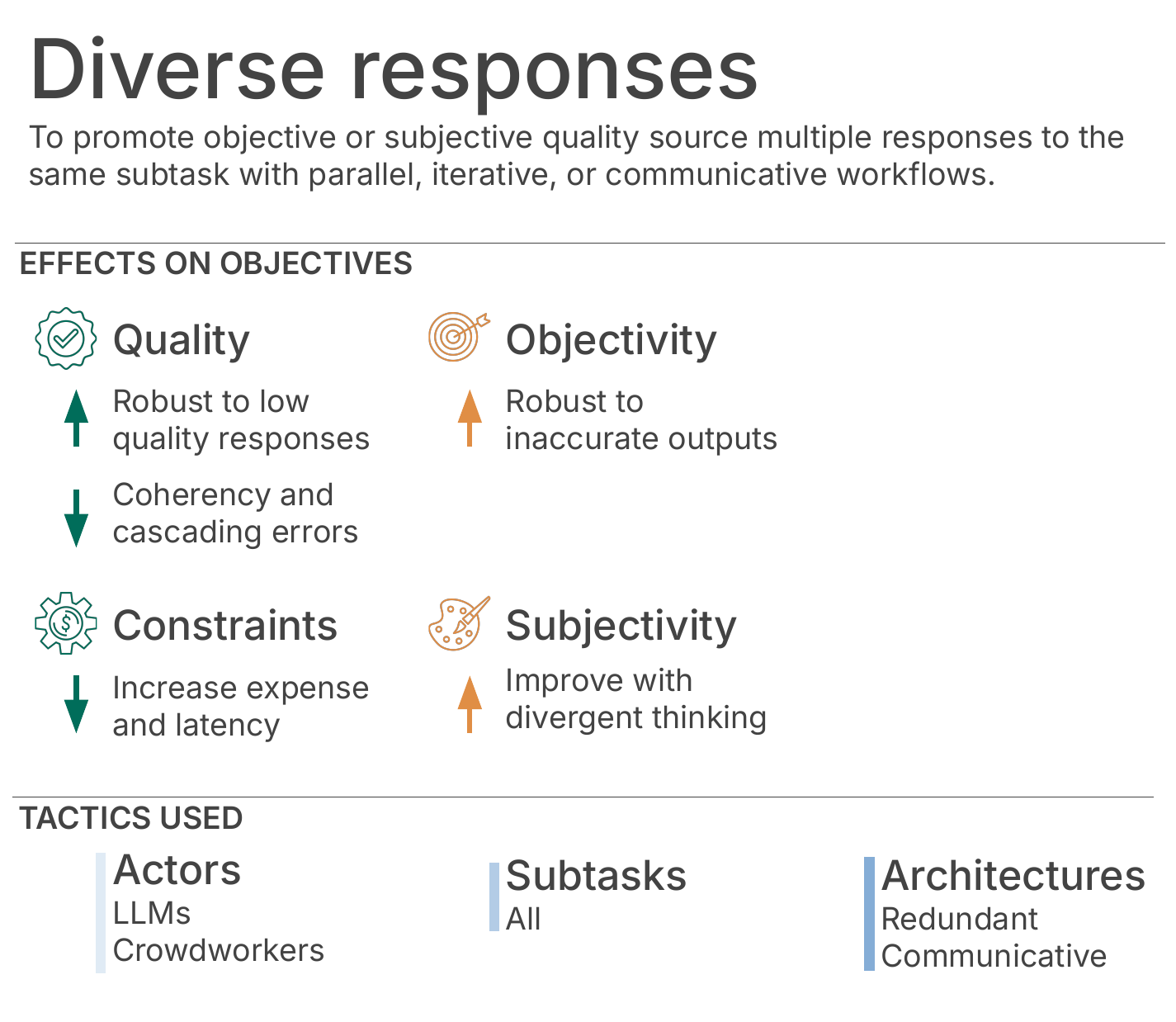}
 \caption{An overview of the tactics used for and the effects of the diverse responses strategy.}
 \Description{Title. Diverse responses, to promote objective or subjective quality, source multiple responses to the same subtask with parallel, iterative, or debate based workflows. Section 1: effects on objectives. Several objectives listed (with their effects). Quality (Robust to low quality responses, Coherency and cascading errors), Constraints (increase expense and latency), Objectivity (Robust to inaccurate outputs), Subjectivity (Improve with divergent thinking). Section 2: tactics used. Actors (LLMs, crowdworkers), Subtasks (all), Architectures (redundant, communicative).}
\label{fig:ds-strat-diverse} 
\end{figure}

\subsection{Diverse responses}
\label{strategies:diverse}
\textbf{To promote objective or subjective quality, source multiple responses to the same subtask with parallel, iterative, or communicative workflows (Figure~\ref{fig:ds-strat-diverse}).}

\vspace{0.5em}
\noindent\textbf{Motivation.}
Individual actors can give low quality outputs, affecting the \emph{quality} of the overall output. Designers may want to improve robustness to low quality outputs, outliers, and spammers to support the accuracy of objective outcomes [C\cite{liem2011iterative, lasecki2013chorus, huang2018evorus, hahn2016knowledge, zaidan2011crowdsourcing, ambati2012collaborative, huffaker2020crowdsourced} L\cite{du2023improving, reppert2023iterated}]. For \emph{subjective} tasks, they may want to encourage divergent and diverse thinking [C\cite{luther2015crowdlines, lasecki2015apparition, nebeling2016wearwrite, huang2017supporting, teevan2016supporting} L\cite{wu2023autogen, liang2023encouraging}].

\vspace{0.5em}
\noindent\textbf{Implementation.}
Getting responses from multiple actors on a subtask can help support these goals. Crowdsourcing workflows and LLM chaining source diverse responses differently. Crowdsourcing does so by querying different people. Chaining leverages the stochastic nature of LLMs (e.g.,~with a higher temperature parameter setting) or uses different models or prompt variants (e.g., role-based prompts) to source a variety of responses [L\cite{wu2023llms, park2022social, arora2022ask}].

\emph{Redundant} architectures are the most direct way of sourcing diverse responses, but \emph{communicative} architectures, such as those that structure debates, can source diverse responses as well. 
Some studies in crowdsourcing systematically study the impact of different architectures (e.g., parallel vs iterative) and architecture parameters (e.g., the number of iterations) on the diversity and \emph{quality} of the outcomes [C\cite{andre2014effects, kittur2011crowdforge, little2010exploring, goto2016understanding}]. However, it is unknown if these findings transfer to similar tactic choices with LLMs. 

\vspace{0.5em}
\noindent\textbf{Secondary benefits.}
There are some benefits of diverse responses that are solely discussed in the crowdsourcing literature, as they are dependent on the variations among people. Crowdsourcing elicits the benefit of having different viewpoints [C\cite{lasecki2013chorus, ambati2012collaborative, chen2019cicero}], filling in other crowdworkers' deficits to create a more rounded final product [C\cite{zhang2012human, kim2016storia}], and sourcing perspectives from expert crowdworkers [C\cite{chilton2014frenzy}] or from crowdworkers with various skill levels [C\cite{mahyar2018communitycrit}].

\vspace{0.5em}
\noindent\textbf{Consequences.}
Sourcing diverse responses is inherently more \emph{costly} because multiple actors complete the same subtask. 
Merging multiple responses can induce \emph{outcome quality} errors: coherency for parallel architectures [L\cite{wu2022promptchainer}] and cascading errors for iterative architectures [C\cite{goto2016understanding}].

\vspace{0.5em}
\noindent\textbf{Examples.}
Examples of the diverse responses strategy include: Apparition [C\cite{lasecki2015apparition}], chains for multi-agent debate [L\cite{liang2023encouraging}], and an exploration of "human computation processes" [C\cite{little2010exploring}].

\vspace{0.5em}
\noindent\textbf{Related strategies.}
Agreement among diverse responses can serve as \emph{validation}. Setting \emph{quality thresholds} can impact the number of diverse responses requested.

\subsection{Validation}
\label{strategies:validation}
\begin{figure}[t]
 \centering 
 \includegraphics[width=\linewidth]{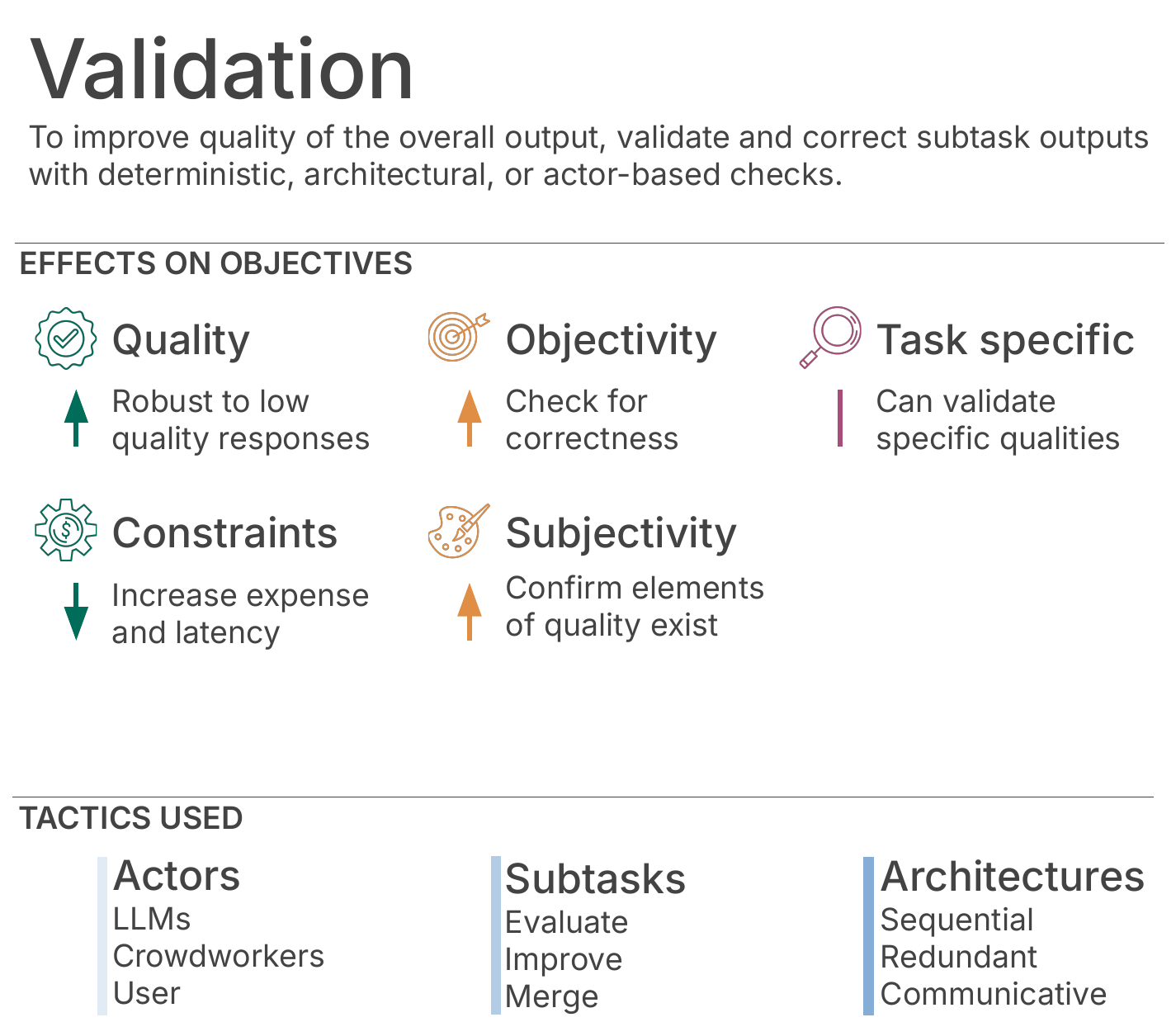}
 \caption{An overview of the tactics used for and the effects of the validation strategy.}
 \Description{Title. Validation, to improve quality of the overall output, validate and correct subtask outputs with deterministic, architectural, or actor-based checks. Section 1: effects on objectives. Several objectives listed (with their effects). Quality (Robust to low quality responses), Constraints (increase expense and latency), Objectivity (check for correctness), Subjectivity (confirm elements of quality exist), Task specific (can validate specific qualities). Section 2: tactics used. Actors (LLMs, crowdworkers, user), Subtasks (evaluate, improve, merge), Architectures (sequential, redundant, communicative).}
\label{fig:ds-strat-validate} 
\end{figure}

\textbf{To improve quality of the overall output, validate and correct subtask outputs with deterministic, architectural, or actor-based checks (Figure~\ref{fig:ds-strat-validate}).}

\vspace{0.5em}
\noindent\textbf{Motivation.}
Individual actors can give low quality responses. Unaddressed, these can lead to cascading errors through the workflow [C\cite{little2010exploring, little2010turkit, goto2016understanding, bernstein2010soylent} L\cite{wu2022promptchainer}]. Therefore, higher quality subtask outcomes throughout the workflow can have myriad benefits on the overall \emph{outcome quality} for both \emph{objectivity} and \emph{subjectivity} focused tasks, increasing robustness to low-quality actor outputs [C\cite{ambati2012collaborative, bernstein2010soylent}].

\vspace{0.5em}
\noindent\textbf{Implementation.} 
The validation process is often \emph{sequential}, in which content is first \emph{generated} before being validated [L\cite{gero2023self}]. 
In parallel and iterative \emph{redundant} workflows, errors from prior passes can be found and fixed or rejected through \emph{improve} subtasks [C\cite{hahn2016knowledge, zaidan2011crowdsourcing}] and differing opinions can be merged [C\cite{luther2015crowdlines, kittur2011crowdforge}] through \emph{merge} and \emph{evaluate} subtasks [C\cite{ambati2012collaborative, kulkarni2012collaboratively} L\cite{wu2023autogen, zharikova2023deeppavlov, wu2023llms, schick2022peer}]. Users can break ties [C\cite{pietrowicz2013crowdband}] and intervene if there are cascading errors [C\cite{kulkarni2012collaboratively}]. Non-worker methods, such as taking the majority vote of the outputs of redundant \emph{evaluate} subtasks can also be effective. Automated methods of flagging inattentive crowdworkers, gold-standard validation steps, or requiring a certain agreement threshold among actors within a \emph{redundant} architecture can also provide validation [C\cite{liem2011iterative, huffaker2020crowdsourced, lasecki2015apparition} L\cite{parameswaran2023revisiting}]. 

Debate-based \emph{communicative} workflows can promote reflection on disagreements [C\cite{kobayashi2018empirical, chen2019cicero} L\cite{liang2023encouraging, du2023improving}]. 
Using different models and roles in debate shows promise [L\cite{du2023improving}], as an LLM judge in a debate was found to more likely agree with its same underlying model [L\cite{liang2023encouraging}].
Other workflows use \emph{adaptable architectures} to make all production always editable so actors can fix errors in an adaptable manner [C\cite{liu2018conceptscape, latoza2014microtask}].

\vspace{0.5em}
\noindent\textbf{Consequences.}
\emph{Costs} increase when validation involves more calls to actors. The literature shows mixed results for an LLM's ability to validate its own responses in \emph{evaluate} subtasks. Some studies suggest that adding a self-validation step improves outcomes [L\cite{madaan2023self, gero2023self, yao2023tree, xie2023decomposition}], while others suggest that LLMs are likely to agree with their previous outputs, even when incorrect, and propose debate \emph{communicative} workflows as an alternative [L\cite{du2023improving, liang2023encouraging}].

\vspace{0.5em}
\noindent\textbf{Examples.}
Examples of the validation strategy include: the ``Verify'' step of Soylent [C\cite{bernstein2010soylent}], a Dual-Pathways structure for transcription [C\cite{liem2011iterative}], and self-evaluation guided reasoning [L\cite{xie2023decomposition}].

\vspace{0.5em}
\noindent\textbf{Related strategies.}
Validation often involves considering the agreement among \emph{diverse responses}.

\subsection{User guidance}
\label{strategies:guidance}
\begin{figure}[t]
 \centering 
 \includegraphics[width=\linewidth]{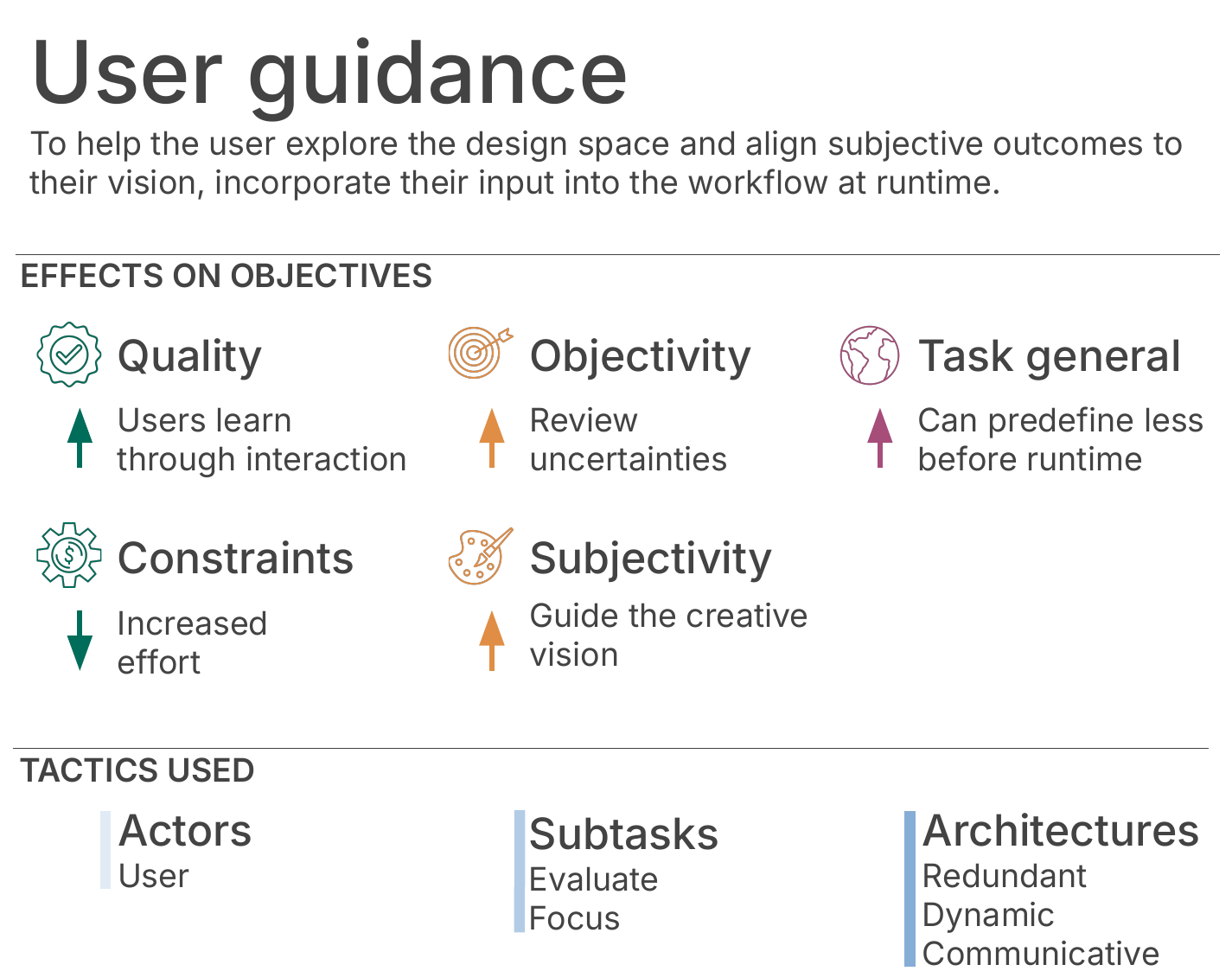}
 \caption{An overview of the tactics used for and the effects of the user guidance strategy.}
 \Description{Title. User Guidance. To help the user explore the design space and align subjective outcomes to their vision, incorporate their input into the workflow at runtime. Section 1: effects on objectives. Several objectives listed (with their effects). Quality (Users learn through interaction), Constraints (Increased
effort), Objectivity (Review uncertainties), Subjectivity (Guide the creative vision), Task general (Can predefine less before runtime). Section 2: tactics used. Actors (user), Subtasks (evaluate, focus), Architectures (redundant, dynamic, communicative).}
\label{fig:ds-strat-user} 
\end{figure}

\textbf{To help the user explore the design space and align subjective outcomes to their vision, incorporate their input into the workflow at runtime (Figure~\ref{fig:ds-strat-user}).}

\vspace{0.5em}
\noindent\textbf{Motivation.}
Users may want to use the workflow to reduce the \emph{effort} of exploring the design space, to align the workflow's output to their creative vision, or some combination thereof [C\cite{kulkarni2014wish, nebeling2016wearwrite, willett2012strategies} L\cite{park2022social}]. If a process includes inherently uncertain data it could benefit from more explicit guidance by the user [C\cite{willett2012strategies, cheng2015flock, chang2017revolt}].

\vspace{0.5em}
\noindent\textbf{Implementation.}
This strategy incorporates the user into a critical role within the workflow.
Users can guide outcomes by dynamically answering questions and providing feedback [C\cite{salehi2017communicating}] or monitor \emph{dynamic} workflows to catch cascading errors [C\cite{kulkarni2012collaboratively}] with \emph{evaluate} subtasks. Users can be called upon when necessary with a ``panic button'' that other workers can invoke [C\cite{kulkarni2012collaboratively}], or automatically when ties occur [C\cite{pietrowicz2013crowdband}]. With a shared \emph{communicative} interface between users and all workers, users can intervene when needed and make real-time suggestions for prototyping directions [C\cite{nebeling2016wearwrite, lasecki2015apparition}]. Users may review at intermediate steps to edit or regenerate sections of the workflow in \emph{evaluate} or \emph{focus} subtasks [C\cite{kulkarni2014wish} L\cite{park2022social, kim2023metaphorian, zhang2023visar}].
A \emph{redundant} iterative architecture may also include multiple iterations with user revisions between each one [C\cite{kulkarni2014wish, cheng2015flock, willett2012strategies} L\cite{kim2023metaphorian, park2022social}]. 
Sometimes disagreements raised with \emph{redundant} architectures may not need to be resolved within the workflow and can instead support an explorable interface for users to directly manipulate [C\cite{chang2017revolt, cheng2015flock, mohanty2019second, willett2012strategies}]. They can sort and filter the returned data [C\cite{cheng2015flock, mohanty2019second, willett2012strategies}], use a slider to control output characteristics like text length [C\cite{bernstein2010soylent}], and consider category labels at multiple levels of precision [C\cite{chang2017revolt}]. 

Crowdsourcing workflows integrate the user in more varied ways than LLM chaining, including providing results without aggregation to sort and filter [C\cite{mohanty2019second, lin2014crowdsourced}], user guidance for prototyping [C\cite{kulkarni2014wish, cheng2015flock, retelny2014expert}], and actor-initiated user revision [C\cite{salehi2017communicating}].

\vspace{0.5em}
\noindent\textbf{Secondary benefits.}
Incorporating the user into the process at runtime can reduce the amount of pre-definition of workflow steps needed, making workflows more \emph{generally} applicable [C\cite{kulkarni2014wish, nebeling2016wearwrite}].

\vspace{0.5em}
\noindent\textbf{Consequences.}
These benefits come at the cost of more \emph{effort} for the user than an entirely automatic workflow.

\vspace{0.5em}
\noindent\textbf{Examples.}
Examples of the user guidance strategy include: Second Opinion [C\cite{mohanty2019second}], Social Simulacra [L\cite{park2022social}], and WearWrite [C\cite{nebeling2016wearwrite}].

\vspace{0.5em}
\noindent\textbf{Related strategies.}
Incorporating user guidance can aid with \emph{validation}.

\subsection{Adaptable architectures}
\label{strategies:adaptable}
\begin{figure}[t]
 \centering 
 \includegraphics[width=\linewidth]{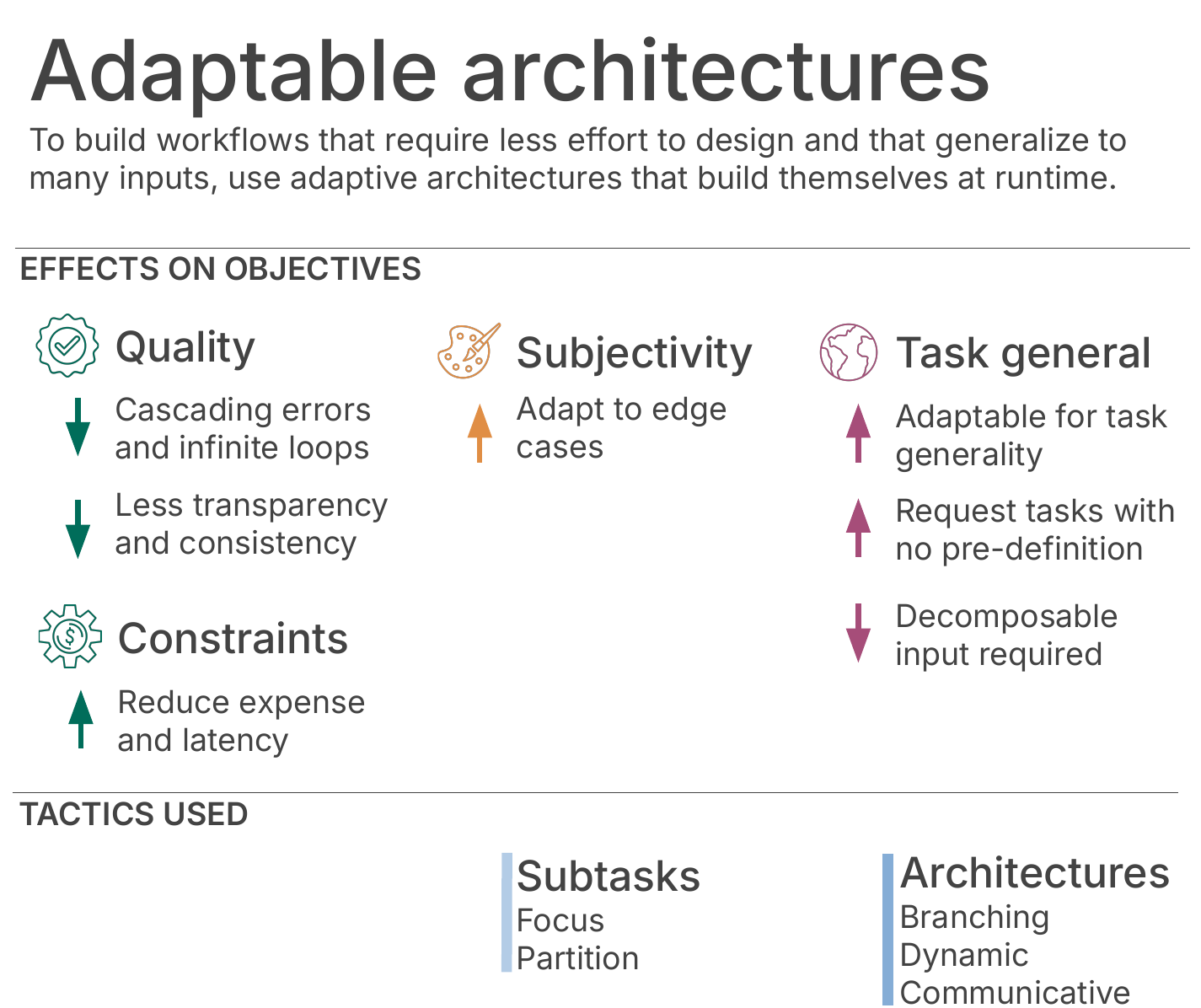}
 \caption{An overview of the tactics used for and the effects of the adaptable architectures strategy.}
 \Description{Title. Adaptable Architectures. To build workflows that require less effort to design and that generalize to many inputs, use adaptive architectures that build themselves at runtime. Section 1: effects on objectives. Several objectives listed (with their effects). Quality (Cascading errors and infinite loops, Less transparency and consistency), Constraints (Reduce expense and latency), Subjectivity (Adapt to edge cases), Task general (Adaptable for task generality, Request tasks with no pre-definition, Decomposable input required). Section 2: tactics used. Subtasks (focus, partition), Architectures (branching, dynamic, communicative).}
\label{fig:ds-strat-adaptable} 
\end{figure}

\textbf{To build workflows that require less effort to design and that generalize to many inputs, use adaptive architectures that build themselves at runtime (Figure~\ref{fig:ds-strat-adaptable}).}

\vspace{0.5em}
\noindent\textbf{Motivation.}
It is challenging and requires high \emph{effort} to design a rigid and pre-defined workflow that can effectively \emph{generalize} to many tasks or that can accommodate for many edge-cases, especially in \emph{subjective} tasks [C\cite{kulkarni2012collaboratively, kittur2011crowdforge, valentine2017flash, kim2016storia}]. 
Instead, designers may want workflows that are adaptable and responsive [C\cite{kulkarni2014wish, valentine2017flash, kim2016storia, chen2019cicero} L\cite{wu2023autogen, creswell2022faithful, nair2023dera}] and to support a variety of inputs without the effort of explicit pre-definition [C\cite{kulkarni2012collaboratively, kittur2011crowdforge} L\cite{creswell2022selection, creswell2022faithful}]. 

\vspace{0.5em}
\noindent\textbf{Implementation.}
Adaptable workflows are not rigid in their progression of subtasks but instead are flexible and generate parts of the workflow architecture at runtime [C\cite{kulkarni2014wish, valentine2017flash, kim2016storia, kittur2011crowdforge, chen2019cicero} L\cite{huang2023pcr, huang2023ai}]. 
Workflows can incorporate subtasks defined by the crowd with \emph{partition} and \emph{focus} subtasks in \emph{dynamic} and \emph{communicative} architectures [C\cite{agapie2015crowdsourcing, retelny2014expert, zhang2012human, ahmad2011jabberwocky} L\cite{reppert2023iterated}]. 
Changes in these adaptable workflows may be made by the worker [C\cite{valentine2017flash, latoza2014microtask, retelny2014expert}], user [C\cite{kulkarni2012collaboratively, nebeling2016wearwrite, valentine2017flash, retelny2014expert}], or an algorithm [C\cite{bragg2013crowdsourcing, dai2013pomdp, chen2019cicero, goto2016understanding}].
Although \emph{branching} architectures are predefined, they also allow for adaptability in which different inputs take different paths.

\vspace{0.5em}
\noindent\textbf{Consequences.}
However, there are tradeoffs. Adaptable workflows can lead to poor quality worker-created subtasks and instructions, causing cascading errors and infinite loops [C\cite{kulkarni2012collaboratively}]. They also are restricted to tasks with an easily decomposable input [C\cite{kittur2011crowdforge}]. 

Chaining papers highly emphasize the importance of transparency [L\cite{wu2023autogen, creswell2022selection, creswell2022faithful, Wu2021AICT, gero2023self, wu2022promptchainer, reppert2023iterated}], increasing the risk of using adaptable workflows which may cause unforeseeable errors as each step is not predefined. The degree to which adaptable workflows degrade this transparency is yet to be explicitly explored.

Some effects of adaptable workflows on worker experiences matter for crowdsourcing, but do not matter for chaining. 
Crowdsourcing work has allowed crowdworkers to chose their own tasks as an incentive to improve task performance [C\cite{chilton2019visiblends, nebeling2016wearwrite, latoza2014microtask, mahyar2018communitycrit, kittur2011crowdforge, zhang2012human}], but the need to incentivize does not apply to chaining. 
One crowdsourcing study found users were frustrated with too much adaptability, an emotion that does not apply to LLMs [C\cite{valentine2017flash}].

\begin{figure}[t]
 \centering 
 \includegraphics[width=\linewidth]{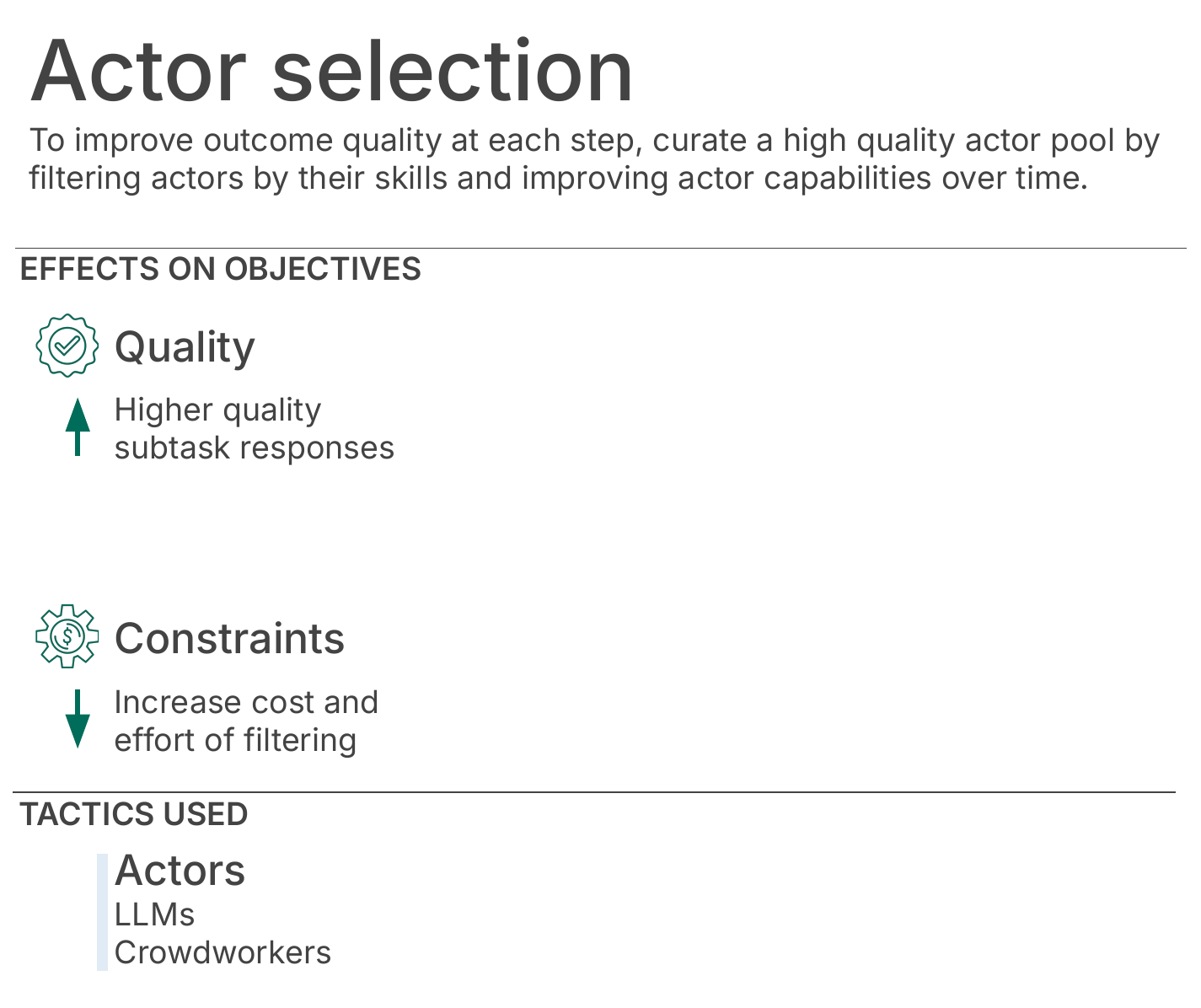}
 \caption{An overview of the tactics used for and the effects of the actor selection strategy.}
 \Description{Title. Actor selection, To improve outcome quality at each step, curate a high quality actor pool by filtering actors by their skills and improving actor capabilities over time. Section 1: effects on objectives. Several objectives listed (with their effects). Quality (Higher quality subtask responses), Constraints (Increase cost and effort of filtering). Section 2: tactics used. Actors (LLMs, crowdworkers).}
\label{fig:ds-strat-recruit} 
\end{figure}

\begin{sloppypar}
\vspace{0.5em}
\noindent\textbf{Examples.}
Examples of adaptable architectures include: map-reduce [C\cite{kittur2011crowdforge}], Flash Teams [C\cite{retelny2014expert}], and microtask programming [C\cite{latoza2014microtask}].
\end{sloppypar}

\vspace{0.5em}
\noindent\textbf{Related strategies.}
Adaptable architectures must support the dynamic creation of \emph{aligned subtasks} that have well defined \emph{context} and worker instructions. Adaptable architectures can also dynamically change course once \emph{quality thresholds} are met. 

\subsection{Actor selection}
\label{strategies:recruitment}
\textbf{To improve outcome quality at each step, curate a high quality actor pool by filtering actors by their skills and improving actor capabilities over time (Figure~\ref{fig:ds-strat-recruit}).}

\vspace{0.5em}
\noindent\textbf{Motivation.}
Actors that provide higher-quality responses support \emph{outcome quality} [C\cite{drapeau2016microtalk,huang2015guardian,ahmad2011jabberwocky}].

\vspace{0.5em}
\noindent\textbf{Implementation.}
In crowdsourcing, selecting high-quality \emph{actors} can mean filtering crowdworkers by pre-specified qualifications or using platforms with more specialized crowdworkers such as Upwork [C\cite{valentine2017flash}]. With support, crowdworkers can improve performance over time [C\cite{chen2019cicero,weir2015learnersourcing,drapeau2016microtalk,kobayashi2018empirical}]. For LLM chains, model choice, specialization, and finetuning impact  abilities [L\cite{creswell2022selection, cobbe2021training, bursztyn2022learning, creswell2022faithful}]. 

\vspace{0.5em}
\noindent\textbf{Consequences.}
Higher quality actors may be more \emph{expensive} [C\cite{huang2015guardian, christoforaki2014step} L\cite{bursztyn2022learning}], although smaller specialized models may reduce \emph{expense} [L\cite{bursztyn2022learning}]. There is also \emph{effort} and \emph{expense} involved in filtering crowdworkers for skill [C\cite{kulkarni2014wish, zaidan2011crowdsourcing, valentine2017flash}] or finetuning models to improve specific skills [L\cite{bursztyn2022learning}]. 

\vspace{0.5em}
\noindent\textbf{Examples.}
Examples that pay particular attention to actor selection include Flash Organizations [C\cite{valentine2017flash}], and chains that finetune models [L\cite{bursztyn2022learning}] or use multiple models [L\cite{liang2023encouraging}].

\vspace{0.5em}
\noindent\textbf{Related strategies.}
Actor selection works hand in hand with \emph{aligning subtasks} to actor capabilities. Actor selection must also be considered when using \emph{adaptive architectures} to define what kind of workers to hire during runtime.

\subsection{Subtask alignment}
\label{strategies:alignment}
\begin{figure}[t]
 \centering 
 \includegraphics[width=\linewidth]{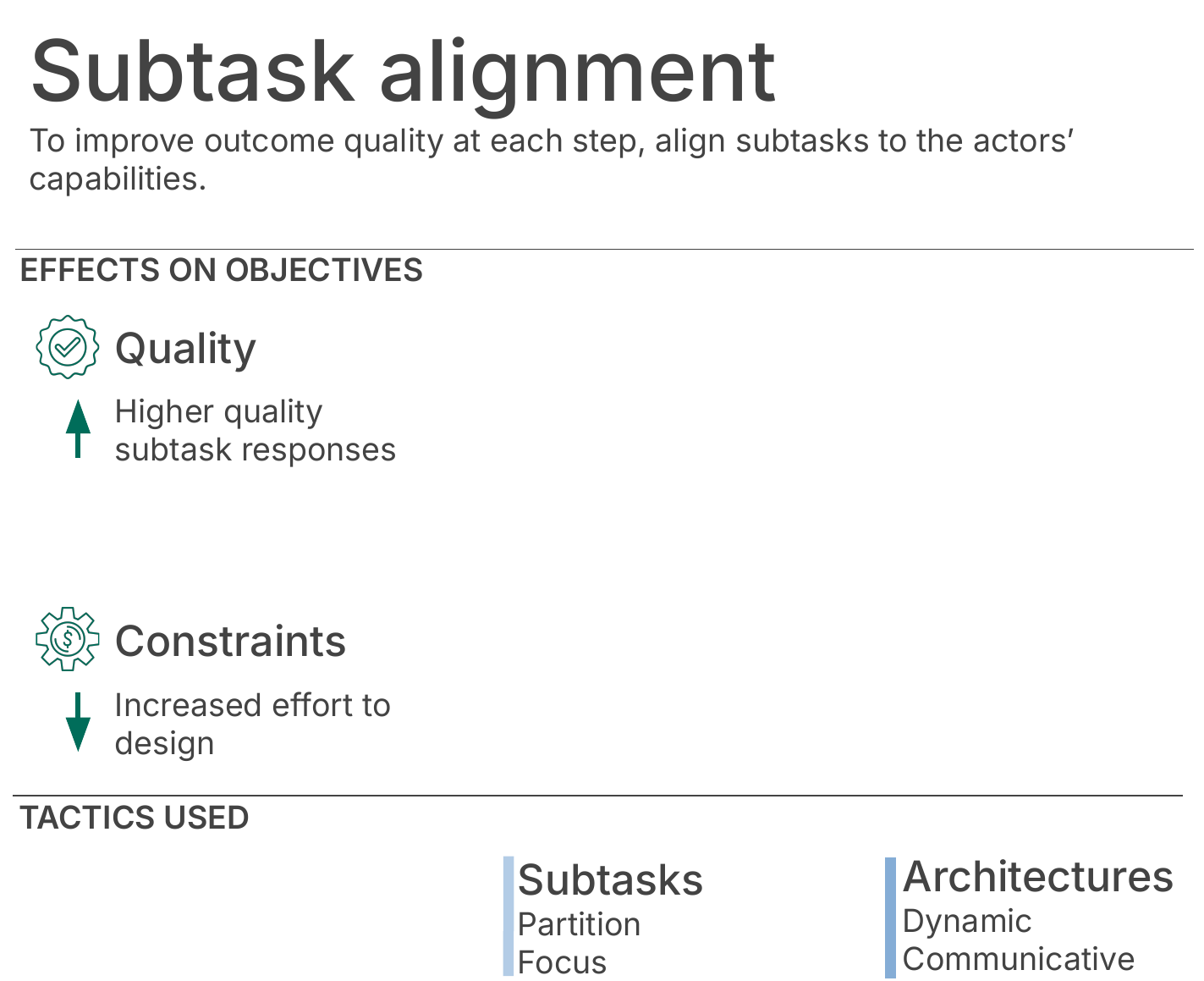}
 \caption{An overview of the tactics used for and the effects of the subtask alignment strategy.}
 \Description{Title. Subtask alignment, To improve outcome quality at each step, align subtasks to actor capabilities. Section 1: effects on objectives. Several objectives listed (with their effects). Quality (Higher quality subtask responses), Constraints (Increased effort to design). Section 2: tactics used. Subtasks (partition, focus), Architectures (dynamic, communicative).}
\label{fig:ds-strat-align} 
\end{figure}

\textbf{To improve outcome quality at each step, align subtasks to actor capabilities (Figure~\ref{fig:ds-strat-align}).}

\vspace{0.5em}
\noindent\textbf{Motivation.}
One critical strategy is ensuring that \emph{subtasks} match the \emph{actors'} capabilities appropriately.
Simplifying tasks can increase \emph{outcome quality} [C\cite{lasecki2013chorus, chilton2013cascade, pietrowicz2013crowdband} L\cite{Wu2021AICT}]. 

\begin{sloppypar}
 
\vspace{0.5em}
\noindent\textbf{Implementation.}
Compared to entire tasks, smaller subtasks are often easier to complete, validate, and correct [C\cite{ambati2012collaborative} L\cite{creswell2022selection, Wu2021AICT, creswell2022faithful, yang2022re3, kim2023metaphorian, schick2022peer}].
Designers can also adapt the subtask content to fit the actor's capabilities [C\cite{pietrowicz2013crowdband, mohanty2019second, chang2017revolt} L\cite{arora2022ask}]. For example, researchers found that crowdworkers were better at generating predictive features than at estimating if a feature is predictive; so they adapted the workflow accordingly [C\cite{cheng2015flock}].
\end{sloppypar}

When the appropriate subtasks are unknown, \emph{dynamic} architectures can use \emph{partition} or \emph{focus} subtasks to use actors' input to break down tasks into an appropriate size at runtime [C\cite{kittur2011crowdforge, latoza2014microtask, kim2017mechanical} L\cite{huang2023ai}]. \emph{Communicative} architectures can create appropriate subtasks for actors when they have a running ``todo'' list from which actors can choose their preferred task of many [C\cite{mahyar2018communitycrit, zhang2012human, lasecki2015apparition}].

\vspace{0.5em}
\noindent\textbf{Consequences.}
The capabilities of LLMs and appropriate subtask content are far from fully understood. Furthermore, it is unknown the degree to which LLMs can judge the appropriateness of subtasks for other LLMs to complete, bringing into question the effectiveness of some \emph{partition} subtask and \emph{communicative} architecture approaches taken when the actors are \emph{crowdworkers}. 
Therefore, testing out these different options to find appropriate subtasks is \emph{resource} intensive in the design process.

\vspace{0.5em}
\noindent\textbf{Examples.}
Examples that pay particular attention to subtask alignment include: Revolt [C\cite{chang2017revolt}], CrowdBand [C\cite{pietrowicz2013crowdband}], and Ask Me Anything [L\cite{arora2022ask}].

\vspace{0.5em}
\noindent\textbf{Related strategies.}
Subtasks must align to the capabilities of \emph{selected actors} and requirements must be communicated with sufficient \emph{context}. Subtask alignment is especially challenging when subtasks are created during runtime with \emph{adaptable architectures}.

\subsection{Context}
\label{strategies:context}
\begin{figure}[t]
 \centering 
 \includegraphics[width=\linewidth]{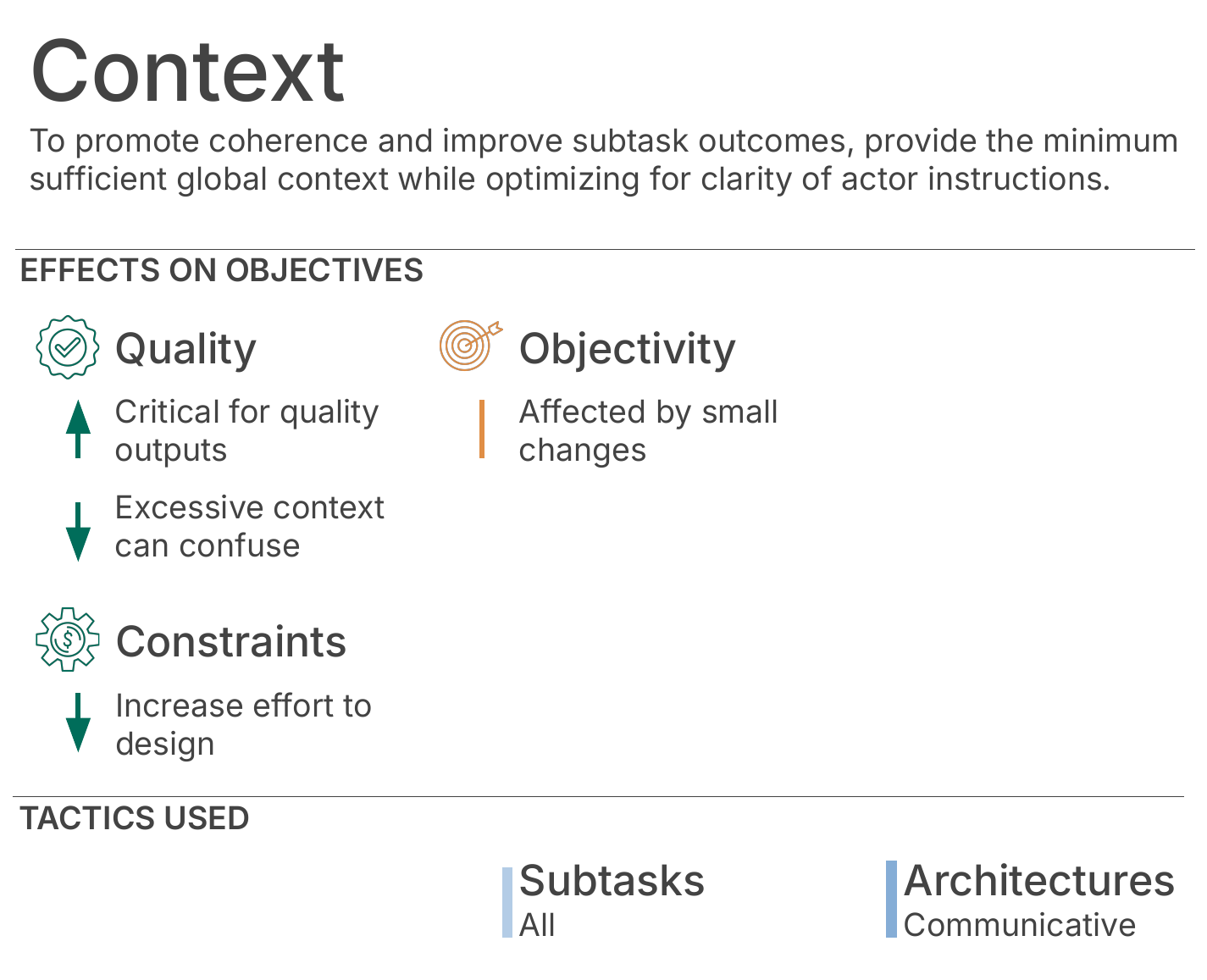}
 \caption{An overview of the tactics used for and the effects of the context strategy.}
 \Description{Title. Context, To promote coherence and improve subtask outcomes, provide the minimum sufficient global context while optimizing for clarity of actor instructions. Section 1: effects on objectives. Several objectives listed (with their effects). Quality (Critical for quality outputs, Excessive context can confuse), Constraints (Increase effort to design), Objectivity (Affected by small changes). Section 2: tactics used. Subtasks (all), Architectures (communicative).}
\label{fig:ds-strat-context} 
\end{figure}

\textbf{To promote coherence and improve subtask outcomes, provide the minimum sufficient global context while optimizing for clarity of actor instructions (Figure~\ref{fig:ds-strat-context}).}

\vspace{0.5em}
\noindent\textbf{Motivation.}
Workflows need to have high \emph{quality} subtask outputs that promote cohesion among disparate parts of the workflow. 

\vspace{0.5em}
\noindent\textbf{Implementation.}
Communicating context to workers can help with improving the quality of the output and creating more coherence among disparate subtasks [C\cite{luther2015crowdlines, chilton2019visiblends, hahn2016knowledge, teevan2016supporting, kulkarni2012collaboratively, bernstein2010soylent, ambati2012collaborative} L\cite{wu2022promptchainer}]. There are many forms of contextual communication, from input text [C\cite{kulkarni2012collaboratively, ambati2012collaborative, bernstein2010soylent}], to workflow goal tracking [C\cite{kim2017mechanical} L\cite{yang2022re3, mirowski2023co}], to explicit communication between users and workers [C\cite{salehi2017communicating}].

Instruction wording is a critical part of all \emph{subtask} designs to support high \emph{outcome quality} [L\cite{parameswaran2023revisiting}]. 
A key difficulty is communicating the user's intent through language. One study reports that crowdworkers tend to need less explicitly defined instruction than LLMs but can still struggle with vague or ambiguous instructions [L\cite{wu2023llms}]. Both crowdworkers and LLMs benefit from examples, namely screening and tutorial tasks for crowdworkers [C\cite{pietrowicz2013crowdband, lasecki2015apparition, lasecki2013chorus, teevan2016supporting, drapeau2016microtalk, bernstein2010soylent, chen2019cicero}] and few-shot examples for LLMs [L\cite{gero2023self, nair2023generating}]. 
Interfaces for crowdworkers can structure their output, while this structure needs to be communicated to LLMs in natural language [C\cite{kim2014crowdsourcing} L\cite{wu2023llms, creswell2022faithful}].

\begin{sloppypar}
 
\vspace{0.5em}
\noindent\textbf{Consequences.}
Although context is generally important, overcomplicating instructions can lead to confusion [C\cite{lasecki2015apparition, chen2019cicero} L\cite{madaan2023self, wu2023llms}], so giving the right amount of context is important [C\cite{kittur2011crowdforge}]. One in-depth crowdsourcing study found that providing context from the user was helpful early on in the workflow if content is poor but becomes detrimental if communicated later or with high-quality content [C\cite{salehi2017communicating}]. Minimizing context seen can also help with preserving privacy [C\cite{bragg2021asl, kaur2017crowdmask}].
For LLMs, context may be included as part of the prompt, although LLMs have the additional barrier of an input token limit [L\cite{mirowski2023co, park2022social}]. Therefore, workflow designs should aim for a minimum amount of sufficient context, to provide coherence while avoiding confusion. 
\end{sloppypar}

For both crowdworkers and LLMs, small changes in instructions or prompts can lead to dramatically different outcomes, impacting the accuracy of the final result [C\cite{mcdonnell2016relevant, kittur2012crowdweaver} L\cite{Wu2021AICT, parameswaran2023revisiting, wu2023llms, gero2023self, wu2022promptchainer, mirowski2023co, arora2022ask, huang2023ai}].
Additionally, different LLMs may respond differently to the same input prompt [L\cite{parameswaran2023revisiting}]. Therefore, figuring out the clearest instructions requires nontrivial \emph{effort} and testing.

\vspace{0.5em}
\noindent\textbf{Examples.}
Examples that pay particular attention to context include Turkomatic [C\cite{kulkarni2012collaboratively}], collaborative writing with microtasks [C\cite{teevan2016supporting}], and a study on communicating context [C\cite{salehi2017communicating}].

\vspace{0.5em}
\noindent\textbf{Related strategies.}
When using \emph{adaptable architectures}, dynamically made tasks can struggle to provide quality worker instructions and context. Additionally, the effectiveness of this strategy is limited if there is improper \emph{subtask alignment}.

\begin{figure}[t]
 \centering 
 \includegraphics[width=\linewidth]{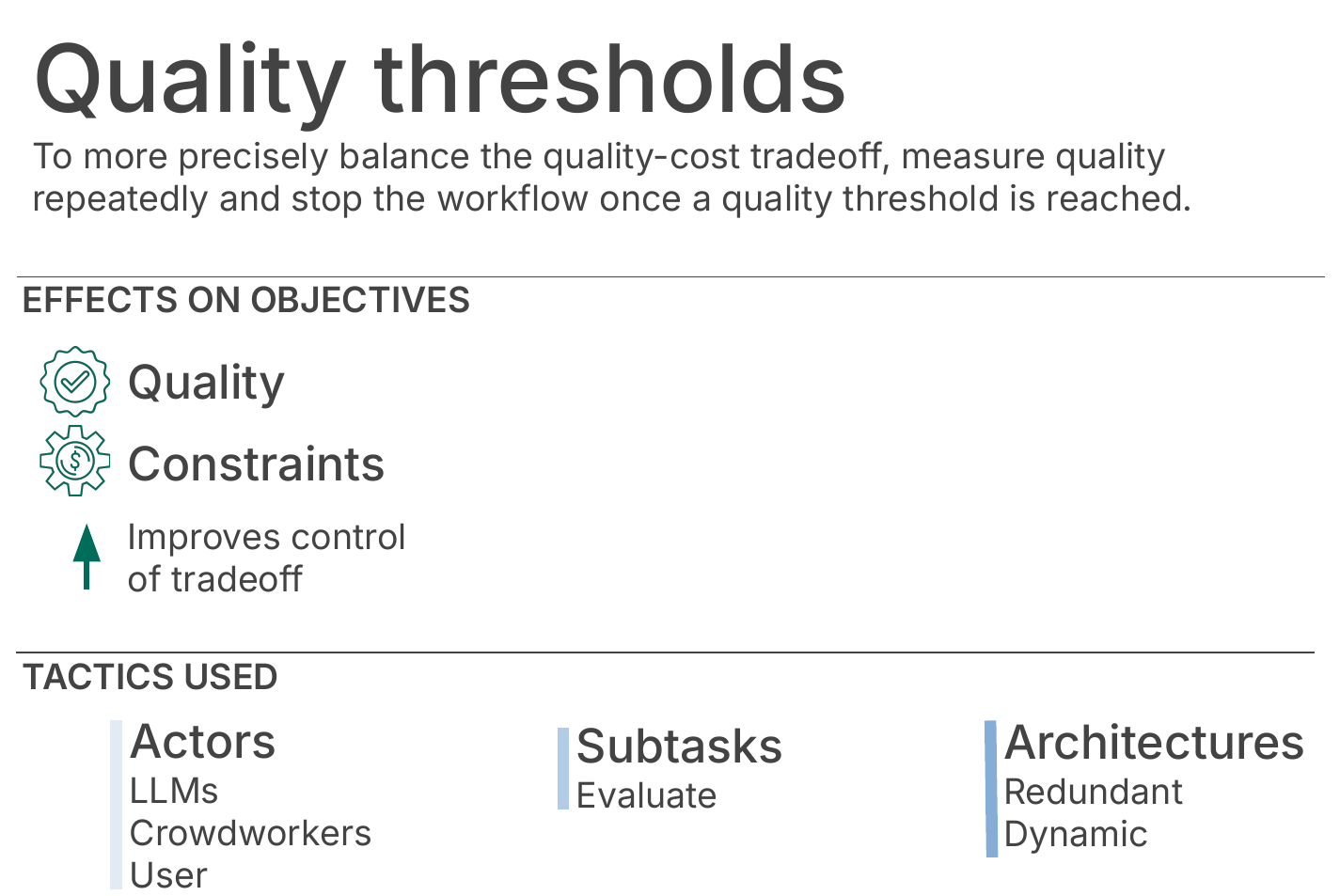} 
 \caption{An overview of the tactics used for and the effects of the quality thresholds strategy.}
 \Description{Title. Quality thresholds, to more precisely balance the quality-cost tradeoff, measure quality repeatedly and stop the workflow once a quality threshold is reached. Section 1: effects on objectives. Several objectives listed (with their effects). Quality and Constraints (Improves control of tradeoff). Section 2: tactics used. Actors (LLMs, crowdworkers, user), Subtasks (evaluate), Architectures (redundant, dynamic).}
\label{fig:ds-strat-thresh} 
\end{figure}

\subsection{Quality thresholds}
\label{strategies:threshold}
\textbf{To more precisely balance the quality-cost tradeoff, measure quality repeatedly and stop the workflow once a quality threshold is reached (Figure~\ref{fig:ds-strat-thresh}).}

\vspace{0.5em}
\noindent\textbf{Motivation.}
Sometimes, designers may not know prior to running the workflow how much work needs to be done to get a high quality output. In those situations, it would be helpful to have the workflow decide how many resources are necessary to use on an input-specific level.

\vspace{0.5em}
\noindent\textbf{Implementation.}
Some workflows give the \emph{actor} control over the number of \emph{redundant} generations using \emph{evaluate} subtasks [C\cite{huffaker2020crowdsourced, drapeau2016microtalk} L\cite{parameswaran2023revisiting, yao2023tree}]. Others allow for tuning \emph{redundant} architectures giving flexible goal precision, recall, and agreement thresholds [C\cite{huffaker2020crowdsourced} L\cite{yao2023tree}], use decision-theoretic choices of the number of responses to obtain in \emph{redundant} workflows [C\cite{bragg2013crowdsourcing, dai2010decision, alshaibani2021pterodactyl, dai2013pomdp}], or \emph{dynamically} choose between workflows by their maximum expected value of utility [C\cite{goto2016understanding}].

\vspace{0.5em}
\noindent\textbf{Consequences.}
Deciding if a threshold is met can require higher \emph{costs} in terms of \emph{effort} and \emph{expense}.

\vspace{0.5em}
\noindent\textbf{Examples.}
Examples of this strategy include POMDP-based control [C\cite{dai2013pomdp}], Pterodactyl [C\cite{alshaibani2021pterodactyl}], and a workflow for detecting emotionally manipulative language [C\cite{huffaker2020crowdsourced}].

\vspace{0.5em}
\noindent\textbf{Related strategies.}
Quality thresholds may be determined by leaning on the \emph{adaptive architectures} and \emph{user guidance} strategies.

%% file: sections/04.4_vignettes.tex
\begin{figure*}[t]
 \centering 
 \includegraphics[width=\linewidth]{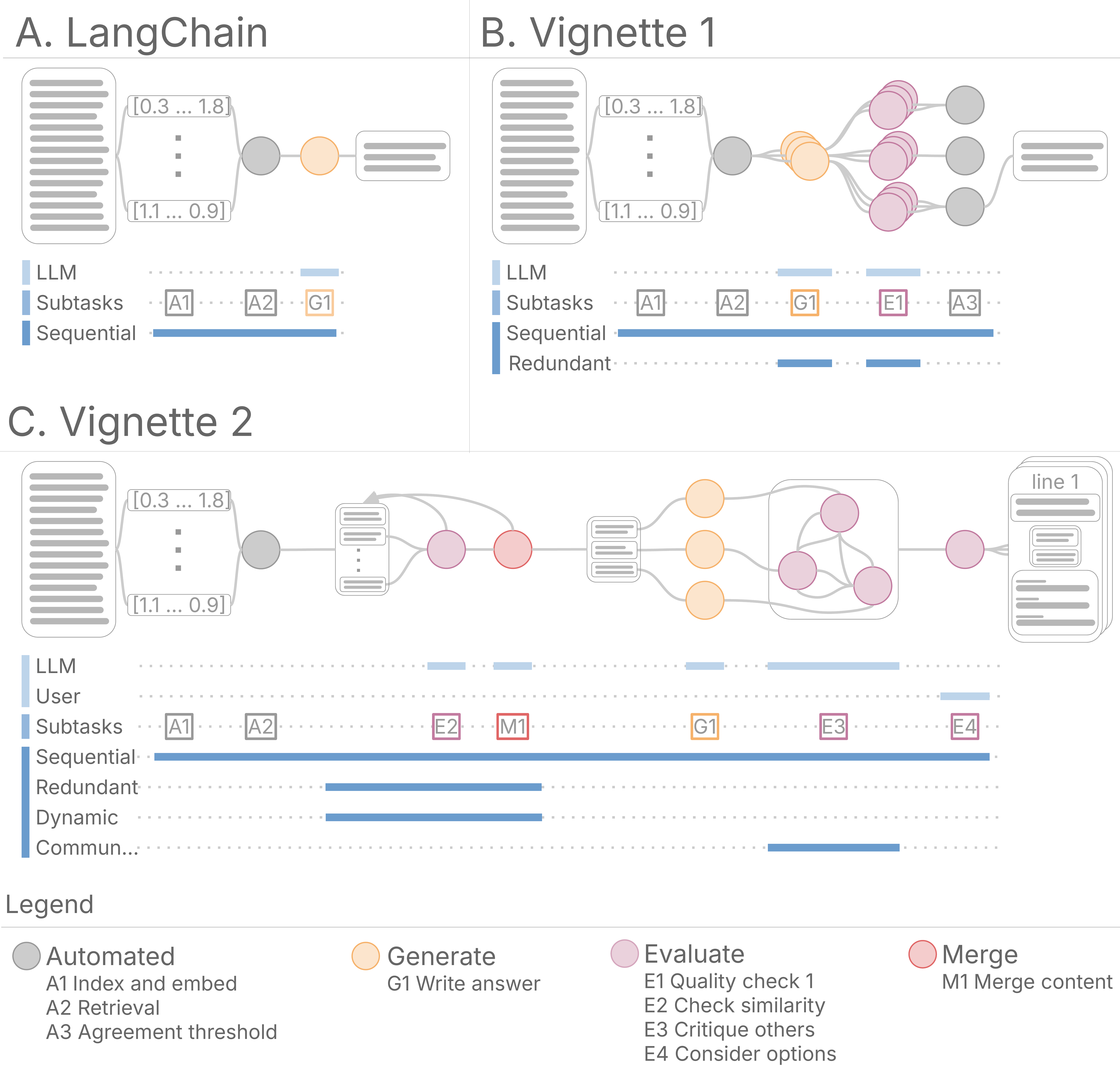}
 \caption{Vignette chains. We show through two vignettes --- both variations of Retrieval Augmented Generation (RAG) --- how our design space can be used to generate novel chain designs. It is common practice for a designer to default to implementing a generic RAG chain, such as (A) the one recommended by the popular chaining software LangChain~\cite{langchain}. While simple, such a design may not fit the specialized needs of a given application. By using our design space a person could be led to consider alternative designs, such as (B) a chain design focused on achieving \emph{objective} answers and a \emph{specific task} with \emph{constrained resources}, or (C) one focused on incorporating \emph{subjectivity} for \emph{general} subjects with high outcome \emph{quality}.
 }
 \Description{Figure shows three flow charts. A) The LangChain index, retrieve, generate chain, 
 B) the chain from vignette 1 and B) the chain from vignette 2. Each follow\new{s} the chains described in the main text. The chains of the two vignettes each contain 8 subtasks (including automated ones), while the baseline chains contain 3 subtasks (including automated ones).}
\label{fig:ds-vig} 
\end{figure*}

\section{Using the design space and strategies}
\label{sec:ds-vig}

 In this section, we show the generative value of the design space through two vignettes. In contrast to the paper's case studies (Section~\ref{sec:case}), which originate with an existing crowdsourcing workflow, here we use the design space to generate the designs of novel workflows from scratch. 
 
After defining our example task of Retrieval-Augmented Generation (Section~\ref{sec:ds-vig-task}), we present two vignettes that implement this task in different contexts (Sections~\ref{sec:ds-vig-vig1} and~\ref{sec:ds-vig-vig2}, Figure~\ref{fig:ds-vig}). These vignettes exemplify how our design space informs which tactic to employ in response to shifting task contexts. 
Section~\ref{sec:ds-vig-takeaways} discusses the function of the design space within the broader design process. 
While we have neither implemented nor evaluated these chains --- and so make no claims that they perform better than existing approaches reported in the literature --- they are (to our knowledge) novel and offer the potential for high performance.

\subsection{Task: Retrieval Augmented Generation}
\label{sec:ds-vig-task}

We consider the task of Retrieval Augmented Generation (RAG), in which an LLM generates responses with the help of a larger corpus from which it can retrieve information. RAG systems are usually implemented with a multi-step chaining process due to context size~\cite{shao2024scaling}. Specifically, we consider a RAG Question-Answering task in which the system must answer questions using information from the corpus. 

Conceptually, we can break the RAG QA task into retrieval and generation steps. The retrieval step inputs the question and the corpus and outputs relevant information from the corpus. The generation step inputs the question and said relevant information, then outputs an answer. The recommendations provided by the LLM chaining software LangChain reflect the standard approach for RAG~\cite{LangChain_RAG}. They recommend chunking and indexing documents in the corpus, using the query to retrieve documents by semantic similarity, then placing the retrieved chunks into an LLM context and generating an answer (Figure~\ref{fig:ds-vig}).

\subsection{Vignette 1: Customer service}
\label{sec:ds-vig-vig1}
For this vignette, we consider implementing RAG as a part of a customer service system, intended to answer questions about company policies and terms of service.

\subsubsection*{Prioritized objective: objectivity}
For customer queries, we will make the assumption that there is just one answer as all information in the corpus should be self-consistent. 
The \emph{diverse responses} and \emph{validation} strategies are especially useful for improving accuracy on \emph{objective} tasks. \emph{Evaluate} and \emph{improve} subtasks support \emph{validation}. Additionally, requiring a high threshold of agreement across \emph{redundant} outputs, with a plan for reconciliation if outputs do not reach that standard, uses the \emph{diverse responses} strategy to \emph{validate} by measuring consistency.

\subsubsection*{Prioritized objective: task specific}
This application of RAG is specific to one corpus about the company policies.
The most helpful strategy for this objective is precise \emph{validation}, since the specificity of the task allows for pre-definition of qualities to be validated. The tactical instantiation of this strategy is  highly specific \emph{evaluate} subtasks.

\subsubsection*{Prioritized objective: resource constraints}
\emph{Latency} and user \emph{effort} lead to customer frustration so are important to minimize in a customer service system. \emph{Expense} and \emph{effort} to design are also desirable to minimize, but less critical in this task context.
To reduce the strain on the latency of the \emph{diverse responses} and \emph{validation} strategies, the architecture should be a parallel \emph{redundant} one, rather than iterative or \emph{sequential}. 
Spending effort \emph{selecting actors}, designing \emph{context}, and \emph{aligning subtasks} reduces latency, expense, and user effort with less impact on quality. Although the \emph{adaptable architectures} and \emph{quality thresholds} strategies can reduce latency, expense, and effort to design, they can induce quality issues best mitigated through \emph{user guidance}. As reducing user effort is prioritized, this strategy should be deprioritized. 

\subsubsection*{Chain design}

An objectives-driven chain for this customer service task pieces together the various suggested tactics into a cohesive chain such as the following.
After indexing and retrieval, the selected chunks are fed into the generation portion of the chain. This portion has multiple \emph{redundant} \emph{generate} subtasks to answer the query based on the retrieved information. They are then passed to \emph{redundant} \emph{evaluate} subtasks, in which the highest scoring generation provides the output. The number and subject of these evaluations should be determined through testing. Finally, if all outputs do not meet an agreement threshold, alternative outcomes are provided such as requesting a rewording of the query or connecting to a representative. See Figure~\ref{fig:ds-vig} for a visual representation.

\subsection{Vignette 2: Fitness corpus}
\label{sec:ds-vig-vig2}

For the second vignette, we consider a RAG system informed by a corpus of scientific papers to answer a wide variety of questions around fitness topics, such as exercise, nutrition, and sleep.

\subsubsection*{Prioritized objective: subjectivity}
\emph{Objectivity} in providing correct answers continues to be critical, but this application must incorporate uncertainty and open-endedness more than the first vignette, as the corpus of literature can support differing conclusions.
The strategies of \emph{diverse responses}, \emph{validation}, and \emph{user guidance} are especially helpful in addressing this uncertainty. To implement these strategies, the \emph{user} can provide input at runtime to \emph{validate} what output is desirable when there is uncertainty in the answer. Measuring uncertainty requires sourcing \emph{diverse responses}, either through measuring variation within a \emph{redundant} architecture or measuring ease of persuasion in the debate \emph{communicative} architecture.

\subsubsection*{Prioritized objective: task general}
The RAG system must generalize to a wider variety of queries and topics than the first vignette.
\emph{Adaptable architectures} provide the flexibility to support such variety. 
As the variety of queries is not pre-defined, \emph{branching} architectures are not a useful tactical instantiation of this strategy. 
Instead, \emph{dynamic} and \emph{communicative} architectures can adapt at runtime to an unknown set of queries and topics.
Finally, incorporating \emph{user guidance} supports flexibility in result interpretation and reduces the need to specify a pre-defined set of topics or queries.

\subsubsection*{Prioritized objective: quality}
Although quality is always desirable, in high-stakes domains like the medical field prioritizing this objective is critical even at higher expense, latency, or effort. For the purposes of this vignette, we assume few resource constraints. 
Of all the objectives, prioritizing an increase in quality has the widest range of options, as nearly all strategies and their tactical implementations improve quality. The exception is that \emph{adaptable architectures} can induce errors and reduce transparency and consistency.
In this health querying system, the \emph{user} needs appropriate \emph{context} to ensure the qualities of transparency and faithfulness. The chain's tactics can attribute outputs to verbatim sections of the corpus to improve the user's context.

\subsubsection*{Chain design}
By combining these prioritized tactics, an objectives-driven chain for this case could look like the following. The initial retrieval step allows for multiple chunks of the corpus to pass through. To accommodate for the potential disagreement from a variety of sources, the next step iteratively merges the corpus chunks into several distinct lines of argument, connected with references to verbatim parts of the corpus. At each step of this iterative \emph{redundant} architecture, a pairwise comparison \emph{evaluate} subtask determines if two chunks support similar takeaways. If they do, a subsequent subtask \emph{merges} the two into a new chunk, collecting the indexes referencing specific parts of the corpus. This iterative loop runs for a \emph{dynamically} determined amount of time until the arguments have converged.
These arguments and their references then pass into the generation portion of the pipeline. Each argument informs a \emph{generation} subtask that answers the query given that set of retrieved information. Then, a \emph{communicative} debate commences in which each debater begins with this generated answer and its justification. Using the ease with which each debater is convinced it is incorrect, the chain outputs the different arguments, their uncertainty, the debate history, and references to the initial corpus for the \emph{user} to \emph{evaluate}. See Figure~\ref{fig:ds-vig} for a visual representation of this chain.

\subsection{Vignette discussion} 

\label{sec:ds-vig-takeaways}
Through these two vignettes, we outline how the design space specializes chains to the task context by explicitly defining objectives and surfacing the cascading implications for strategies that then recommend beneficial tactic choices. In this section, we discuss how the process described in these vignettes fits into the broader design process.

\begin{figure}[t]
\centering
\includegraphics[width=\linewidth]{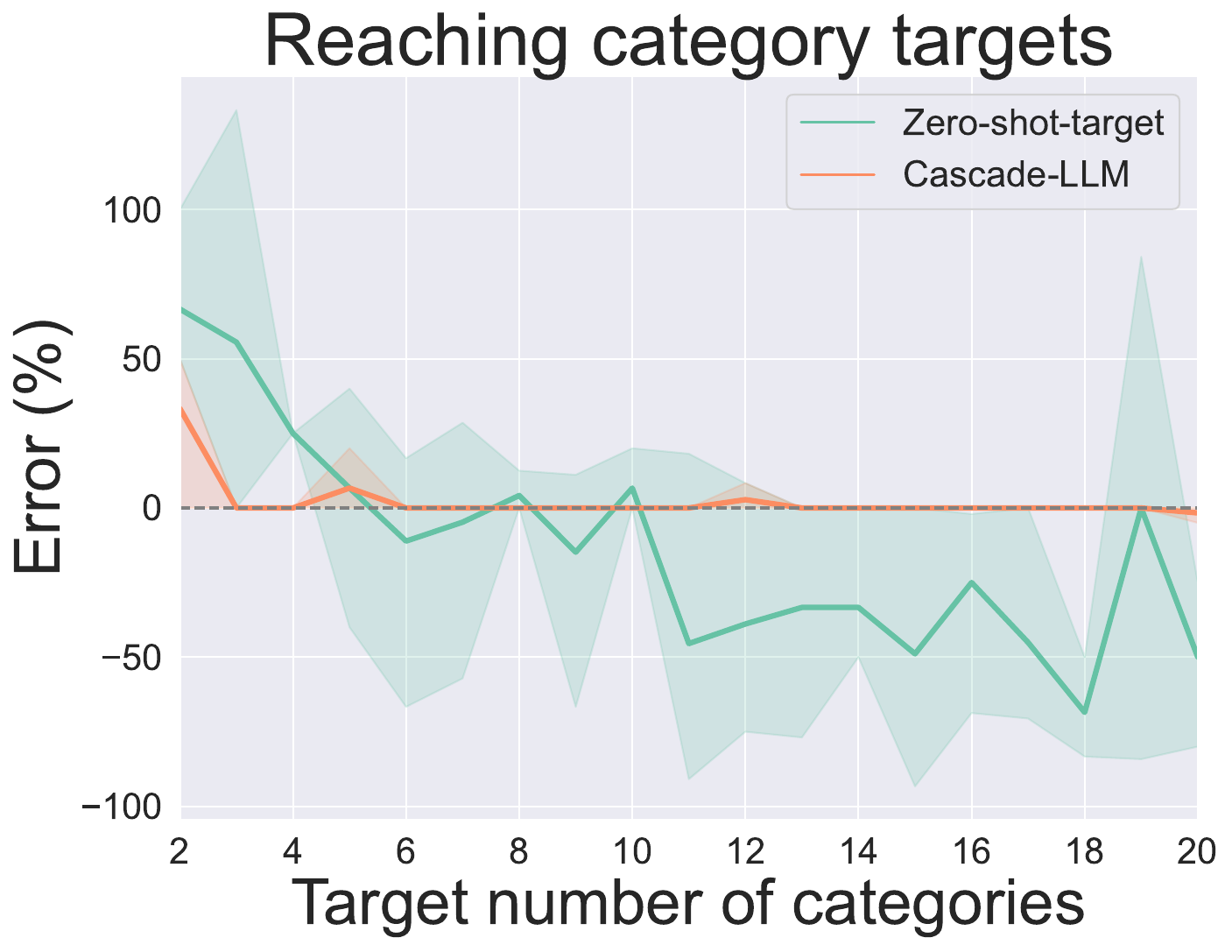}
\caption{Controllability of Cascade-LLM versus zero-shot taxonomies. Manipulating Cascade-LLM outputs through given parameters can more precisely achieve a wide range of category sizes than indicating a target value within a zero-shot prompt. Error (\%) refers to the percent error between the method's closest option and the target value. Across three datasets, Cascade-LLM exhibits low deviation from the target number of categories, while prompting with zero-shot gives noisy responses, often with too few categories. The shaded region indicates the 95\% confidence interval.}
\Description{A line graph showing the target number of categories on the X-axis from 2-20, and The percentage error on the y-axis from -1 to 1, in which 0\% error means exactly hitting the target. Two lines are shown. One line, for Cascade-LLM, stays much closer to the 0\% error line and has a narrower 95\% confidence interval than the second line, zero-shot-target. At 11 or more categories, zero-shot-target has a negative percentage error rate.}
\label{fig:Cascade-numcatall} 
\end{figure}

\subsubsection*{Tactic suggestions respond to task context.}
Standard practice applies the same steps for any task context: retrieval and generation. With our design space, a designer can flexibly accommodate unique combinations of objectives for varied applications, as seen in the differing vignette designs.

\subsubsection*{Incorporating non-LLM tools can reduce costs.}
The costs of using LLMs for all steps can be prohibitive and unnecessary. A designer could replace the semantic similarity retrieval step with LLMs by chunking the corpus and using \emph{focus} subtasks to determine relevant parts of each chunk. Although the potential for LLMs for processing complex unstructured data is an active area of exploration~\cite{shankar2024docetl}, using LLMs for this step would cost much more than standard retrieval with a large corpus. 
Our design space focuses on the LLM chaining literature, but there is immense opportunity and future work in incorporating tools beyond LLMs into these chains. 

\begin{figure*}[h]
\centering
\includegraphics[width=\linewidth]{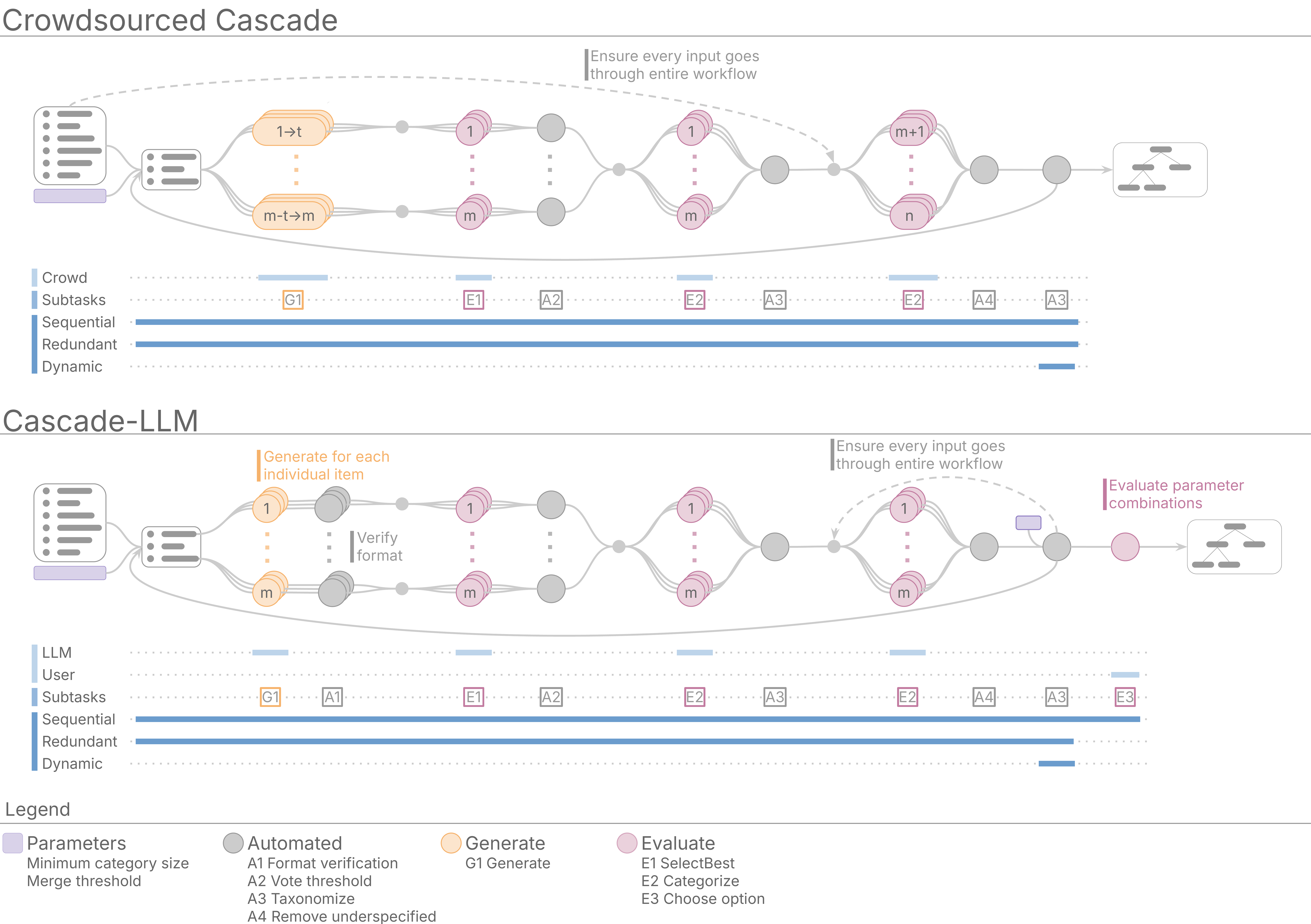}
\caption{Case study 1: Cascade~\cite{chilton2013cascade} for taxonomy creation. Crowdsourced Cascade uses various actors, subtasks, and architectures from the design space, as defined under the workflow diagram. Although automated events are not defined as subtasks in our design space, we indicate them here for clarity. Crowdsourcing workflows can inspire LLM chains, but the differing capabilities of crowdworkers and LLMs require changes to the design. We adapt the crowdsourcing workflow Cascade with elements from our design space to update the use of tactics to better work for LLMs. For example, we found that generating responses for individual items, rather than sets of items, is a more appropriate subtask because it creates better \emph{response diversity}. Through these adaptations, we can control the output taxonomy shape with parameters.} 
\Description{Figure shows two flow charts, one for crowdsourcing Cascade and one for LLM chain Cascade. Both follow the workflow described in the main text. Several adaptations are highlighted in the LLM chain: the generate step with only one item-category comparison, and the categorize step only looking at the m current items, instead of the entire list.}
\label{fig:case-casc} 
\end{figure*}

\subsubsection*{The design space filters and inspires, rather than prescribes.}
Using the design space does not prescribe a definitively optimal chain, but it does serve the design process through filtering and generating ideas.
Through the vignettes, we see that the design space prioritizes subsets of options within a task context. Given the enormous space of possible designs, this filtering is valuable guidance. 
Furthermore, using the design space allowed us to create unique chains. Due to the comprehensiveness of the design space over two literatures, it serves as inspiration during the brainstorming process for a variety of potentially unintuitive options validated through prior work.

\subsubsection*{The design space provides an informed starting point, rather than a finished design.}

The design space serves as a launching pad for other design decisions. Further testing can compare performance within the subset of tactic choices recommended by the design space (e.g., which type of \emph{verification} performs best). Many of these changes depend on model performance (e.g., are diverse generations necessary, or is fluency high enough to have confidence in a single generation). 
Additionally, the design process will need to test specific tactical design decisions (e.g., what model to choose, prompt engineering, how many iterations, and for which specific qualities to evaluate).
Although these chain designs may update after testing on specific corpora and models, they provide a starting point motivated by findings from prior work. 
Future work is needed to improve the granularity of design choice recommendations and develop evaluation mechanisms for comparing among design choices (Section~\ref{sec:disc-designprocesses}).

%% file: sections/05_case.tex
\section{Case studies}
\label{sec:case}

In order to better assess how crowdsourcing methods transfer to LLM chain development, we adapt crowdsourcing workflows to LLM chains in three case studies. This process of adapting workflows supports two investigations. First, it allows us to compare the performance of chaining to both zero-shot prompting and crowdsourcing. Second, through the process of adapting these workflows from crowdsourcing to chaining, we can see how the workflow design changes to accommodate for LLMs. These changes, found through practice, directly compare what \textbf{tactics} support the same \textbf{strategies} and \textbf{objectives} when using \emph{crowdworkers} or \emph{LLMs}.

To ensure we are suitably focused, we explore one strategy in more depth: supporting quality control through \emph{user guidance} in which the user can directly manipulate workflow outputs. We focus on this strategy as it is implemented in crowdsourcing work, but it is underexplored in the LLM chaining space. Prior non-chaining LLM work explores controllability as a helpful attribute~\cite{macneil2023prompt,kim2023cells}, so here we seek to incorporate it within chains.
With directly manipulable outputs, users can make ``rapid, incremental, reversible operations whose impact on the object of interest is immediately visible''~\cite{shneiderman1982future}, increasing control over output elements and helping users explore alternative results~\cite{hutchins1985direct}. 
By adapting this subset of the user guidance strategy from crowdsourcing, we explore how and if such a concept can transfer to chains, what modifications are necessary, and its relative advantages compared to controlling the outputs through zero-shot prompting.

We choose three popular crowdsourcing tasks that incorporate different degrees of the \emph{objectivity} and \emph{subjectivity} design space objectives. We implement Cascade~\cite{chilton2013cascade} to build taxonomies with manipulable structures, Soylent~\cite{bernstein2010soylent} to shorten text to a variety of target lengths, and Mechanical Novel~\cite{kim2017mechanical} to control the elements of a story. Diagrams of the workflows, as well as our adaptations to use LLMs, can be found in Figures~\ref{fig:case-casc},~\ref{fig:case-soylent}, and~\ref{fig:case-mechnov}.

For each case study, we first document how the workflow was changed to incorporate LLMs and map those changes to the design space. 
Common limitations we faced include LLM context lengths, inconsistent output formatting, lack of diversity in model responses, and models struggling to perform the workflow-defined subtasks. We adapted to these limitations with more granular subtasks, more diverse prompts, comparative tasks, and hard-coded validation steps. Across all three case studies, we use different prompts\footnote{https://github.com/madeleinegrunde/designingchains} at each parallel generation step to increase \emph{response diversity}, update worker instructions and \emph{context} through trial and error to get the right output formatting, and structure voting \emph{evaluation} subtasks to choose among numbered options to minimize hallucinations.

We compare our LLM workflows to zero-shot baselines in terms of output control precision, salient qualities for each task, and the latency and number of API calls.
To prioritize human-preferred LLM outputs, we use an instruction-tuned model: OpenAI's API gpt-3.5-turbo with default parameters~\cite{ouyang2022training}.
Compared to zero-shot prompting, chained workflows use more resources but can improve output quality.
These workflows also provide a greater precision of output control than parameterized zero-shot baselines.

\input{sections/05.1_cascade}
\input{sections/05.2_soylent}

\input{sections/05.3_mechnov}

%% file: sections/05.1_cascade.tex
\begin{figure*}[t]
  \centering 
  \includegraphics[width=\linewidth]{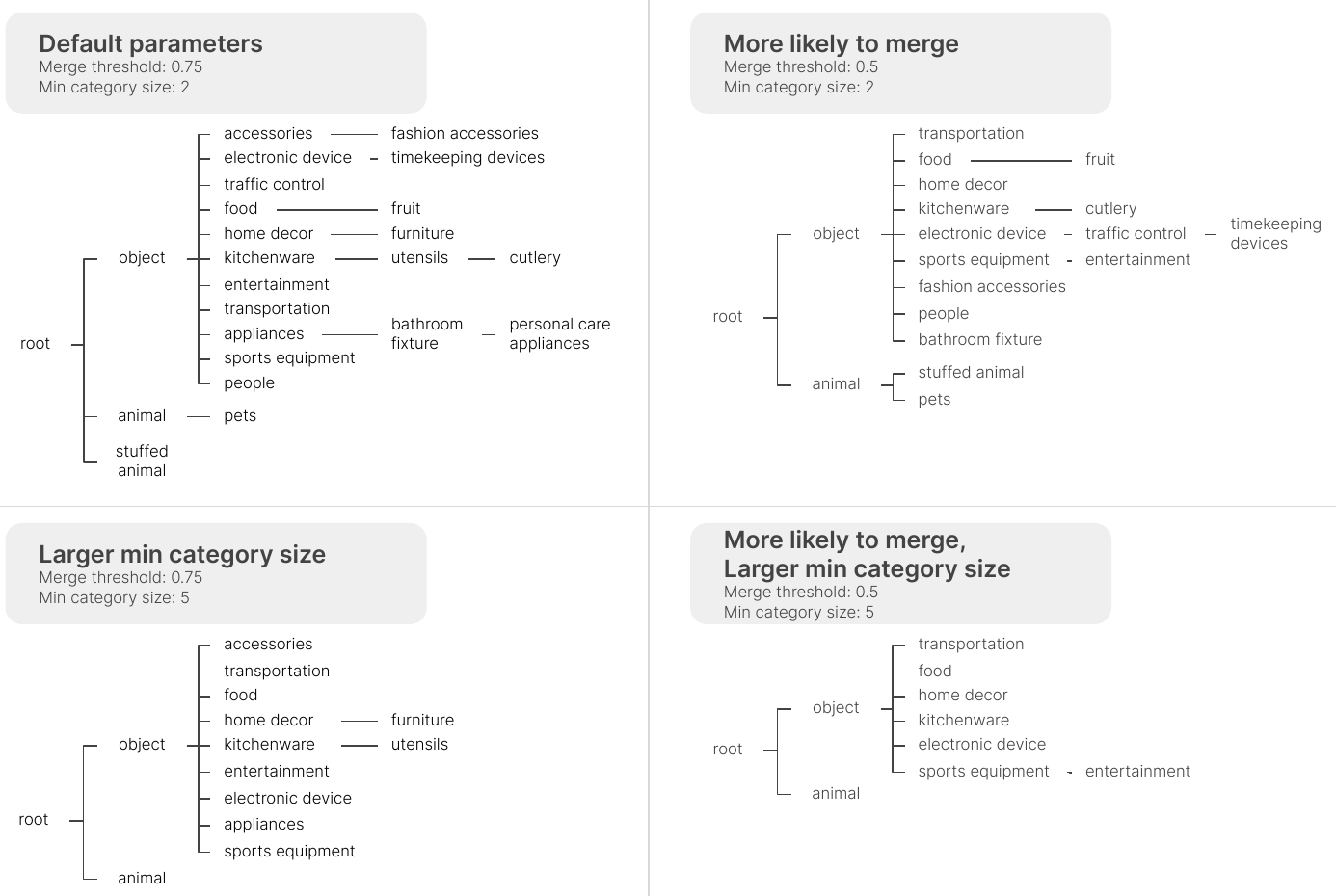}
  \caption{Taxonomy examples using Cascade-LLM on MSCOCO labels~\cite{lin2014microsoft}. Categorization into a taxonomy is a challenging task because the same item may be categorized with different levels of precision. Our implementation of the Cascade crowdsourcing workflow~\cite{chilton2013cascade} enables manipulation of the similarity threshold needed to merge categories, along with the minimum category size. By changing these values, we can sort data into taxonomies of different levels of precision and compare options as the final stage of the workflow. This direct manipulation allows the user not only to validate any inaccuracies but also to make opinionated decisions, such as if ``stuffed animal'' should or should not be a subcategory of ``animal.''
  }
  \Description{Four taxonomies. The first with default parameters (merge threshold 0.75, minimum category size 2), contains a hierarchy of 4 layers and 23 categories. The second that is more likely to merge (merge threshold 0.5, min category size 2), contains 4 layers and 18 categories. The third is a larger minimum category size (merge threshold 0.75, min category size 5) contains 3 layers and 13 categories). The fourth is both more likely to merge and has a larger min category size (merge threshold 0.5, min category size 5), contains 3 layers and 9 categories).}
\label{fig:case-casc-tax} 
\end{figure*}

\subsection{Case study 1: Taxonomy creation}
\label{sec:case-casc}

For our first case study, we explore taxonomy creation, the task of organizing a set of items into a category tree to aid sensemaking. This task's outcome qualities are primarily \emph{objective}, with some degree of \emph{subjectivity}: categories need to be correct, but most items can fit into multiple categorizations of different levels of precision. We focus on the ability to control the granularity of these categorizations by controlling the taxonomy shape (Figure~\ref{fig:case-casc-tax}).

\subsubsection{Workflow design}
\label{sec:case-casc-workflow}

We adapt the Cascade~\cite{chilton2013cascade} crowdsourcing workflow (Figure~\ref{fig:case-casc}), which takes as input a list of items and outputs a tree mapping items to categories. The workflow begins by selecting a subset of 32 items, \emph{generating} category suggestions for each item, \emph{evaluating} by selecting the best categories among the generated options, and \emph{evaluating} by confirming which among all categories works for each item. The next step turns the item-category assignments into a taxonomy, deleting small categories and merging similar categories. The workflow then updates the categorizations for the entire list of items with this new assignment. The workflow \emph{redundantly} iterates with new subsets until all items are categorized, building a growing set of categorizations.

To build Cascade-LLM, our version of Cascade powered by LLMs, we made multiple changes to the original workflow.
We highlight some of the most salient modifications.
To achieve diverse categorizations, we improved \emph{response diversity} by giving different prompts to parallel generations and ensuring that every item is input into a category \emph{generation} step.
At first, the categories were too broad. Specifying category qualities via targeted prompts proved unsuccessful, so instead we added a programmatic \emph{validation} step to delete categories that included  ``and,'' ``or,'' commas, or slashes. As this hard-coded \emph{validation} step worked, it was not necessary to include another LLM call.
The original workflow gives actors sets of eight items for which to generate eight categories. However, the LLM would try categorizing multiple items at once, consistent with prior findings that LLMs struggle to disambiguate multiple subtasks in a prompt~\cite{wu2023llms,madaan2023self}. To increase diversity, we \emph{aligned subtask} design to fit model capabilities with one item categorization.
Finally, the original workflow also requires parameters for the minimum category size and a merge threshold. To support \emph{user guidance} through direct manipulation, we dynamically apply these parameters at the last taxonomization stage. To investigate different taxonomies with varied levels of precision, a user can manipulate the parameters of 1) the minimum category size and 2) the similarity threshold determining if two categories will merge (Figure~\ref{fig:case-casc-tax}).

\subsubsection{Evaluation setup}
\label{sec:case-casc-eval}

We compare Cascade-LLM against zero-shot baselines on three lists of nouns.

\input{tables/cascade-qual}

\vspace{0.5em}
\noindent\textbf{Datasets.}
Our experiments use the labels from the scenes in MIT Indoor Scenes~\cite{quattoni2009recognizing}, from object categories in MSCOCO~\cite{lin2014microsoft}, and from CIFAR100~\cite{krizhevsky2009learning}. These label sets are of size 67, 80, and 100, respectively.
The original Cascade paper explores three datasets sourced from Quora.com question responses.
Our LLM workflow is capable of processing these larger datasets, as only individual items need to be within the model's context size; however, the Quora.com data exceeds the context limit for the zero-shot baselines.

\vspace{0.5em}
\noindent\textbf{Baselines.}
We compare our workflow to two baselines.
The \emph{zero-shot} baseline directly prompts the model to create a taxonomy. 
This baseline establishes how the model completes the task given no restrictions.
The \emph{zero-shot-target} baseline further specifies the target number of categories 
in the prompt. With this baseline, we compare the ability to control workflow outputs in a single prompt.

\subsubsection{Workflows enable direct manipulation}
\label{sec:case-casc-control}

To measure the controllability of the taxonomy's shape, we measure how well methods can change the taxonomy size (the number of categories). 
Given a target number of categories (range: 2--20), we measure the percent error between the method's closest option and the target value.

As seen in Figure~\ref{fig:Cascade-numcatall}, one run of Cascade-LLM almost always achieves a precise match with the desired number of categories. 
In contrast, zero-shot-target is close but usually unable to produce a taxonomy with the target number of categories. 
The zero-shot condition produces small taxonomies the size of which is not controllable (MIT Indoor Scenes: 2, MSCOCO: 8, CIFAR100: 4).

\subsubsection{Workflows reduce hallucinated and forgotten items}
\label{sec:case-casc-quality}

We also evaluate the quality of the outputs through qualitative evaluation. Similar to the original paper, we hand-code taxonomies without knowledge of which method created it. 
First, we look for errors in items across all baselines. Specifically, we see if the output hallucinates or forgets any items. 
We then code the zero-shot and default-parameter Cascade-LLM baselines for the same mistakes considered in the original paper: duplicate categories, missing parent-child relationships, and incorrect parent-child relationships. 

Cascade-LLM does not forget any items, but 83.72\% of zero-shot taxonomies do not include all items. Cascade-LLM cannot hallucinate new items, while zero-shot baselines hallucinate 51 items across the 129 taxonomies.
Cascade-LLM taxonomies have similar error rates to zero-shot baselines' taxonomies on duplicate categories and missing/incorrect category-item and parent-child pairs, yet are richer taxonomies with more categories and depth (Table~\ref{tab:cascade:qual}).

\subsubsection{Workflows have higher costs}
\label{sec:case-casc-cost}

We measure generation time and the number of API calls to an LLM.
Cascade-LLM takes more time (mean: 2,642.58 sec., std: 655.81) than the zero-shot (mean: 24.35 sec., std: 5.04) and zero-shot-target (mean: 21.69 sec., sd: 4.49) baselines. Crowdsourced Cascade took 43 hours and 3 minutes to taxonomize three item sets~\cite{chilton2013cascade}. 
Cascade-LLM makes more model calls (mean: 18,898.33, std: 4472.61) than the one call from the zero-shot and zero-shot-target baselines. 
Most of these calls are short in terms of token length, as they compare one item to one category. Crowdsourced Cascade reports 1760 calls to crowdworkers per iteration.
For our goals, the increased cost in time and computation enables direct manipulation of more detailed categorizations without item forgetting or hallucination.

%% file: tables/cascade-qual.tex
\begin{table*}[t]
\caption{Cascade-LLM versus zero-shot performance across three datasets. Though the quality of categorizations is similar between methods, the Cascade-LLM generated taxonomies include more categories and have greater taxonomy depth.}
\label{tab:cascade:qual}
\centering
\resizebox{0.8\linewidth}{!}{
\begin{tabular}{lllllll}
\hline
                        & \multicolumn{2}{c}{MIT Indoor Scenes~\cite{quattoni2009recognizing}} & \multicolumn{2}{c}{MSCOCO~\cite{lin2014microsoft}} & \multicolumn{2}{c}{CIFAR100~\cite{krizhevsky2009learning}} \\
                        & Zero-shot                          & Cascade-LLM                        & Zero-shot                          & Cascade-LLM                         & Zero-shot                           & Cascade-LLM                         \\ \hline
\# of Categories        & 2                          & 21                       & 8                          & 24                        & 4                           & 19                        \\
Tree Depth              & 1                          & 3                        & 1                          & 4                         & 1                           & 3                         \\
Duplicate Categories    & 0                          & 0                        & 0                          & 0                         & 0                           & 0                         \\
Missing Category-Item   & 0                          & 2                        & 1                          & 3                         & 7                           & 2                         \\
Incorrect Category-Item & 0                          & 0                        & 1                          & 2                         & 2                           & 2                         \\
Missing Parent-Child    & N/A                        & 0                        & N/A                        & 1                         & N/A                         & 0                         \\
Incorrect Parent-Child  & N/A                        & 0                        & N/A                        & 1                         & N/A                         & 1                         \\ \hline
\end{tabular}
}
\end{table*}

%% file: sections/05.2_soylent.tex
\begin{figure*}[t]
\centering
\includegraphics[width=\linewidth]{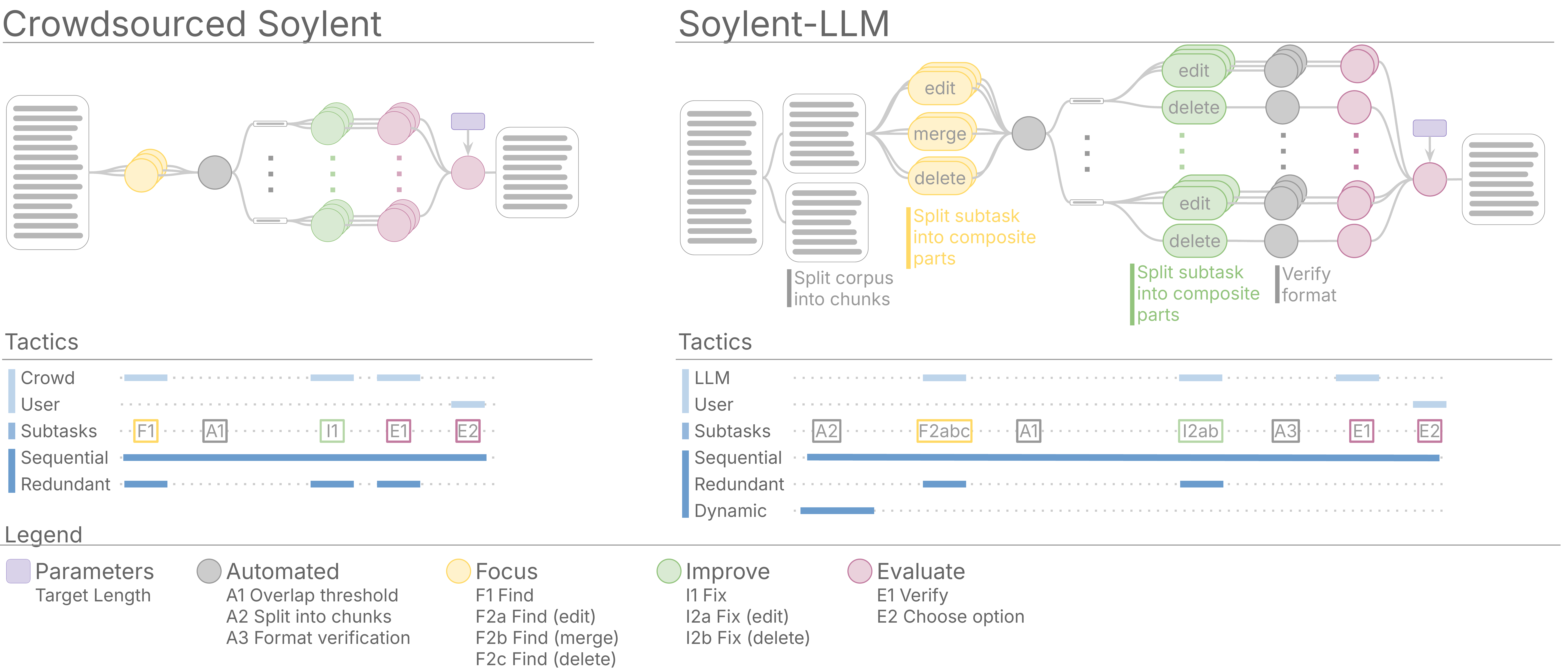}
\caption{Case study 2: Soylent~\cite{bernstein2010soylent} for text shortening. When using LLMs, the crowdsourcing workflow Soylent initially outputs low quality responses and runs over context limits. In response, we built Soylent-LLM with updated tactics to have appropriately specified subtasks, to work within context limits, and to verify there were no hallucinated quotes.} 
\Description{Figure shows two flow charts, one for crowdsourcing Soylent and one for LLM chain Soylent. Both follow the workflow described in the main text. Several adaptations are highlighted in the LLM chain: 1) the initial text splits into smaller chunks, 2) the Find and Fix steps are split into parallel edit, merge, and delete steps, and 3) automated format validation is added after the fix step.

}
\label{fig:case-soylent} 
\end{figure*}

\subsection{Case study 2: Text shortening}
\label{sec:case-soy}

The next case study is the task of shortening text. Unlike summarization, shortening the text involves making edits to create a condensed version of the text that attempts to preserve the full meaning.
Shortening is a task primarily on the \emph{subjectivity} side, but requiring elements of \emph{objectivity}. Shortened versions must stay faithful to the original text, and there are many options for what phrases to shorten and how to shorten them. By implementing a workflow to generate many shortening options, an end user can control the final length of a paragraph dynamically. 

\subsubsection{Workflow design}
\label{sec:case-soy-workflow}

To shorten text we adapt the Find-Fix-Verify workflow (Figure~\ref{fig:case-soylent}) introduced in Soylent~\cite{bernstein2010soylent}. 
In the ``Find'' step, crowdworkers select areas that might be shortened.
If at least two actors agree on a text span, that span progresses to the ``Fix'' step.
In the ``Fix'' step, crowdworkers make edits to shorten the text.
In the final ``Verify'' step, more crowdworkers verify the possible changes by rejecting by majority vote any edits deemed grammatically incorrect or that change the meaning of the sentence.
These steps respectively map to the \emph{focus}, \emph{improve}, and \emph{evaluate} subtask types in our design space.
The workflow outputs edit options for multiple spans in the paragraph, which when combined create multiple possible outcome texts. 
Soylent defaults to showing the shortest possible option, but it also provides a length slider for the user to adjust the final output as desired. The slider enables the user to test different shortening phrasing options and lengths through a simple interaction.

We made multiple adjustments to produce our Soylent-LLM workflow.
First, we ran into context limits and parallelization constraints with longer texts. In response, we apply a \emph{dynamic} approach that splits the workflow into chunks of 10 sentences and applies the workflow to each. 
Second, Soylent allows actors to fix inputs by editing, merging, and deleting phrases. LLMs performed better when we split these tasks into separate prompts, so we further decompose Find and Fix into finer grained editing, merging, and deleting steps to get an \emph{aligned} subtask size to the model's capabilities. 
Finally, in the Find step, the LLM would sometimes ``hallucinate'' phrases to shorten that do not exist in the text or would change quotes.
To mitigate these errors, we add several hard-coded \emph{validation} steps: we ensure that the phrase to replace exists in the input text, that the replacement is shorter, and that no quotes were altered or added.

Recent work also implemented three unique variations of a Find-Fix-Verify workflow using LLM chains~\cite{wu2023llms}. 
Their evaluation, using student preferences, found mixed results on effectiveness for creating better shortenings than a zero-shot baseline.
We evaluate our workflow along different dimensions: controllability of output length, fluency, grammaticality, what type of content is deleted, and workflow resource usage. 
Unlike this prior work, we also retain the parallel generation and validation steps from the original paper.

\subsubsection{Evaluation setup}
\label{sec:case-soy-eval}

For evaluating Soylent-LLM, we use texts from newspapers and from the original Soylent paper~\cite{bernstein2010soylent}. We compare Soylent-LLM with three zero-shot baselines.

\vspace{0.5em}
\noindent\textbf{Dataset.}
We use the XSUM summarization dataset~\cite{narayan2018don}. We only use the input texts because our task is shortening, not matching gold-standard summarizations. We use 100 randomly sampled texts from XSUM, with the requirement that examples are more than 500 words long.
For qualitative analysis, we use the texts from the Soylent evaluation~\cite{bernstein2010soylent}. Through this dataset, we can compare the Soylent-LLM shortenings to Crowdsourced Soylent shortenings.

\vspace{0.5em}
\noindent\textbf{Baselines.}
The \emph{zero-shot} baseline prompts the model to shorten the paragraph directly.
For the \emph{zero-shot-target} baseline, we specify the target output length in a zero-shot prompt.
In the \emph{zero-shot-ffv} baseline, we ask the LLM to explicitly perform Find-Fix-Verify steps in one prompt.

\begin{figure*}
\centering
\includegraphics[width=0.85\linewidth]{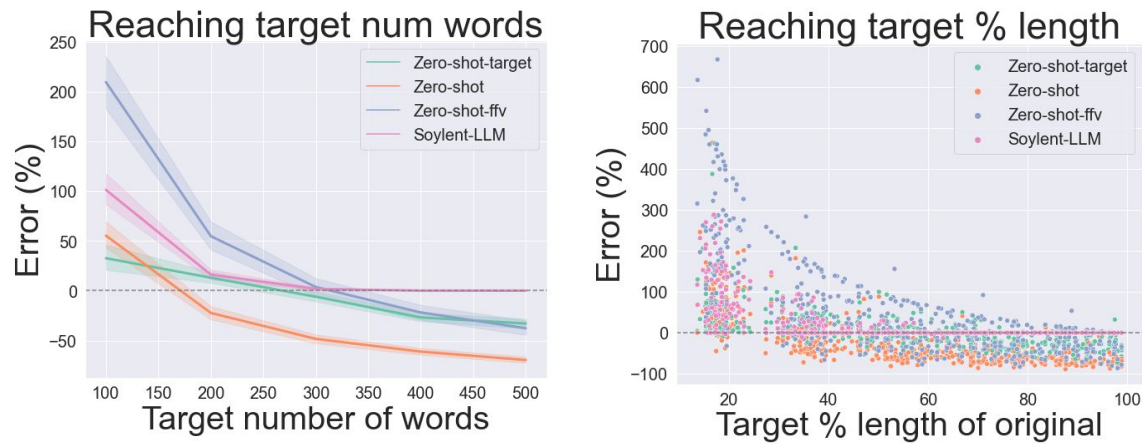}
\caption{
Text shortening results for Soylent-LLM and zero-shot baselines. 
A 0\% error indicates outputs that exactly match a target length, and the shaded region (left) indicates the 95\% confidence interval. 
Na\"ive zero-shot prompting without length guidance averages between 100-200 words, regardless of the desired target length.
With length guidance added to the prompt, zero-shot-target does substantially better at reaching all targets, especially when the target length is short. 
Soylent-LLM struggles to hit short targets but is quite precise at longer lengths (left) and when reducing only a short amount relative to the original length (right).
The zero-shot-ffv baseline, which instructs the model to perform Find-Fix-Verify steps within one prompt, behaves more similarly to the Soylent-LLM workflow but is consistently less accurate.
}
\Description{
Two graphs side by side. The first graph is a line graph graphing the target number of words on the X axis from 100-500 and percent error on the y axis from -75\% error to 250\% error. The graph contains 4 lines. On short texts, the zero-shot-target gets the lowest error followed by zero-shot, Soylent-LLM, and zero-shot-ffv. On longer texts, Soylent-LLM has approximately 0\% error. Zero-shot has the highest error, and zero-shot-target and zero-shot-ffv have approximately the same error at 50\%/
The 95\% confidence intervals are narrow. The second graph is a scatter plot with the X axis the target percent length of the original ranging from 0 to 100 and with the Y axis the percentage error from -100 to 500\%. There are 4 colors of points. The highest percent errors are for small target lengths across all 4 baselines. As the target percentage increases, Soylent-LLM converges to close to zero percent error, while the three zero-shot baselines trend negative.
}
\label{fig:soylent-words} 
\end{figure*}

\subsubsection{Workflows enable length manipulation}
\label{sec:case-soy-control}
We first compare how well the various LLM approaches reach a desired output length.
We represent target lengths through word counts of 100-500 words.
In initial experiments, we found that zero-shot baselines perform more competitively when asked to shorten to a given word count rather than to a length percentage and hence use word counts in targeted prompts. 
Nevertheless, we also measure accuracy in terms of a target percentage of length.
We calculate the target percent length by dividing the target number of words by the input text's number of words, and we calculate the actual percent length as the output text's number of words divided by the input text's number of words.
We compare the percent difference (i.e., error) between the actual and target percent lengths.

We find that as text length increases, Soylent-LLM can more precisely meet the target length (Figure~\ref{fig:soylent-words}). 
When shortening text, the zero-shot prompt condition consistently outputs approximately 100 word long shortenings (mean: 126.14, std: 52.63). Thus, for longer target lengths, this method achieves lower performance.
Zero-shot-target is similarly best at lengths close to 100 words, but it still shows a lack of precision in reaching longer target lengths. 
Soylent-LLM and zero-shot-ffv both struggle to hit shorter targets.
This result matches the Soylent paper, in which output texts were between 78-90\% of the original text length~\cite{bernstein2010soylent}. 
Soylent-LLM can more precisely hit the larger target lengths than any of the baselines.

We find similar trends measuring percent error with the percent length reduction required to meet the target length.
Soylent-LLM and zero-shot-ffv can perform poorly with large reductions, in which the target \% is a small fraction of the input text, while zero-shot and zero-shot-target perform worse with smaller reductions.
In summary, Soylent-LLM does a better job of tightening long texts but is less well-suited to extreme shortenings that cross over into summarization.
In addition, Soylent-LLM is unique in providing options among all targets without having to requery. Moreover, at each target length, there are often multiple possible output texts.

\subsubsection{Workflow outputs have abrupt transitions but lower perplexity}
\label{sec:case-soy-quality}

We measure the average perplexity of output texts as a proxy for fluency. We use a different language model (Facebook's OPT-1.3b) than the actor model (OpenAI's gpt-3.5-turbo) for an external evaluation.
We also perform a qualitative analysis of the shortening procedure. We use the five texts used for evaluation in the original Soylent paper~\cite{bernstein2010soylent}. We compare the edits made with the shortest output from Soylent-LLM, the baseline conditions, and the crowdsourced condition's output provided by the authors of the original paper. 
Without knowing the condition, we coded the outputs for fluency and grammatical errors, following the qualitative evaluation procedure of the original paper.
We additionally coded what content changed during shortening.

Soylent-LLM has the better result of a lower perplexity (mean: 11.69, std: 3.77) than zero-shot (mean: 15.37, std: 5.73), zero-shot-target (mean: 12.31, std: 3.91), and zero-shot-ffv (mean: 16.19, std: 5.68).
This result surprised us, as zero-shot is fully generated by a large language model fine-tuned to provide fluent responses. 

As with the original crowdsourcing workflow, some shortenings diminished fluency or introduced grammatical errors. Zero-shot, zero-shot-ffv, and Soylent-LLM shortenings all included passive voice but were otherwise grammatical. However, Soylent-LLM and zero-shot-ffv both could contain abrupt endings or transitions. This abruptness is due to separately editing parts of the input text.
Crowdsourced Soylent induced grammatical errors as well as abrupt transitions.
We found that zero-shot and Soylent-LLM both shorten by removing unnecessary phrasing and deleting examples and justifications. Zero-shot more successfully merges sentences than Soylent-LLM, while Soylent-LLM changes less text and more successfully retains the user's voice, rather than formalizing it.
Soylent-LLM's changes affect a larger portion of the text than Crowdsourced Soylent. In sum, Crowdsourced Soylent made 42 edits to phrases and 12 to entire sentences, while Soylent-LLM made 12 edits to phrases and 36 to entire sentences. 

\begin{figure*}
\centering
\includegraphics[width=\linewidth]{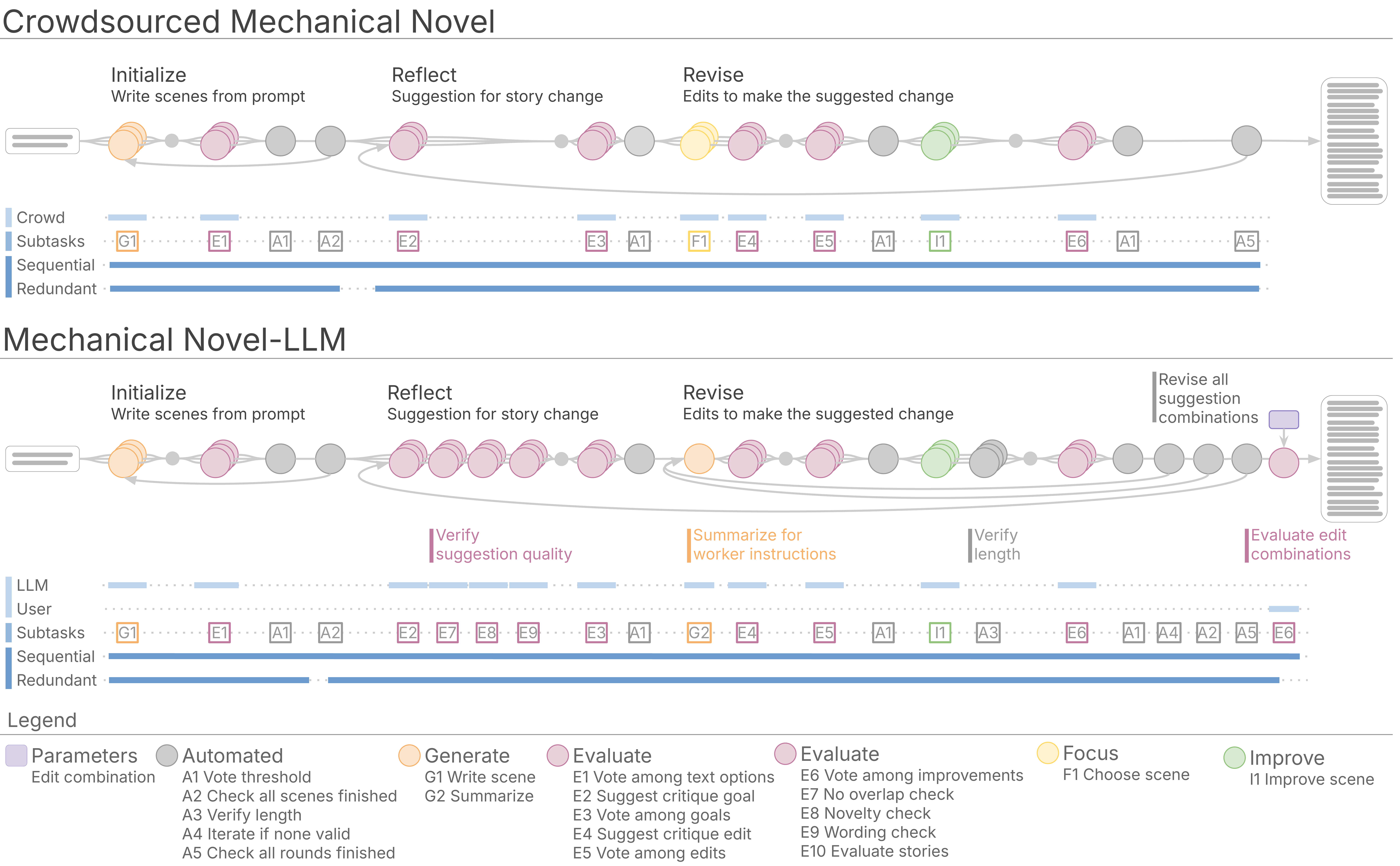}
\caption{Case study 3: Mechanical Novel~\cite{kim2017mechanical} for short story generation. We alter the tactics used in the crowdsourcing workflow Mechanical Novel for LLMs. These adaptations include a new summarization step, more validation steps, and iterations over different input subsets. These changes fit the task within context limits, encourage diverse outputs, and enable control of plot elements.} 
\Description{Figure shows two flow charts, one for crowdsourcing Mechanical Novel and one for LLM chain Mechanical Novel. Both follow the workflow described in the main text. Several adaptations are highlighted in the LLM chain: 1) additional validation steps in the reflect part of the workflow, 2) an automated length validation step in the revise part of the workflow 3) extra iterations through the scenes of the text as well as with different seeded texts.}
\label{fig:case-mechnov} 
\end{figure*}

\subsubsection{Workflows have higher costs than zero-shot prompting}
\label{sec:case-soy-cost}

We measure the time it takes to generate the outputs, as well as the number of API calls to the model. The number of API calls varies depending on the length of the input text and the number of items that are selected at each step.

Soylent-LLM takes much longer (mean: 105.44 sec., std: 60.16) than the zero-shot (mean: 9.09 sec., std: 2.92), zero-shot-target (mean: 13.65 sec., std: 5.47), and zero-shot-ffv (mean: 43.55 sec., std: 23.16) baselines. Crowdsourced Soylent takes the longest with a reported range of 2,640--29,340 sec. (44-489 min.)~\cite{bernstein2010soylent}.
Compared to the one call to the model for the zero-shot baselines, one Soylent-LLM run averages 127.11 calls to the model (std. 63.56). 
Crowdsourced Soylent reports incorporating 158--362 crowdworkers.

\subsubsection{Chaining demonstrates tradeoffs with Chain of Thought}
\label{sec:case-soy-cot}
\emph{Zero-shot-ffv} is a Chain of Thought (CoT)~\cite{wei2022chain} baseline, in which the LLM is prompted to use the steps of the Find-Fix-Verify chain all in one prompt-response set. 
As previously discussed, \emph{zero-shot-ffv} follows the quality trends of Soylent-LLM more closely than those of the \emph{zero-shot} baseline. Compared to Soylent-LLM, the \emph{zero-shot-ffv} baseline's quality and cost are lower. 
Therefore, the relative utility of doing a chain or a CoT depends on the restrictions on cost and the importance of high quality. 
We did not do a CoT baseline for the other two case studies, as context lengths did not enable the chain to be represented in one prompt and response. Therefore, another benefit of chains over CoT is the capability to do tasks and follow logical flows outside the context limit of an LLM. These restrictions will change as the context limits of LLMs grow.

%% file: sections/05.3_mechnov.tex
\subsection{Case study 3: Short story generation}
\label{sec:case-mechnov}

The final case study concerns story generation given a short text prompt.
This task lies on the open-ended \emph{subjectivity} side of the spectrum, as short stories have a wide design space of possible plots, characters, messages, imagery, and lessons.
In this task, \emph{user guidance} of the story involves higher level control of different plot elements, with the ability to see counterfactuals (e.g., if a story element is included or not).
We implement a workflow that, compared to zero-shot approaches, enables this exploration more precisely, while also creating stories that are preferred by crowdsourced raters for imagery, originality, and style.

\subsubsection{Workflow design}
\label{sec:case-mechnov-workflow}

We adapt the Mechanical Novel (MN)~\cite{kim2017mechanical} workflow for this task. This workflow involves three main steps (Figure~\ref{fig:case-mechnov}). The first \emph{generate} subtask ``initializes'' a story with multiple scenes based on a given story prompt (e.g.,~``Malcolm finds himself alone in a runaway hot air balloon and
accidentally travels to a city in the sky.''). Then, the ``reflect'' and ``revise'' steps are iterated five times in a loop. In the ``reflect'' step, actors consider what changes might helpfully adjust the story.
In the ``revise'' step, actors implement those changes.
Each step has subtasks for parallel \emph{generation} of suggestions and edits, as well as \emph{evaluate} subtasks voting for the best generation. Crowdworkers additionally \emph{focus} editing attention on the most important scenes.

We adapted the Crowdsourced MN to MN-LLM by incorporating fine-grained \emph{user guidance} with manipulatable outputs.
At the conclusion of the workflow, instead of receiving just one or more separate short stories, the user can mark check-boxes on an interface selecting different combinations of edits to view stories with different qualities and compare counterfactuals on editing decisions. 
To enable these outputs, we run the ``revise'' stage on texts of all combinations of suggestions from the ``reflect'' stage.

Beyond introducing control of story elements, the largest challenge was managing token context limitations.
Getting \emph{response diversity} with iterative improvement on texts often increases the text length~\cite{little2010exploring}.
To account for context length, we generate stories with four scenes rather than six and \emph{iterate} through scenes rather than informing actors to select scenes. In our design space this change maps to using a \emph{redundant} iterative architecture rather than a \emph{sequential} architecture with a \emph{focus} subtask.
At the beginning of the ``revise'' step, we add a \emph{generation} subtask that summarizes the current story to provide abbreviated context in the \emph{worker instructions}.
We included a hard-coded \emph{validation} step with an \emph{evaluate} subtask to disqualify edits that were more than 1.5 times the input text length. Unfortunately, this disqualified many highly descriptive texts and sometimes meant that no edits were made because all suggestions were too long.

\input{tables/mn-control}
\input{tables/mn-suggs}

Our initial pilots generated editing suggestions that were very generic (e.g.,~``Add more detail to the story'') and were repetitive across iteration rounds.
To improve the \emph{diversity} of suggestion outputs, we tried changing the \emph{worker instructions} to be contrastive by providing previously used ``banned'' suggestions. 
However, the outputs anecdotally seemed more likely to follow those of prior rounds after this change.
We hypothesize that the length of the context and request to do two tasks in one (generate a suggestion and avoid the prior suggestions) led to a lack of diversity. 
Instead, we add several \emph{validation steps} to promote suggestion diversity and quality. This \emph{subtask} granularity worked better.
In these \emph{evaluate} subtasks, we prompted the LLM to ensure that 1) the suggestion did not overlap with prior suggestions or two seeded generic suggestions (\emph{``Generic recommendation to introduce conflict.'' and ``Generic recommendation to add detail.''}), 2) the suggestion has not already been implemented in the story, and 3) that the suggestion is worded as a suggestion, not new text for the story.

\subsubsection{Evaluation setup}
\label{sec:case-mechnov-eval}
To evaluate MN-LLM, we use the story prompts and evaluation procedure from the original paper~\cite{kim2017mechanical} and compare MN-LLM's results to those of two baselines.

\input{tables/mn-amt}

\vspace{0.5em}
\noindent\textbf{Dataset.}
Our dataset consists of the five short story prompts in the Mechanical Novel paper~\cite{kim2017mechanical}.

\vspace{0.5em}
\noindent\textbf{Baselines.}
The \emph{zero-shot} baseline method prompts the model to generate five suggestions, then write one story that incorporates these suggestions. 
The \emph{zero-shot-combo} baseline method prompts the model to generate five suggestions, then write different versions of a story based on combinations of those suggestions.

\subsubsection{Workflows enable control of plot elements}
\label{sec:case-mechnov-control}

We assess each output to evaluate the number and quality of suggestion combinations that emerge from MN-LLM and zero-shot-combo.
To measure quality, we calculate the length of texts, as well as the precision and recall of the included suggestions.
To measure precision, we calculate what percentage of included suggestions were supposed to be included.
To measure recall, we calculate what percentage of suggestions that were supposed to be included were included.
As MN-LLM is iterative, some later suggestions reference elements initialized from earlier suggestions, so when coding, we consider the parts of each suggestion that differ from one another.

As shown in Table~\ref{tab:mn-control}, MN-LLM generates more, longer, and higher quality suggestion combinations than the zero-shot baseline. MN-LLM does not always generate the 33 story options afforded by the initial story and combinations of 5 suggestions, as sometimes no generated suggestions passed our validation steps, or all generated fixes were too long. However, it still generates more than 20 options on average, in comparison to zero-shot's average of 3.8.

\subsubsection{Workflows generate a wider variety of suggestions}
\label{sec:case-mechnov-suggs}

Without knowing the condition, we coded the story suggestions in the MN-LLM and both zero-shot conditions by theme.
MN-LLM suggestions were more varied in type than the zero-shot suggestions (Table~\ref{tab:mn-suggs}). The MN-LLM validation steps removed nearly half of the generated suggestions and most often voted for adding plot, characters, or emotional details as the final suggestions.

\subsubsection{The MN-LLM workflow improves imagery, originality, and style}
\label{sec:case-mechnov-quality}

Following the evaluation methods of the MN paper, for each story prompt we gave the output from MN-LLM and the zero-shot baseline, in randomized order, to 200 crowdworkers on Amazon Mechanical Turk using EasyTurk~\cite{krishna2019easyturk}.
We asked crowdworkers to vote between stories from the Soylent-LLM and zero-shot baselines along seven story writing dimensions~\cite{kim2017mechanical}.
We paid crowdworkers \$1.50 USD, the equivalent of \$18 per hour. 

The Mechanical Novel workflow improved the quality of the output narratives, although sometimes at the expense of a concise plot.
We find that MN-LLM stories were more highly rated along the Imagery, Originality, and Style, and Overall dimensions. Zero-shot-generated stories gained more votes in the Coherency, Plot, and Technical dimensions (Table~\ref{tab:mn-amt}).

\subsubsection{Workflows can struggle with coherence}
\label{sec:case-mechnov-coherency}

Without knowing the condition, we also reviewed the MN-LLM and zero-shot conditions for plot irregularities. 
Long-form plot coherence of creative LLM outputs is difficult to achieve~\cite{mirowski2023co,yang2022re3}, and we find some plot irregularities in the MN-LLM outputs. Three stories revealed the same information multiple times. Two plots had disjointed elements, one in which the same pronoun ``she'' referred to two characters in confusing ways and one with a time and location skip due to a scene that was never selected for editing. 
Such plot irregularities occur in the MN-LLM outputs because of the iteration through scenes and the use of a story summary as context.
Both of these design choices were made to avoid exceeding LLM context limits.

\subsubsection{Workflows have higher costs}
\label{sec:case-mechnov-cost}

We measure the time and number of API calls required to run the workflow compared to running the baselines.
Compared to zero-shot (mean: 21.88 sec., std: 5.89), MN-LLM takes much longer in seconds to complete (mean: 1,452.97 sec., std: 930.26). 
Compared to zero-shot's one call, MN-LLM makes 2,398 LLM calls on average. The crowdsourced Mechanical Novel paper does not report latency or the number of crowdworkers~\cite{kim2017mechanical}.

%% file: tables/mn-control.tex
\begin{table}[t]
\caption{Controllability of story plot elements with MN-LLM and zero-shot baselines.  We find that the MN-LLM workflow creates more combinations of long story options. Both approaches incorporate intended story elements at similar rates (Recall), but the zero-shot baseline more often includes unintended story elements (Precision).}
\label{tab:mn-control}
\centering
\resizebox{0.85\linewidth}{!}{
\begin{tabular}{lll} 
\hline
                                & Zero-shot     & MN-LLM           \\ \hline
Avg. \# combinations & 3.80          & 20.40         \\
Avg. length in words     & 276.38  & 642.93 \\
Precision                       & 77.05 & 97.74\\
Recall                          & 88.68 & 90.00       \\ \hline
\end{tabular}
}
\end{table}

%% file: tables/mn-suggs.tex
\begin{table}[t]
\caption{MN-LLM and zero-shot generate suggestion types for story elements. The MN-LLM workflow first generates a suggestion, which must pass a series of verification steps. The workflow then votes on valid suggestions to decide which to incorporate. At all stages, MN-LLM provides a wider variety of story suggestions than zero-shot.
}
\label{tab:mn-suggs}
\centering
\resizebox{\linewidth}{!}{
\begin{tabular}{lllll}
\hline
Suggestion type                            & Zero-shot & MN-LLM & Passed & Voted \\ \hline
Add plot events                      & 25       & 30     & 13     & 6      \\
Introduce character                  & 14       & 18     & 13     & 3      \\
Introduce significant object         & 6        & 1      & 1      & 0      \\
Add plot twist                       & 4        & 11     & 10     & 3      \\
Add detail to interactions & 1        & 6      & 3      & 1      \\
Add emotional detail                 & 0        & 17     & 8      & 3      \\
Add background details               & 0        & 8      & 5      & 0      \\
Add a different perspective          & 0        & 5      & 2      & 2      \\
Add a moral lesson                   & 0        & 2      & 1      & 1      \\
Add story detail                     & 0        & 2      & 1      & 0      \\ \hline
\end{tabular}
}
\end{table}

%% file: tables/mn-amt.tex
\begin{table*}[t]
\caption{Crowdworker preferences between MN-LLM and zero-shot baseline generated stories. More crowdworkers voted for MN-LLM stories in terms of imagery, originality, style and overall preference.
However, crowdworkers judged the zero-shot stories to have a more coherent and complete plot, reflecting challenges with outputting coherent answers from workflows.}
\label{tab:mn-amt}
\centering
\resizebox{0.8\linewidth}{!}{
\begin{tabular}{llllllll}
         & \textbf{Imagery} & \textbf{Coherency} & \textbf{Plot} & \textbf{Originality} & \textbf{Style} & \textbf{Technical} & \textbf{Overall} \\ \hline
MN-LLM   & 124              & 97                 & 93            & 137                  & 118            & 97                 & 112              \\
Zero-shot & 76               & 103                & 107           & 63                   & 82             & 103                & 88              
\end{tabular}
}
\end{table*}

%% file: sections/06_discussion.tex
\section{Discussion}
\label{sec:disc}

The approach of decomposing a task into a series of subtasks is conceptually the same for both crowdsourcing workflows and LLM chains. Through overlap in the literature and our case study observations, we find that the same \textbf{objectives} and \textbf{tactics} apply to both, with the exception of the choice of workers. We also find that the \textbf{strategies} found in crowdsourcing apply to chaining, as do their impact on \textbf{objectives}. 
Despite these similarities, LLMs and people have different abilities, strengths, and weaknesses, so we sometimes need different approaches for employing \textbf{tactics} to effectively implement the same \textbf{strategies}. 

\textbf{Thus, strategies and their effect on objectives transfer directly from crowdsourcing to LLMs. However, the tactic designs to support strategies can vary significantly and could benefit from improved development and testing processes.}
Understanding how to effectively use \textbf{tactics} to implement \textbf{strategies} is not only a more focused goal than the connection between \textbf{tactics} and \textbf{objectives}, but it is also more generalizable. For example, learning how to effectively get LLMs to validate their own outputs would be applicable to many different tasks and objectives.

In this section, we discuss further the similarities and differences between LLM chains and crowdsourcing workflows, lessons that exploring this intersection has taught us about using \textbf{tactics} to support \textbf{strategies} in LLM chain design, and unknowns that mark promising avenues for future discovery.

\subsection{LLM stochasticity is different than human diversity}

One of the key benefits of crowdsourcing workflows is that they incorporate multiple people, such that crowdworkers can fill in each other's deficits~\cite{huang2017supporting, nebeling2016wearwrite, kim2016storia, zhang2012human}. 
Therefore, requesting multiple crowdworkers for the same or sequential subtasks inherently leads to differing outputs and considerations.
The diversity of perspectives happens inherently by querying different actors but is limited to available crowdworkers and thus affected by demographic and geographic disparities on crowdsourcing platforms~\cite{difallah2018demographics}.

LLMs also have the capability to output different responses, as their training data includes information produced by many people on the internet~\cite{bender2021dangers}.
To induce diversity, prior work has tried methods like adjusting prompts, temperature, and model types~\cite{parameswaran2023revisiting, park2022social, du2023improving, arora2022ask}. In some cases, one can automatically calculate the optimal mix of LLMs for a task~\cite{Madaan2023AutoMixAM}.
Our case studies suggest that LLMs need more variation in prompting and explicit intervention to get \emph{diverse responses}, although we did not test this systematically.
Intuitively, this makes sense because there is great diversity in the human population, but models are known to mimic the majority voice to minimize their objectives, ignoring marginalized opinions that are a smaller percentage of the training data~\cite{gordon_jury2022}.

Using a high temperature and prompting the LLM with a variety of different instructions can improve diversity somewhat but is unlikely to match that of crowds.
Although this difference between actors may at some level seem obvious, it impacts the tactics needed to effectively use diverse responses. 
Diverse responses are used both to source variety and to check agreement.
In crowdsourcing, the underlying tactic behind both of these goals is identical: run the same subtask with multiple actors. However, different tactics can better explicitly address these applications for LLMs.

\vspace{0.5em}
\noindent\textbf{Variety.}
Diverse responses are often used to support tasks with a variety of subjective outputs. 
Consider one step in Mechanical Novel that generates suggestions for updating the story. As this is an iterative workflow, we want a variety of story update suggestions both at each iteration and across iterations. Unfortunately, even with variety in generation prompt wording, we found a low diversity of suggestion content (Table~\ref{tab:mn-suggs} ``Zero-shot'' column). 
After repeated challenges in prompt engineering to increase the variety of suggestions, we eventually intervened with an added \emph{validation} step of multiple \emph{evaluate} subtasks to ensure diversity of story updates. This approach was effective, but inefficient, as it filters out many \emph{generated} responses. 
Role-based \emph{communicative} architectures are also promising in giving prompts extra power to increase variety. Although giving identity based roles should treated with caution~\cite{wang2024large}, giving other roles to LLMs could help induce a variety of responses~\cite{park2022social}.
Many roles in crowdsourcing workflows were based on organizational positions for relevant domains such as journalism~\cite{agapie2015crowdsourcing} or project development~\cite{valentine2017flash}.

To source variety, the best choice of \emph{redundant} architecture can differ when using crowdworkers or LLMs. 
Prior crowdsourcing work has found that parallel and iterative architectures source different degrees of diversity and quality of responses in crowdsourcing depending on the task~\cite{goto2016understanding, little2010exploring, kittur2011crowdforge, andre2014effects}. For example, iterative architectures have less variation and a higher average quality of outputs but have a lower maximum quality output~\cite{little2010exploring}.
Some of these tradeoffs will hold with LLMs, but differences may emerge as well.
For instance, some work suggests that, relative to crowdworkers who experience anchoring bias on iterative architectures, LLMs can perform divergent thinking to generate something explicitly different than provided examples~\cite{wu2023llms,wu2023autogen}.
Similar studies on LLM chains would better inform the effect of different redundant architectures on variety and quality.

\vspace{0.5em}
\noindent\textbf{Agreement.}
Crowdsourcing work has used agreement among crowdworkers as a proxy for confidence vs uncertainty~\cite{liem2011iterative,chang2017revolt}. This agreement can be used to \emph{validate} the quality of outputs, so understanding how agreement relates to quality is of critical importance. Here, we focus on agreement among outputs of identical tasks, but the difference between a different person validating the outputs of another and a model sequentially validating its own outputs should also be considered.

There are many potential methods for measuring confidence for LLMs, including a confidence score, checking consistency across small prompt variations, and querying multiple times for taking the majority answer~\cite{xie2023decomposition, parameswaran2023revisiting, liang2023encouraging, du2023improving}.
The effect of each of these approaches on the quality of answers is not well understood~\cite{xie2023decomposition, parameswaran2023revisiting, liang2023encouraging, du2023improving}. It is unclear the degree to which just the consistency in one LLM's outputs is a proxy for confidence and accuracy, versus consistency across outputs from varied prompts or models. 

This more tenuous connection between agreement and correctness may have some impact on chosen tactics.
For example, \emph{communicative} architectures in crowdsourcing were often based on the premise that agreement among people mattered, but that may not transfer to LLMs. Consider debate-based \emph{communicative} architectures. Some crowdsourcing studies only bring up debate if there’s disagreement~\cite{chen2019cicero, drapeau2016microtalk, chang2017revolt}, but LLMs may benefit more from inducing debate for all inputs, rather than if two outputs disagree.

Therefore, the differences between human diversity and LLM stochasticity impact the design of \emph{redundant} and \emph{communicative} architectures to support divergent thinking or agreement.

\subsection{LLMs validate response quality less reliably than crowdworkers}

\emph{Validation} in crowdsourcing workflows often involves asking a different actor to verify if errors or desirable qualities exist in outputs. However, both the literature and our case studies show mixed results on the effectiveness of this strategy for chaining. 

While some studies show that prompting a model to reflect can help achieve higher accuracy outputs for \emph{objective} tasks~\cite{madaan2023self,xie2023decomposition,gero2023self,yao2023tree}, others caution against the same model validating its own outputs, as LLMs exhibit high confidence for incorrect prior outputs~\cite{liang2023encouraging,du2023improving}. 
Some approaches for more robust validation include: validating with a different \emph{LLM}~\cite{liang2023encouraging}, debate-based \emph{communicative} structures~\cite{liang2023encouraging,du2023improving}, and structural validation such as similarity across parallel \emph{redundant} streams~\cite{liem2011iterative}. One paper found that the ``ease of persuasion,'' or how quickly an LLM relented in a debate, was more indicative of factual accuracy than other measures of confidence~\cite{du2023improving}.

Through our case studies, we found several helpful practices for adapting validation steps to work with LLMs. 

First, validating ``quality'' is open-ended, so it is helpful to specify aspects of quality to check instead of using one general ``quality'' validation step.
Crowdsourced Soylent already split the validation into separately checking for ``grammar'' and for ``meaning'' changes~\cite{bernstein2010soylent}. However, we needed to make these checks even more targeted to adapt to LLM-specific flaws.
Crowdsourced Soylent did not need to check that the rewritten text was shorter, but we added a length-checking step to Soylent-LLM after struggling to ensure a shorter length through prompting alone. We also explicitly validated that the model did not change direct quotes, a nuance that is understood in crowdworkers' common sense. However, the subphrases from Soylent-LLM were for the most part fluent, deprioritizing the importance of the grammaticality check. 
In another example, crowdsourced Mechanical Novel involves one \emph{evaluate} step in which crowdworkers choose the ``best'' suggestion. We added three validation checks to MN-LLM before this step to account for common errors in chosen suggestions.
We explicitly check that LLM responses 1) do not overlap with prior suggestions, 2) are not already implemented in the story, and 3) are worded as a suggestion rather than as new text.
Although we did not specify any in our case studies, LLMs may require additional validation of output structures, as text-based responses cannot be restricted by interface design the way crowdworker interfaces can.

Validation should address specific elements of quality that are aligned with actor capabilities. In our case studies, we determined what elements of quality to validate by evaluating intermediate outputs in a trial and error process. Determining these qualities can be tricky, especially as the capabilities of LLMs rapidly change.
To determine what elements of quality require validation, intuition needs to be tempered by empirical testing and subtask evaluation. 

Even after the set of qualities under consideration has been determined, we found LLMs could still struggle with successfully validating these qualities.
To adapt to these issues, we used several approaches. 
First, when possible we made \emph{evaluate} steps programmatic rather than LLM-based. This approach is most reliable, but there are a limited number of domains where programmatic checks are feasible. Second, we continued to break qualities down into even simpler parts when needed to \emph{align subtask} difficulty with actor capabilities. Third, we \emph{simplify context} to the minimum needed. Considering the story suggestion quality checks above, we found that including multiple prior suggestions, or including the entire story when not necessary, reduced the quality of validation checks.

Some types of validation remain difficult for LLMs. Anecdotally, we found that some more abstract validation tasks, such as identifying changes in meaning, were challenging to enact with LLMs. In Crowdsourced Soylent, the ``Verify'' step evaluates if the fix changes the meaning of the statement. This requires nuance for what elements of the meaning are necessary.
Overall, supporting specificity and simplicity in validation steps were key for adapting these workflows to use LLMs.

\subsection{Chains for subjective tasks can use crowdsourcing approaches to enable users to navigate the space of outputs}

Some tasks have a large \emph{subjective} space of possible correct outputs, due to uncertainty or creativity, and could benefit from the user giving direct input. Under these conditions, it is helpful to draw on longstanding work on mixed-initiative user interfaces~\cite{horvitz1999principles}. Such an approach allows for parts of the workflow to be automated while others involve user input.

By enabling user initiative on workflow direction or outputs, users can navigate the space of outputs---either to explicitly take control or to learn through the process of exploration. 
For example, this could be helpful for determining classification label sets. Instead of predefining all possible rules in instructions, workflows can surface uncertain areas for users to make final decisions~\cite{chang2017revolt}. 
In more creative ventures like prototyping, incorporating user feedback to align the output to their vision can avoid technically correct but irrelevant outputs~\cite{kulkarni2014wish}.

User control through prompting is brittle~\cite{arora2022ask}, so workflows can enable more fine grained control. LLM chains can draw from user interaction methods from crowdsourcing workflows. Several large types of user interaction emerge from the literature. 

First, the user may act as a step in the workflow for validation or for clarifying context when triggered by ties~\cite{pietrowicz2013crowdband} or crowdworkers~\cite{kulkarni2012collaboratively, salehi2017communicating}. 
The main design decisions for this approach are deciding what and when the user will validate. The structure and timing of user interaction can matter, for example when supplying workers with needed context~\cite{salehi2017communicating}.

Second, the user and crowdworkers communicate through a shared interface in a relatively unstructured cycle of interactions while creating a shared creative output. Hurdles here include making user review of outputs \emph{easier} and reducing \emph{latency}~\cite{nebeling2016wearwrite, lasecki2015apparition}. Metaphorian is a chaining paper that follows a structured version of this in which users work on an interface to develop metaphors, but are limited to a few predefined actions~\cite{kim2023metaphorian}.

Third, a user may review options between iterations of the workflow~\cite{cheng2015flock, mirowski2023co}. This approach can help with exploring design spaces~\cite{park2022social, kulkarni2014wish, liu2023improving} and determining when the final quality threshold is met~\cite{willett2012strategies}. 

Fourth, a user may directly manipulate the output to see the breadth of user responses and to make final decisions about the output with less effort than doing the task directly~\cite{chang2017revolt,mohanty2019second,bernstein2010soylent,cheng2015flock}. This approach enables users to surface uncertainty in the data and express control over the final output. 

\begin{sloppypar}
  
Many chaining papers discuss a generic possibility of a user replacing steps or intervening if needed~\cite{schick2022peer, yang2022re3, wang2022iteratively, wu2023empirical}, but few actually implement this with people~\cite{park2022social, kim2023metaphorian, mirowski2023co}. We can see from crowdsourcing that there are many ways of incorporating the user that bring myriad benefits. In our case studies, we take direct manipulation from crowdsourcing and apply it to chaining, increasing the controllability of LLM outputs. 
\end{sloppypar}

\subsection{On-demand availability of LLMs reduces barriers to real-time crowdsourcing architectures}

Unlike crowdworkers, LLMs are available to be queried at all times with low \emph{latency}.
While, in one sense, this observation is obvious, it has deep implications on tactics.

The unpredictable availability of crowdworkers presented challenges for workflows in which crowdworkers must stay available for extended periods of time or must be called on-demand~\cite{bernstein2011crowds}.
While incentive schemes could provide some real-time response, incentive tuning and implementation added significant complexity~\cite{lasecki2013chorus, latoza2014microtask, weir2015learnersourcing}. As a result, several crowdsourced workflows --- while achieving impressive results on paper --- were too complex for widespread deployment~\cite{chen2019cicero,valentine2017flash}. The availability of LLMs reduces the barriers to these workflow designs and their benefits.

Methods like retainer models that incentivize keeping crowds were important for crowdsourcing workflows that involve real-time interaction for users~\cite{bernstein2011crowds} and for \emph{communicative} workflows like debate in which actors must remain online as other actors respond to them~\cite{drapeau2016microtalk,chen2019cicero}. These workflows improve ~\emph{latency} for users and improve quality and ~\emph{accuracy} for objective tasks, respectively, but the challenges of retaining crowdworkers increased barriers in crowdsourcing. Similarly, for mixed-initiative interfaces, delayed crowdworker responses increased the \emph{latency} of interactions~\cite{lasecki2015apparition}. Reduced latency of worker responses can make user interactions more real-time.

In other workflows, such as Flash Organizations~\cite{valentine2017flash}, crowdworkers recruit other crowdworkers to complete subtasks that are dynamically created. Finding the right crowdworkers for the job took a median of 14 minutes with a light screening process and 15 hours with a more intense screening process~\cite{valentine2017flash}. In contrast, the Orchestrator LLM of the AutoGen~\cite{wu2023autogen} chain ``recruits'' with other calls to LLMs, which can happen flexibly and quickly~\cite{wu2023autogen}. 

Therefore, using LLMs as workers reduces barriers to using powerful \emph{communicative} workflow regimes.

\subsection{Instructions should provide the minimum necessary context}

Providing global \emph{context} to subtasks is widely agreed to be a critical element of subtask design and output coherence, but it can increase instruction complexity~\cite{wu2022promptchainer, salehi2017communicating, ambati2012collaborative, luther2015crowdlines}. Crowdsourcing research more broadly has studied the process of improving worker instruction, including by ensuring comprehension through tests~\cite{Liu2016EffectiveCA} and communicating misconceptions to requesters~\cite{Bragg2018SproutCT}. However, findings from the literature survey and case studies suggest that LLMs can respond poorly to long, complicated instructions~\cite{du2023improving, parameswaran2023revisiting, madaan2023self, wu2023llms}. This context limitation should be re-evaluated over time as LLM context windows grow.

The nuances of context inclusion appeared in our case studies. 
For example, we wanted to provide MN-LLM the entire short story thus far as context, as occurs in crowdsourced Mechanical Novel. To generate \emph{diverse} suggestions for improvement we gave the LLM previous suggestions as context for what it should avoid. However, adding this requirement to the overall prompt anecdotally increased the output's chance of anchoring to prior suggestions. Therefore, we made a separate \emph{evaluate} subtask that determined if the new suggestion matched previous suggestions out of context from the rest of the story.

We recommend designers take into consideration that LLMs may not adequately process all the nuances of complicated instructions. Simplifying the subtask size in turn simplifies instructions and, often, the amount of necessary context.

\subsection{LLMs need different, often simpler, subtasks}
Throughout our case studies, we often needed to break crowdsourcing \emph{subtasks} into multiple, more specified \emph{subtasks}. Multiple factors necessitated these changes: natural language is difficult to structure (e.g., thus, we query for one item category at a time rather than eight in Cascade), LLMs have difficulty balancing multiple requirements in a subtask~\cite{wu2023llms, madaan2023self, hsieh2023tool} (e.g., in response, we split Soylent's ``Fix'' step into separate edit, merge, and delete queries), and LLMs do not understand length-related constraints (e.g., to address this, we add a hard-coded \emph{validation} step after a \emph{generate} subtask in Mechanical Novel to check for length). 

In our case studies, LLMs performed better on highly specified subtasks for chains rather than on fewer, more complicated subtasks. However, this may change as newer, more powerful LLMs are introduced.

\subsection{Chains can be useful despite increased cost}

Although chains bring many improvements over zero-shot performance in our case studies, they also incur sizeable increases in expense and latency. Without thoughtful design, AI support tools can make easy tasks easier and hard tasks harder~\cite{simkute2024ironies}.
When should one choose between chaining or a zero-shot approach for potential applications?
We consider the effects of cognitive load, the degree of user agency, the interruptibility of the overall workflow, and limitations in model capabilities.

The increased effort and latency of a chain's runtime can be offset by quicker and less effortful interpretation of the outputs. For instance, in writing editing, it is easier to review highlighted minor edits rather than entirely new large blocks of text.

Chains may be helpful, despite higher costs, when the goal is spurring users' innovation, learning, and design, rather than giving responses outright. For example, 
Metaphorian allows users to create scientific metaphors in an interactive interface~\cite{kim2023metaphorian}. The goal of such a system is not only to output the metaphors but also to be exploratory and spur thinking. Compared to zero-shot creation of metaphors, the paper finds users retain a greater sense of agency, metaphors increase in originality and coherence, and user satisfaction is higher when using an interactive chain. 

The impact of increased latency on the utility of chains lessens when the broader workflow can be interrupted without major disruption. Chains that do not require user guidance, or only require it in spaced-out intervals, can run in the background while the user performs other tasks. Some tasks, such as in-line coding assistance, are not conducive to this type of interruption. However, others, such as short-story generation, can incorporate a long runtime, thus reducing the impact of longer latency on the user experience. 

Finally, another reason to use chains despite the cost is if the task cannot be completed without a chain. At the moment, long-form writing and generating large code repositories are outside the capabilities of models and require a chain. It is worth noting that these capabilities will change over time. 
Thus, thoughtfulness about objectives and the broader task context can motivate the use of chains when a higher utility offsets increased costs.

\section{Future opportunities for chaining}

In this section, we highlight practices in crowdsourcing that we consider likely to be worthwhile to transfer to chaining. We encourage future work on these topics.

\subsection{Subjective tasks with dynamic and communicative architectures}
Crowdsourcing studies currently support a wider range of tasks for \emph{subjective} work with a wider range of approaches than chaining studies. We expect these approaches to transfer to chaining and bring myriad benefits.

Future work could attempt \emph{subjective} tasks that are under-explored in chaining, such as: prototyping~\cite{lee2017sketchexpress, valentine2017flash, kulkarni2014wish}, categorizing~\cite{huffaker2020crowdsourced, bragg2013crowdsourcing, chang2017revolt}, topic concept maps~\cite{liu2018conceptscape, hahn2016knowledge, luther2015crowdlines}, project development~\cite{mahyar2018communitycrit}, or more specific tasks like itinerary creation~\cite{zhang2012human}, researching a purchase~\cite{kittur2011crowdforge}, and conceptual blends~\cite{chilton2019visiblends}.

Crowdsourcing work also uses a broader range of \emph{communicative} and \emph{dynamic} architectures to support \emph{subjective} work. Giving actors roles during \emph{communication} can assist both crowdsourcing workflows~\cite{valentine2017flash, agapie2015crowdsourcing, retelny2014expert} and LLM chains~\cite{wu2023autogen,li2024camel}. In crowdsourcing work, some studies go beyond giving a role in instructions and instead create free-form role-based~\cite{nebeling2016wearwrite,valentine2017flash,zhang2023ecoassistant}, hierarchical~\cite{agapie2015crowdsourcing,retelny2014expert}, and state-sharing~\cite{lasecki2015apparition,lee2017sketchexpress,nebeling2016wearwrite,lasecki2013chorus} \emph{communicative} architectures. 

\emph{Dynamic} architectures provide more flexibility when contingencies arise~\cite{kim2016storia, valentine2017flash, kittur2011crowdforge,kulkarni2012collaboratively} and adapt to edge cases~\cite{nair2023dera,kulkarni2012collaboratively,chen2019cicero}. LLM chains rarely implement \emph{dynamic} architectures for \emph{subjective} creative work~\cite{yao2023tree,wu2023llms,kim2023metaphorian,zhang2023visar}.
Crowdsourcing work applies \emph{dynamic} architectures to a wider variety of \emph{subjective} creative tasks~\cite{lasecki2015apparition, retelny2014expert,chilton2019visiblends,kittur2011crowdforge} and includes more complex dynamic workflows such as multi-level map-reduce~\cite{kittur2011crowdforge} and dynamically choosing between workflows using their maximum expected value of utility~\cite{goto2016understanding}. Decision theoretic techniques, such as Partially Observable Markov Decision Processes, yield great efficiencies in crowdsourcing~\cite{dai2010decision,dai2013pomdp}, but have thus far barely been considered for LLM chaining~\cite{Madaan2023AutoMixAM}. 

We recommend pursuing a broader range of \emph{subjective} creative tasks with more \emph{communicative} (e.g.~establishing a hierarchy of roles) and \emph{dynamic} (e.g.~multi-level map-reduce) architectures.

\subsection{Integrating the user}
Although many chaining papers discuss involving humans into the chain hypothetically~\cite{schick2022peer, yang2022re3, wu2023autogen, wang2022iteratively, wu2023empirical}, few implement this user oversight~\cite{stretcu2023agile,toubal2024modeling}.
Our case studies showcase one example of interactions through directly manipulable outputs, but the varied ways of incorporating the user in crowdsourcing workflows could continue to inspire similar approaches in chaining.

Explorable interfaces surface disagreements to the user rather than resolving them in the workflow itself~\cite{willett2012strategies,chang2017revolt,cheng2015flock,mohanty2019second}. Worker-initiated guidance allows workers to define when to alert the user, focusing user attention only on what matters to the workers~\cite{salehi2017communicating,kulkarni2012collaboratively}. In \emph{subjective} tasks, quick iteration and choosing among uncertain options reduces the effort it takes for the user to prototype creative outputs and allows the user to guide ideation~\cite{nebeling2016wearwrite,lasecki2015apparition}. For example, unstructured text can be interpreted in many ways depending on an analyst's research questions, so recent work such as Concept Induction allows analysts to steer the process of LLM-aided text analysis~\cite{lam2024concept}.

We can also consider how our case study examples could be expanded to incorporate the user differently. With inspiration from the crowdsourcing workflow Revolt~\cite{chang2017revolt}, a user of Cascade-LM could review edge cases that only match one category label of two that were merged. Soylent-LLM could replace the verify step with a user, as the user may recognize subtle undesirable changes in fixed phrases that are difficult to pre-define through prompts. Mechanical Novel-LLM could implement the \emph{quality thresholds} strategy by having the user review after each iteration if the generation process should continue. The user could also give more oversight on the edit suggestions chosen. Incorporating the user in more ways reduces \emph{expense} by calling LLMs fewer times and by directing the calls of LLMs that do occur towards more important needs. However, it does require more \emph{effort} on the user's part and could increase \emph{latency} if the user is slower than the equivalent call to an LLM.

We recommend research into the opportunities and limitations of user guidance, especially for subjective tasks, as initial findings 
conflict on if using generative AI tools leads to divergent thinking~\cite{chiou2023designing} or design fixation~\cite{wadinambiarachchi2024effects}. 

\subsection{Studying chains more systematically}

\begin{sloppypar}
    
More systematic studies on optimal chain design would be valuable. Several studies already approach questions around model choice~\cite{Madaan2023AutoMixAM,du2023improving}, incorporating roles into prompts~\cite{wu2023autogen,park2022social}, and using self-reflection or debate to validate outputs~\cite{du2023improving, liang2023encouraging}. Although the outcomes may be different, systematic crowdsourcing studies could provide inspiration for some of these explorations. Prior crowdsourcing studies investigate the effect of context timing, architecture, and content on task output~\cite{luther2015crowdlines,salehi2017communicating}. Others measure tradeoffs of iterative and parallel workflows (e.g.,~effect on latency~\cite{luther2015crowdlines}, quality averages and variance~\cite{little2010exploring,kittur2011crowdforge,goto2016understanding}, risk of cascading errors~\cite{goto2016understanding}) in various task contexts, as well as good practices for each (e.g.,~appropriate number of branches or iterations~\cite{andre2014effects,huffaker2020crowdsourced}).
\end{sloppypar}

By specifying a strategy of interest, the design space can orient studies for understanding optimal tactical design decisions. For example, studying debate designs could look at debate to establish \emph{adaptability} for tasks with many edge cases, debate to \emph{validate} outputs, or debate to encourage a \emph{diversity} of creative ideas for a \emph{subjective} task. Evaluating the use of tactic designs within the context of specific strategies will lead to more generalizable findings than evaluating solely on task performance.

We recommend systematic tests for tactical design decisions informed by prior crowdsourcing work, objectives, and strategies.

\subsection{LLM workflow design processes are unstructured and design support is needed}
\label{sec:disc-designprocesses}

Since LLMs are cheaper and faster to query~\cite{Terwiesch2023}, there are more opportunities to find optimal designs through systematic testing. 
While the lower barrier to entry may allow LLM chain designers to iterate more quickly, it doesn't guarantee that they focus their attention on the right things while iterating. Since LLMs display unpredictable outputs in response to low-level changes, designers may end up fixating on minor tweaks like prompt rewordings rather than exploring the design space of strategies and tactics. To avoid potential design fixation that tunnels into technical detail rather than exploring higher-level design directions, LLM chain designers would benefit from tools that scaffold their design and iteration process.
Although running an LLM chain is cheaper and faster than running a crowdsourcing workflow, the design process is unstructured. There is a need for design support.

We see many promising directions to help build chains efficiently and effectively, given that iterations are often based on extensive qualitative analysis and enlightened trial-and-error.

As there are many overlapping tactics among workflows, the flexibility and reusability of modules can support the development of future workflows~\cite{josifoski2023flows}. A repository of ``battle-tested'' workflow operators could allow for sharing of quality parts for reuse in new workflows. Such a platform could include operator-level test suites to assist with validation that would monitor operator output quality in new content areas or with newly released LLMs. The platform design could draw on crowdsourcing to create such test suites. Additionally, corresponding workflow design tools could suggest different options to achieve a goal by collecting multiple approaches for sub-parts of the workflow.

Other work engineers more optimal LLM prompts by iteratively generating and scoring test prompt candidates using mathematical or natural language optimization functions~\cite{zhou2022large,yang2023large}. Future work could adapt these approaches within the context of chaining, for example to optimize for diverse responses.

Comparing many workflow designs can be difficult because of the requirement that the output of a previous step match the input of the next step. Therefore, removing or adding steps can be a non-trivial process. More formalized input and output indicators, potentially with matching prompts, could facilitate easier chain development. This approximates the approach of Flash Teams which documents the inputs and outputs of steps~\cite{retelny2014expert}.

Sketching aids practitioners in rapidly instantiating ideas to explore the possible solution space rather than committing too early to one design direction~\cite{buxton2010sketching, dow_parallelPrototyping,yang_sketchingNLP}. Incorporating sketching could aid LLM chain designers in arriving at better solutions, rather than getting stuck in local optima. 
Sketches are quick, but they also should be low-fidelity and minimal in detail. While LLMs afford quick instantiation, they produce high-fidelity output that can distract designers from their primary prototyping goals~\cite{jiang_PromptMaker, lam2023modelSketching}.
Opportunities to support this designing perspective include providing interfaces amenable to rapid chain construction and reconfiguration~\cite{wu2022promptchainer,arawjo2023chainforge} and suggesting strategies and tactics when a user may be tunneling on technical details.

We recommend continued work on chain design support tools such as repositories of workflow operators, prompting optimization methods for chaining, formalized input-output indicators, and sketching-inspired workflow design approaches.

%% file: sections/07_conclusion.tex
\section{Limitations}

Our work addresses crowdsourcing workflows and LLM chains, but it is limited in scope beyond these fields and in discussion of broader impacts. Additionally, our recommendations are contingent upon an imperfect understanding of LLM capabilities, and we support the utility of our case studies through theory instead of a user study.

\vspace{0.5em}
\noindent\textbf{Scope.} 
Our work targets LLM chain design by adapting crowdsourcing workflow techniques. However, there are opportunities to look beyond LLMs to multi-modal models and other tools as the workers~\cite{gupta2023visual,suris2023vipergpt,liu2022opal} and beyond crowdsourcing to other task decomposition methods as a source of inspiration~\cite{anya2015bridge,andreas2016neural,mao2019neuro}. 

Additionally, there may be limitations to the scope of work covered by our review of the LLM chaining literature. Since LLM chaining is a relatively new area, researchers have not converged upon established terminology as in the crowdsourcing domain. Our keyword search may not have captured some work that did not use one of our keyword search terms.

\vspace{0.5em}
\noindent\textbf{Broader impacts.}
We focus on the ability of workflows to serve a beneficial role in supporting various tasks. However, there are potential risks in using LLMs and crowdworkers as support tools that we do not explicitly consider. Several papers in the design space cite concerns around a loss of jobs and unintended consequences of using LLMs, especially for creative tasks~\cite{park2022social,mirowski2023co,wu2023autogen}. Crowdsourcing work similarly raised concerns around replacing expert workers and around labor rights for crowdworkers~\cite{kittur2013future}. Broader discussions about privacy~\cite{wu2023autogen,peris2023privacy}, ownership~\cite{kim2023metaphorian,mirowski2023co}, and perceived sense of agency we also leave to future work. Finally, although workflows can improve output quality, results will still be imperfect, so errors from implemented chains could still have harmful impacts~\cite{eggmann2023implications,greshake2023not}.

\vspace{0.5em}
\noindent\textbf{Rapid changes of LLM capabilities.} The capabilities of LLMs are rapidly changing, so researchers' understanding of and subsequent recommendations to work with these capabilities may fall out of sync with new model developments. We recommend future work to monitor the efficacy of chain designs as new models are released, with potentially differing capabilities and failure modes.

\vspace{0.5em}
\noindent\textbf{Lack of user evaluation.} 
We build on longstanding HCI theory and practice that demonstrates the benefit of direct manipulation interfaces~\cite{shneiderman1982future,hutchins1985direct}.
However, our work does not conduct a user evaluation to explicitly test the benefits and drawbacks of our direct manipulation case study implementations. Future work may more deeply explore how users interact with LLM chains via direct manipulation and whether these interactions positively impact user control and task outcomes.

\section{Conclusion}
\label{sec:conclusion}

By revisiting the established field of crowdsourcing workflows, we find inspiration and direction for the nascent area of LLM chaining. 
To facilitate adaptation, we contribute a design space for LLM chain designers that incorporates findings from the crowdsourcing workflow and LLM chaining literatures. This design space covers the objectives supported by and the tactical building blocks of prior workflows and chains. We also surface strategies employed by designers to use tactics to support objectives. 
Through three case studies, we build an understanding of how crowdsourcing workflows can be adapted for LLM chains. 

Finally, we synthesize our findings to discuss tactical design decisions for LLM chains through the context of their comparison with crowdsourcing workflows. 
We find LLM chains need additional support in subtask design and validation due in part to differences between LLM stochasticity and human diversity. On the other hand, the comparatively low cost and latency of LLM calls allow for faster iteration and for workflow techniques that were studied but never practical with human crowds. We hope these findings guide future work towards more approaches to user guidance and subjective tasks, the systematic discovery of good practices for tactic design, and artifacts to support the chain design process.

LLM chains present remarkable flexibility, which comes along with both promising capabilities but also substantial complexity. Our work helps LLM chain designers navigate this vast design space by building on the established field of crowdsourcing. 

%% file: sections/08_acks.tex
\begin{acks}
We thank Michael Bernstein and Joy Kim for providing outputs from the original Soylent and Mechanical Novel papers. Thank you to Clement Zheng for your input on creating corpus tables similar to prior work~\cite{bae2022designing}. We also thank the Allen Institute for Artificial Intelligence for access to Amazon Mechanical Turk resources. We thank OpenAI for offering research credits to the GPT-3.5 API. This research was supported in part by NSF award IIS-1901386, ONR grant N00014-21-1-2707, and the Allen Institute for Artificial Intelligence (AI2).
\end{acks}

%% file: sections/09_appendix.tex
\appendix

\section{Paper categorization criteria}

To show the design space's completeness in representing the existing literature, we revisited each paper and categorized it within the design space. We describe the nuances of that process here.

We record the year of publication, such that if we collected a paper on arXiv in 2023 and it was published in a conference in 2024, we record the year as 2024. 

We show the quality-cost objective spectrum as four columns, in order to show the increasing and decreasing of quality and cost. A paper may both increase and decrease quality or cost at once (e.g., the output is higher quality, but there are reported quality issues with cascading errors). We count the ability to do a task at all as an increase in quality, and an equal match with baseline capabilities as neither an increase nor a decrease. Another challenge for comparison is that the baseline differs across papers. For instance, some papers compare quality to expert outputs, others to novice crowdworkers, and others to zero-shot LLM outputs. We label changes in quality and cost as they are discussed in the paper, rather than use a consistent cross-paper standard. For instance, if the chain could logically be expected to be more expensive than a single call to an LLM, but this is not explicitly discussed in the paper, we do not mark it as increasing costs. We take this approach to avoid speculation and to show trends along with what effects are considered worth reporting. 

We used our discretion in classifying papers as only \emph{subjective}, if the main focus of the task is highly subjective, such as with creative writing.  There may still be \emph{objective} elements, such as wanting a generated story to be grammatically correct.  This subjectivity in the categorization itself is why we express it as a spectrum in the design space. 

In the design space text, we mention some automated elements that perform similar functions (e.g., rule based heuristics for \emph{evaluate}, sourcing outside material for \emph{generate}, and algorithmic \emph{merges}). We include these in the text descriptions to reference how workflows may use \emph{actors} or automated elements to complete tasks. However, we do not label a paper as including a subtask of these types if it is only deterministically automated. 

Finally, if there are multiple workflows tested in the paper, we include takeaways and elements from all in the coding.